\documentclass[12pt,a4paper]{article}
\usepackage{etoolbox}
\usepackage{dcolumn}
\usepackage{dcolumn,subfig}
\usepackage{setspace}
\usepackage{bm}
\usepackage{dirtytalk}
\usepackage{blindtext}
\usepackage{parskip}
\usepackage[bottom]{footmisc}
\usepackage[
colorlinks=true,
urlcolor=blue,
linkcolor=red,
citecolor=blue
]{hyperref}
\usepackage{cite}
\usepackage[dvipsnames]{xcolor}
\usepackage[export]{adjustbox}
\usepackage{xcolor}
\usepackage{color,soul}
\usepackage{amsmath,amsfonts,amssymb}
\usepackage{rotate,wrapfig}
\usepackage{color}
\usepackage{float}
\usepackage{epsfig,epsf}
\usepackage{bm}
\usepackage{dcolumn,subfig}
\usepackage[top=1in, bottom=2.5cm, left=2cm, right=1.5cm]{geometry}
\usepackage{array}
\usepackage{relsize}
 \usepackage{color}
\usepackage{multirow}
 \usepackage{caption}
 \usepackage{float}
 \usepackage{relsize}
 \usepackage{cancel}
 \usepackage{makecell}
 \usepackage{placeins}
 \usepackage{graphicx}
 \usepackage{authblk}
 \usepackage[normalem]{ulem}
 \usepackage{comment}
\def\Babar{{\mbox{\slshape B\kern-0.1em{\smaller A}\kern-0.1em B\kern-0.1em{\smaller A\kern-0.2em R}}}}
\renewcommand*{\thefootnote}{\fnsymbol{footnote}}

\renewcommand*{\thefootnote}{\fnsymbol{footnote}}
\title{\bf Prospects of Gluino searches in multi-lepton channels in the light of ongoing LHC RUN-III}

    \author[1]{Abhi Mukherjee \thanks{\href{abhiphys18@klyuniv.ac.in}{abhiphys18@klyuniv.ac.in}}}
    \affil[1]{Department of Physics, University of Kalyani, Kalyani 741235, India}
    \author[2]{Saurabh Niyogi \thanks{\href{saurabhphys@gmail.com}{saurabhphys@gmail.com}}}
    \affil[2]{Gokhale Memorial Girls' College, 1/1 Harish Mukherjee Road, Kolkata 700 020, India}
    \author[3]{Sujoy Poddar \thanks{\href{sujoy.phy@gmail.com}{sujoy.phy@gmail.com}}}
    \affil[3]{Department of Physics, Diamond Harbour Women's University, Diamond Harbour Road, Sarisha, South 24 Parganas, West Bengal 743368, India}
    \author[1]{Jyoti Prasad Saha \thanks{\href{jyotiprasadsaha@gmail.com}{jyotiprasadsaha@gmail.com}}}
\begin{document}

\maketitle

\begin{abstract}

    This study investigates the prospect of discovering strongly interacting gluinos in different multi-lepton channels with lepton multiplicity greater than or equal to 2 at LHC RUN-III, considering several pMSSM scenarios. The effectiveness of the Multivariate Analysis (MVA) method with the Boosted Decision Tree (BDT) algorithm is explored in order to obtain a better significance for different models. Promising results are obtained for the 3-lepton channels, indicating that the use of MVA methods can improve the sensitivity of the search for gluinos at LHC RUN-III. The study probes the multi-lepton signatures arising from gluinos via intermediate eweakinos and sleptons at an early stage of the LHC RUN-III. The heavier eweakinos can give rise to three or four lepton signals in which the squark hierarchy between L and R types plays a crucial role. The study considers two sets of benchmark points that satisfy all the collider constraints obtained from the LHC RUN-II data. Moreover, these sets of benchmark points are mostly consistent with WMAP/PLANCK data and the muon (g-2) constraint. The corresponding results from the MVA technique demonstrate that even for an integrated luminosity of 270 $fb^{-1}$, the 5 $\sigma$ discovery prospect of $3l+ jets + \cancel{E_T}$ for $M_{\widetilde{g}}=1.8$ TeV in the wino type model is promising. The study also presents the other various models that may show up at the early stage of LHC RUN-III. The wino type models in the scenario where left squarks are light and right squarks are heavy, exhibit the best prospect of discovering gluinos in the multi-lepton channels in the LHC RUN-III experiment. This paper's findings provide crucial insights into the potential discovery of gluinos in multi-lepton channels.
    
\end{abstract}

\setcounter{footnote}{0}
\renewcommand*{\thefootnote}{\arabic{footnote}}

\section{Introduction}
Supersymmetry (SUSY) \cite{Nilles:1983ge,bagger1992theory,Lykken:1996xt,Martin:1997ns,Chung:2003fi,drees2005theory,Fayet:1974pd,Fayet:1977yz,Fayet:1978qc,Fayet:1978rb,Baer:2006rs} 
is the most promising beyond standard model (BSM) theory  which resolves many of the lacunae of the Standard Model (SM) \cite{Gildener:1976ih,PhysRevD.14.1667,Witten:1981nf,Dimopoulos:1982af,Dimopoulos:1995mi,Arkani-Hamed:1998jmv,Ellis:1990wk,Amaldi:1991cn,Jungman:1995df,Lahanas:2003bh,Freedman:2003ys,Roszkowski:2004jc,Bertone:2004pz,Olive:2005qz,Baer:2008uu,Drees:2012ji,Arrenberg:2013rzp}.  This elegant theory has been extensively studied for last few decades both theoretically and experimentally. Such theory is currently being probed through various search channels in the ongoing Large Hadron Collider (LHC) experiment. Unfortunately, no appreciable deviation from the SM predictions in the form of a statistically significant excess of events have been found. Therefore, the negative results from such searches during RUN-I \cite{aad2014search,ATLAS:2014kpx,ATLAS:2014eel,Aad:2013wta,CMS:2015adc,aad2015search,javurek2016search} and RUN-II \cite{Aad:2016eki,Aad:2016qqk,Aaboud:2016zdn,Aad:2016tuk,Aaboud:2017bac,Aaboud:2017ayj,Aaboud:2017dmy,Aaboud:2017vwy,Aaboud:2017hrg,Aaboud:2018ujj,Aaboud:2018mna,Aad:2019ftg,Aad:2020nyj,Aad:2021egl,Aad:2021jmg} of the LHC have imposed stringent bounds on masses of the sparticles (supersymmetric counterpart of the SM particles). 

Due to the large number of free parameters in SUSY models particularly, the soft SUSY breaking parameters \cite{Fayet:1977yc,Inoue:1983pp,Ibanez:1983di}, it is challenging to interpret the experimental results unless certain assumptions are made. For this reason, the  ATLAS/CMS collaborations interpret their results in terms of simplified models in which certain sparticles of relevance are considered in the decay topology keeping others decoupled. These assumptions may not always be possible to realize within the context of the Minimal Supersymmetric Standard Model (MSSM). In this study, we consider the phenomenological Minimal Supersymmetric Standard Model (pMSSM), a reduced version of MSSM with 19 free parameters.

Strong sector of the SUSY comprising of gluinos and squarks are the best probes for searching SUSY signals because of their large production rate. In RUN-II of the ATLAS experiment of the LHC, the gluino mass is excluded approximately up to 2.3 TeV in the channels involving $jets+\cancel{E_T}$ \cite{Aad:2020aze}, $1l+jets+\cancel{E_T}$ \cite{Aad:2021zyy}, and $2l+jets+\cancel{E_T}$\cite{ATLAS:2022zwa}, assuming simplified models. In these analyses \cite{Aad:2020aze,Aad:2021zyy,ATLAS:2022zwa}, all sparticles except for the gluino, lighter chargino (second lightest neutralino for ref. \cite{ATLAS:2022zwa}), and the lightest supersymmetric particle (LSP) (here the lightest neutralino) have been assumed to be decoupled. The bound has reached almost at the kinematic edge of the LHC. However, in ref. \cite{Mukherjee:2022kff}, the authors have discussed the effect of the inclusion of second lightest neutralino ($\widetilde{\chi}_2^0$) into the decay chain of gluinos along with the left and right squarks mass hierarchy for the final state comprising of $jets+\cancel{E_T}$ and $1l+jets+\cancel{E_T}$.  The inference of such study is that it is possible to bring down the bound on gluino mass by an appreciable amount. Compressed SUSY scenarios are the other examples where such relaxation of bounds are possible \cite{LeCompte:2011fh,Dreiner:2012gx,Bhattacherjee:2012mz,Bhattacherjee:2013wna,Cohen:2013xda,Mukhopadhyay:2014dsa,Low:2014cba}.

The next step would be to see the effects of the inclusion of heavier electroweakinos (in short, eweakinos) and sleptons/sneutrinos into the decay chain of gluino, thereby lengthening the decay cascade even further. It must be noted that the gluino searches have not been performed in more than two lepton final states. Therefore, the idea is to infuse more leptons in the final state coming from the decays of heavier eweakinos via the sleptons/sneutrinos. This obviously will generate more clean signal comprised of multi-leptons at the LHC environment. Thus the impetus of this work is to set the bounds on gluino in the multi-lepton final states and to explore the possibility of discovering it in the latest LHC RUN-III at 13.6 TeV of center-of-mass energy.

It is worth mentioning that the study of direct production of heavier eweakinos have been performed in \cite{Chakraborti:2017vxz} via multi-lepton channel using LHC RUN-II data. Additionally other phenomenological studies \cite{Chakraborti:2014gea,Chakraborti:2015mra,Datta:2016ypd,Datta:2018lup,Adam:2021rrw} for probing eweakinos through have also been performed in the light of ATLAS/CMS data \cite{Aad:2014vma,Aad:2014nua,Aad:2014yka,Aad:2015jqa}. The multi-lepton\footnote{There are a few phenomenological works \cite{Sabatta:2019nfg,Buddenbrock:2019tua,Hernandez:2019geu} which point towards multi-lepton excess at the LHC.} ($nl+jets+\cancel{E_T}$) signals for $n \ge 3$ are in general the typical characteristic signatures of the heavier eweakinos. 
The usual searches for gluinos apart from multijet $+ \cancel{E_T}$ channel are focused on the final state comprising  up to $2l+jets+\cancel{E_T}$ at the LHC RUN-II. However, if sleptons and heavier eweakinos are assumed to be lighter than the gluino, the following situations may arise:
\begin{itemize}
\item Lighter eweakinos and sleptons (i.e. lighter than $M_{\widetilde{g}}$) would change the LHC RUN-II exclusion plot for gluinos. However, we do not intend to study this feature in this work.
  
\item  If heavier eweakinos appear in the decay cascade of gluinos and the lighter eweakinos and sleptons constitute a compressed spectrum, the qualitative features of the signal change significantly. As a result, multi-lepton signals may become potential discovery channels for probing the strong sectors of the pMSSM through gluino searches \cite{Baer:1986au,Barnett:1987kn,Baer:1995va,Baer:2003wx}. The main objective of this study is to explore the potential of these multi-lepton channels in gluino searches at the LHC RUN-III.
\end{itemize}
      
Bringing heavier eweakinos below the gluino mass has other theoretical motivation. Low higgsino mass parameter $\mu$, which governs heavier eweakino masses, is believed to be favorable from the perspective of low fine tuning \cite{Sakai:1981gr,Kaul:1981hi,Barbieri:1987fn,Feng:1999mn,Arkani-Hamed:2006wnf,GAMBIT:2018gjo}. Bringing heavier eweakinos into the gluino searches not only helps in collider searches but also favors in explaining the WMAP/PLANCK \cite{Huang:2018xle,Planck:2018vyg} measured relic density of Dark Matter (assuming the LSP i.e. the lightest neutralino ($\widetilde{\chi}_1^{0}$) emerges as a stable dark-matter candidate) and the precise measurement of the anomalous magnetic moment of muon \cite{Chakraborti:2021dli,Chakraborti:2022sbj,Chakraborti:2022vds,Chakraborti:2022wii}. 

Many variants of pMSSM are considered in this work. They are broadly categorised into two classes – wino type and higgsino type scenarios. Since the decays of gluinos to electroweak sector depend upon squark mixing, we vary left (L) and right (R) squark compositions and make subcategories for them. For each type of models, two benchmark points (BPs) are chosen. The BPs considered in this study are consistent with the LHC RUN-II data from the gluino searches in different final states. Due to the presence of electroweakinos and sleptons in the decay cascades, the BPs chosen must have to also consistent with their mass bounds as well obtained from the LHC RUN-II. We have also checked that the BPs satisfy the WMAP/PLANCK data\cite{Huang:2018xle,Planck:2018vyg} and the latest muon (g-2) data obtained from the Fermilab~\cite{Muong-2:2021ojo}.

In this work we employ Multivariate Analysis (MVA) technique for better signal-to-background ratio. These techniques are widely used in various collider analysis, such as particle identification, event reconstruction, signal discrimination etc. MVA methods are used to classify particle interactions based on various kinematic input variables, called features. Some of the common MVA techniques used in collider physics are Boosted Decision Trees (BDT), Neural Networks (NN), Random Forest (RF), Support Vector Machines (SVM) etc. In this study we use BDT which takes a set of input features and splits input data iteratively based on those features.

The paper is organized as follows. In sec. \ref{sec2} we construct and discuss the characteristic features of the models considered for our analysis. Sec. \ref{sec4} covers the overall methodology adopted for this analysis. The constraints from LHC RUN-II data, as well as muon (g-2) and constraint on relic density of DM from PLANCK data, are briefly described in this section. The simulation strategy, along with the selected BPs are also touched upon in this section. Sec. \ref{sec3} includes a comprehensive overview of pair production of gluinos along with its decay modes and the qualitative discussion on yield of multi-lepton events in respect of our models. In sec. \ref{sec5}, the cut and count analysis (CCA) method for a particular signal topology and the details of the MVA method for all signals are presented. Additionally  the results obtained from the MVA method are also discussed in this sec. Finally, we conclude our overall findings in sec. \ref{sec6}.


\section{Model} \label{sec2}

We consider pMSSM scenario where the masses of squarks of first two generations are greater than the mass of the gluino ($m_{\widetilde{q}} > m_{\widetilde{g}}$). We study the pair production of gluinos in the proton-proton collision at 13.6 TeV center-of-mass energy of the LHC. The gluino mass is fixed by SU(3) gaugino mass parameter $M_3$. The gluinos further decay to eweakinos along with a pair of light quarks ($u,d,c,s$) via off-shell squarks.  Obviously, the left(L)/right(R) squark composition is important in determining the decays of gluino. The following two cases might arise:
\begin{itemize}
    \item The squark with dominant SU(2) (i.e. left) component takes the gluino to wino-type eweakinos with a pair of accompanying quarks.
    \item The squark with dominant U(1) (i.e. right) component facilitates final state with bino-type LSP with two quarks.
\end{itemize} 

The electroweak sector in the R-parity conserving minimal supersymmetric standard model (MSSM) consists of the mixture of spin-1/2 partners of the U(1), SU(2) gauge bosons and the Higgs bosons. The corresponding mass eigenstates are referred to the charginos $(\widetilde{\chi}_j^{\pm}, j = 1, 2)$ and the neutralinos $(\widetilde{\chi}_i^{0}, i = 1,2,3,4)$, respectively. The increasing order of indices conventionally implies increasing masses of the eweakinos. The four parameters which determine the masses and compositions of these sparticles are: $M_1$ (the U(1) gaugino mass parameter), $M_2$ (the SU(2) gaugino mass parameter), $\mu$ (the higgsino mass parameter) and $\tan\beta$ (the ratio of the vacuum expectation values of the two neutral Higgs bosons). In this paper, we assume the parameter $\tan\beta$ to be  equal to 30. This choice is motivated by the fact that larger values of $\tan\beta$ provide a better fit to the data on the parameter space allowed by the $(g-2)_{\mu}$ constraint \cite{Kosower:1983yw,Yuan:1984ww,Chakraborti:2014gea}. It also allows the mass of the SM-like Higgs boson to be as large as possible at the tree level. The lightest neutralino $(\widetilde{\chi}_1^{0})$, which is the LSP acts as a tenable dark matter candidate simply because, it is the stable one and escapes detection at the LHC in the R-parity conserving SUSY scenario.

Sleptons are associated with either L- or R-type leptons, while sneutrinos are the superpartners of neutrinos. In this paper, it is assumed that all flavors of L(R)-type sleptons are degenerate in mass. Here, the masses of the sneutrinos  and the corresponding charged sleptons are non degenerate due to the small contribution of D-term. This always results in sneutrinos to be lighter than the charged sleptons and hence, can not produce leptons in their decay.

In this analysis, the Higgs sector does not play any role and hence, is kept decoupled except the Higgs boson of the standard model consistent with current experimental data.

\begin{figure}[h!]
    \centering
    \includegraphics[width=0.5\linewidth]{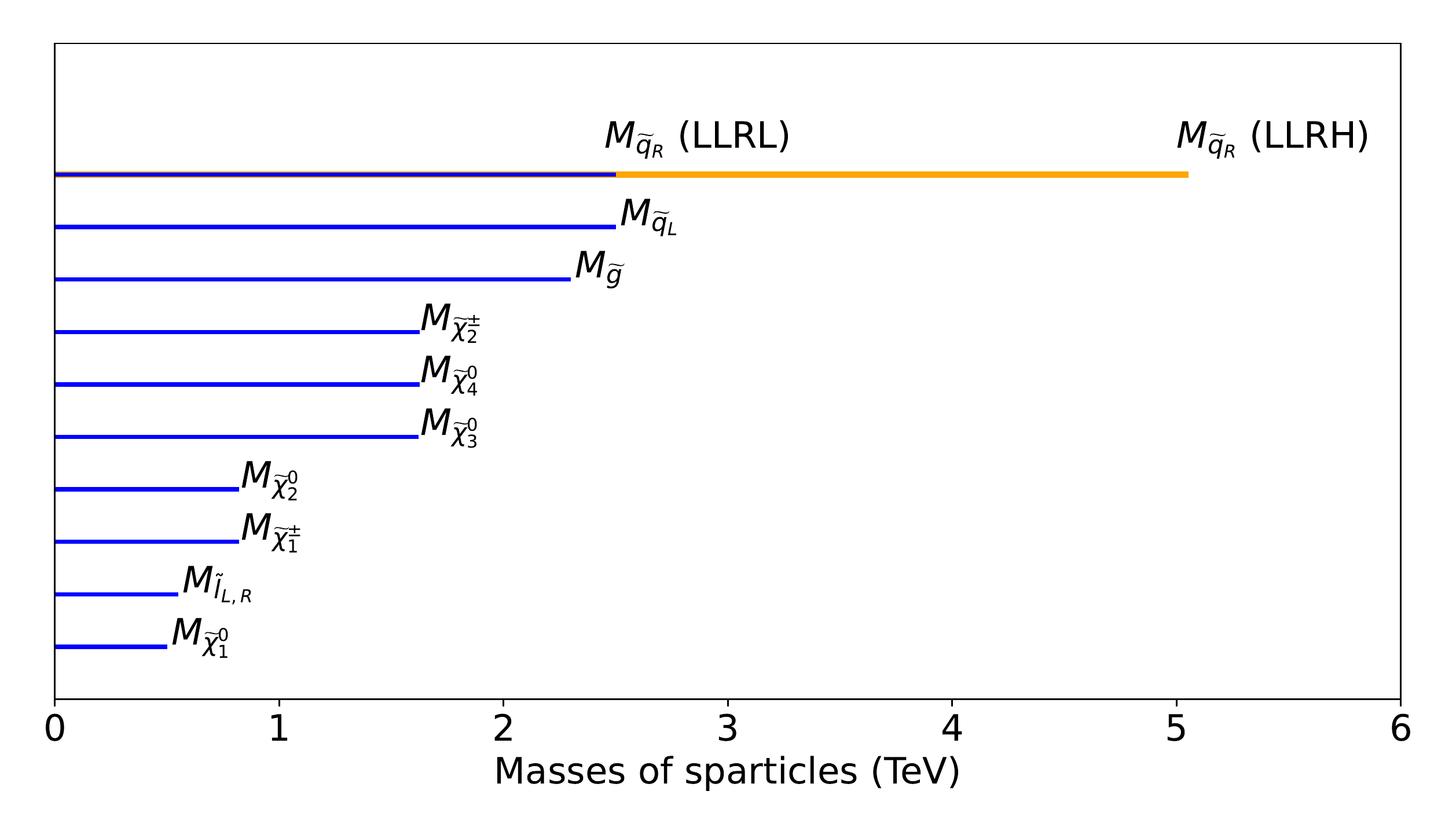}
    \caption{The sparticle mass hierarchy and their respective values corresponding to BP1 have been shown in this figure for the Light Electroweakino Light Slepton Wino (LELSW) models. The blue colored horizontal bars correspond to scenario with Left squark Light and Right squark Light (LLRL). The orange horizontal bar corresponds to Left Light and Right Heavy squark (LLRH) scenario in which all other mass parameter are the same as LLRL case.}
    \label{fig:LELSWparam}
\end{figure}

Therefore, to sum up the interplay of various factors, we segregate our analysis into three scenarios described in the following:

1) {\bf The Wino type model}:  The wino type model bears resemblance to the simplified models considered by the LHC collaboration. In pMSSM, we specify the wino type model by setting $M_1 < M_2 < \mu$. In this model, $\widetilde{\chi}_1^{\pm}$ and $\widetilde{\chi}_2^{0}$ (which are nearly mass degenerate), are predominantly made up of wino components, whereas, $\widetilde{\chi}_1^{0}$ is bino like. On the other hand the heavier eweakinos ( $\widetilde{\chi}_2^{\pm}$ , $\widetilde{\chi}_3^{0}$, and $\widetilde{\chi}_4^{0}$) are higgsino-like with masses governed by $\mu$. In order to include the heavier eweakinos in the gluino decay chain, their masses must be smaller than that of the gluino. For these reasons, we must keep the value of $\mu$ smaller than $M_3$. 
We have kept the slepton mass in between the $\widetilde{\chi}_1^{0}$ and $\widetilde{\chi}_1^{\pm}/\widetilde{\chi}_2^{0}$ so that all the eweakinos can decay via sleptons, thereby increasing the size of the leptonic signals. We refer this model as the Light Eweakinos Light Slepton Wino type models (LELSW). 
In short, the hierarchy looks like
$$
M_{\widetilde{g}} > M_{\widetilde{\chi}_4^{0}}, M_{\widetilde{\chi}_3^{0}},  M_{\widetilde{\chi}_2^{\pm}} > M_{\widetilde{\chi}_2^{0}}, M_{\widetilde{\chi}_1^{\pm}} > M_{\widetilde{l}} > M_{\widetilde{\chi}_1^{0}}
$$
The pictorial representation of the mass hierarchy of relevant sparticles for this model is shown in Fig. \ref{fig:LELSWparam}. 

\begin{figure}[h!]
    \centering
    \includegraphics[width=0.5\linewidth]{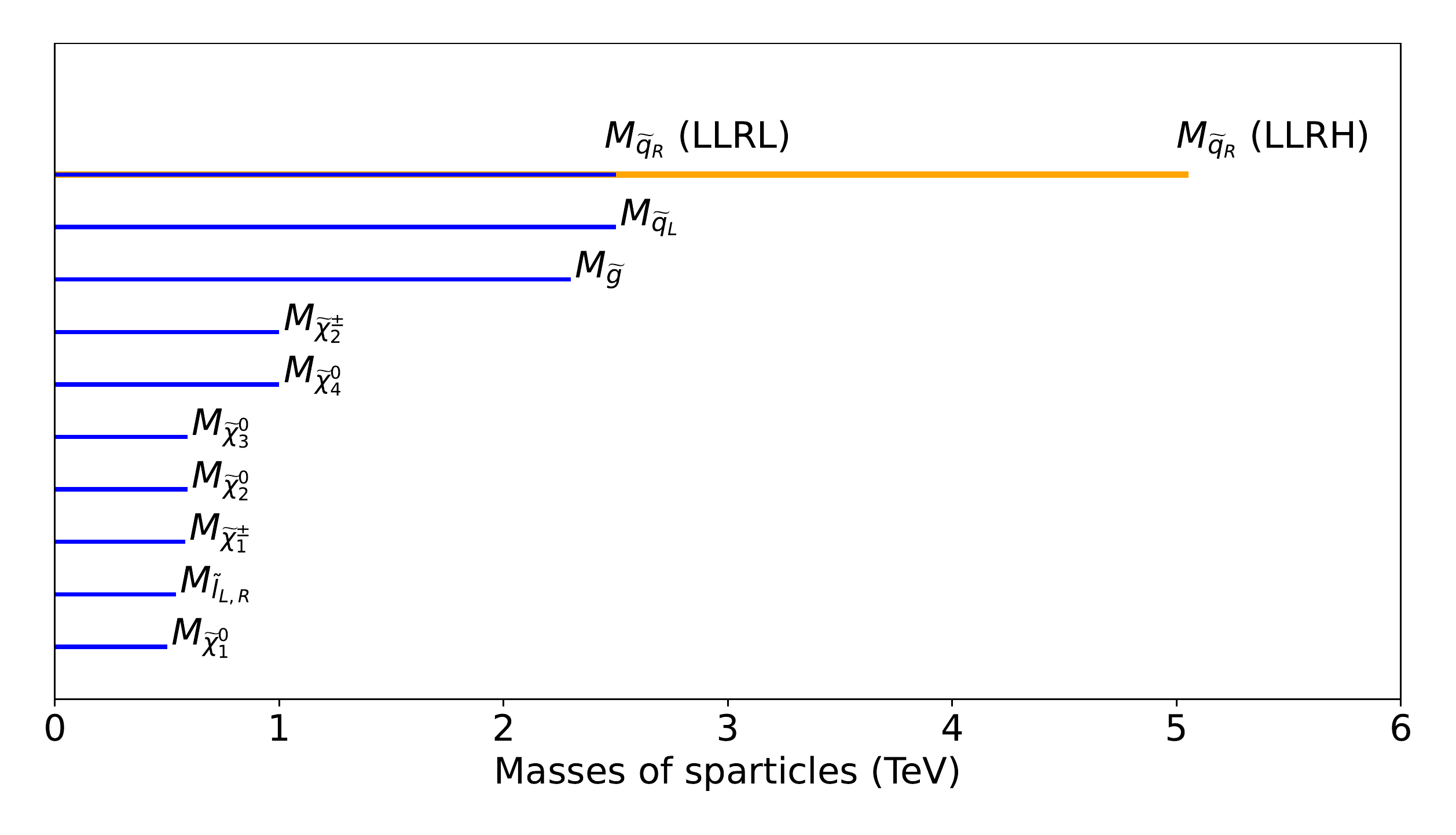}
    \caption{The sparticle mass hierarchy and their respective values corresponding to BP1 have been shown in this figure for the Light Electroweakino Light Slepton Higgsino (LELSH) models. The colors and conventions are same as in Fig. \ref{fig:LELSWparam}}
    \label{fig:LELSHparam}
\end{figure}

2) {\bf The Higgsino type models}: In this model, we primarily focus on scenarios where the $\widetilde{\chi}_1^{\pm}$ , $\widetilde{\chi}_2^{0}$ and $\widetilde{\chi}_3^{0}$ contain large higgsino components and have closely spaced masses determined by $\mu$, while the LSP is an admixture of bino and higgsino components. The mass hierarchy is determined by gluino mass parameter $M_3$.
In order to keep all the eweakinos lighter than the gluino one must choose $M_1, M_2,\mu < M_3$. Further, the condition, $\mu \simeq M_1$ is required to enhance the multi-lepton signatures with lepton multiplicity greater than 2. The two heavier eweakinos, $\widetilde{\chi}_2^{\pm}$ and $\widetilde{\chi}_4^{0}$, are of wino type. Their masses  are approximately equal to $M_2$, with $M_2 > \mu$. 
In this model the sleptons masses are kept in between $\widetilde{\chi}_1^{0}$ and $\widetilde{\chi}_1^{\pm}$ so that all the eweakinos can decay into the sleptons.
We refer these models as Light Eweakinos Light Slepton Higgsino type models (LELSH). 
$$
M_{\widetilde{g}} > M_{\widetilde{\chi}_4^{0}},  M_{\widetilde{\chi}_2^{\pm}} > M_{\widetilde{\chi}_3^{0}}, M_{\widetilde{\chi}_2^{0}}, M_{\widetilde{\chi}_1^{\pm}} > M_{\widetilde{l}} > M_{\widetilde{\chi}_1^{0}}
$$
In Fig. \ref{fig:LELSHparam} we present the mass hierarchy among relevant sparticles.

\begin{figure}[h!]
    \centering
    \includegraphics[width=0.5\linewidth]{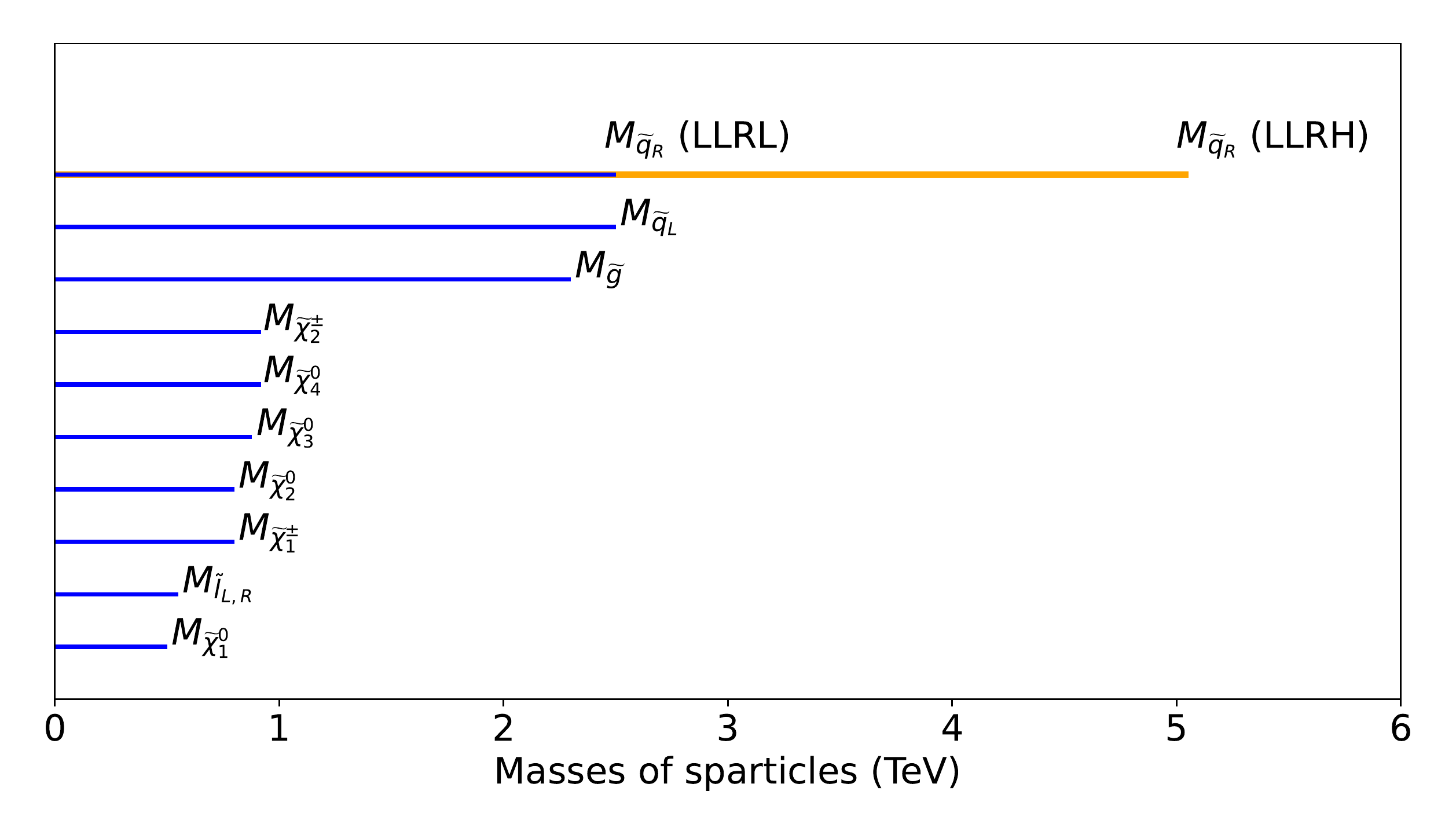}
    \caption{The sparticle mass hierarchy and their respective values corresponding to BP1 have been shown in this figure for the Wino Higgsino mixed (Wiggsino) models. The colors and conventions are same as in Fig. \ref{fig:LELSWparam}}
    \label{fig:Wiggsinoparam}
\end{figure}

3) {\bf The Wino Higgsino Mixed type (Wiggsino) models}: In the mixed type models, the eweakinos (excluding the LSP) are combinations of higgsino and wino components and have closely spaced masses (i.e., $\mu \simeq M_2$). In this scenario the LSP is predominantly bino type, with its mass controlled by $M_1$. The sleptons are placed in between the eweakinos and the LSP to obtain lepton-rich final states. The models are termed as the Wino- Higgsino mixed type models (Wiggsino) in our analysis. The mass hierarchy of relevant sparticles are displayed in Fig. \ref{fig:Wiggsinoparam}.

The above models are further subcategorised on the basis of the squark mass hierarchy as mentioned at the beginning of this section. The squarks contribute to the gluino production via t/u-channel diagrams. Most importantly, as the squarks are heavier than the gluinos in our analyses, the gluino decays various eweakinos are dictated by the off-shell squark propagation. 
Thus, following two major scenarios appear: 
\begin{itemize}
    \item In the left light right light squark mass models (LLRL), both L- and R-squarks have similar mass just above the gluino mass i.e. ($M_{\widetilde{g}} < M_{\widetilde{q}_{L,R}} \simeq 2.5$ TeV).
     \item In the left light right heavy squark mass models (LLRH), the L-squarks are light and positioned just above the gluino mass in the mass scale ($M_{\widetilde{q}_{L}} \simeq 2.5$ TeV). However, the R-squarks are made much heavier $M_{\widetilde{q}_{R}} > 5$ TeV and hence decoupled.
\end{itemize}
If the L-squark is decoupled, regardless of the mass of the R-squark, the gluino most likely to decay into $q \Bar{q} \widetilde{\chi}_1^{0}$, giving multijet final states and hence, is not an interesting case for this study. 

\section{The Methodology} \label{sec4}  
In this section, we first mention various constraints on pMSSM parameters that are the guiding principles in selecting the benchmark points of various models we have considered in this work. Next, we describe two benchmark points chosen for this study. Finally, we give an outline of the simulation techniques used in this work, particularly emphasising the MVA method.

\subsection{Constraints}   \label{sec4.1}

Supersymmetry has been put to test for long time. Hence, most of the avenues are constrained by experimental data. In this section we mention, in brief, various relevant constraints from the LHC and other low energy experiments. We have abided by these bounds while selecting the benchmark points. 

\subsubsection{Constraints on masses of the Gluinos and Squarks at the LHC RUN-II}
The ATLAS collaboration have explored several search channels to constrain the masses of the gluino and squarks. In RUN-II, the limits on the mass of the gluino (squark) for an integrated luminosity of 139 $fb^{-1}$ for various final states are almost on the edge of the kinematic reach of the LHC. The constraints on the gluino (squarks) mass for {\it negligible LSP mass} are summarized in Table \ref{tab:strongconstraints}. 

It can be noted from Fig. 13 (14) of ref. \cite{Aad:2020aze} that there is practically no bound on the mass of $\widetilde{g}$ ($\widetilde{q}$) for $M_{\widetilde{\chi}_1^0} \gtrsim$ 1.1 (0.8) TeV for $jets +\cancel{E_T}$. One can further infer from Fig. 8 of ref. \cite{Aad:2021zyy} that  the bound on the mass of $\widetilde{g}$ ($\widetilde{q}$) for $M_{\widetilde{\chi}_1^0} \gtrsim$ 1.26 (0.7) TeV does not exist in the final state comprising of $1l+jets+\cancel{E_T}$. Similar observation holds for $M_{\widetilde{\chi}_1^0} \gtrsim$ 1.4 (0.9) TeV in the $2l+jets+\cancel{E_T}$ final state  which can be seen from  Fig. 16 of ref. \cite{ATLAS:2022zwa}. More details can be found in the given references.
\FloatBarrier
\begin{table}[t!]
    \centering
    \resizebox{0.8\textwidth}{!}{
    \begin{tabular}{||c|c|c|c||}
    \hline
        Pair Production & Signal topology  & \makecell[c]{Bounds on gluino (squarks)  \\ for negligible mass of LSP (GeV)} & Ref.\\
        \hline
        $\widetilde{g}$ ($\widetilde{q}$) & $jets+\cancel{E_T}$  & 2300 (1850) & \cite{Aad:2020aze} \\
        \hline 
        $\widetilde{g}$ ($\widetilde{q}$) & $1l+jets+\cancel{E_T}$  & 2200 (1400) & \cite{Aad:2021zyy} \\
        \hline
        $\widetilde{g}$ ($\widetilde{q}$) & $2l+jets+\cancel{E_T}$ & 2250 (1550) & \cite{ATLAS:2022zwa}  \\
        \hline
     \end{tabular}}
    \caption{The bounds on gluino (squarks) masses from various search channels at LHC RUN-II.}
    \label{tab:strongconstraints}
\end{table}

\subsubsection{Constraints on the masses of Eweakinos and Sleptons at the LHC RUN-II}
During LHC RUN-II, the ATLAS collaboration have conducted searches for SUSY through electroweak sparticle pair production. The searches are carried out in different multi-lepton channels, such as $2l+\cancel{E_T}$ \cite{ATLAS:2019lff} for lighter eweakinos ($\widetilde{\chi}_1^+$, $\widetilde{\chi}_1^-$) and sleptons, and $3l+\cancel{E_T}$ \cite{ATLAS:2021moa} for lighter eweakinos ($\widetilde{\chi}_1^{\pm}$, $\widetilde{\chi}_2^0$). The results have been interpreted in terms of simplified models. In an integrated luminosity of 139 fb$^{-1}$, no significant excess of signal events over SM backgrounds have been observed. The exclusion limits for the masses of eweakinos and sleptons are summarized in Table \ref{tab:weakconstraints}. 

From Fig. 7(c) in \cite{ATLAS:2019lff} one can infer that the bound on the mass of sleptons evaporate if $M_{\widetilde{\chi}_1^0} \gtrsim 420$ GeV in $2l+\cancel{E_T}$ final state. Further from Fig. 7(b) of  ref. \cite{ATLAS:2019lff} one can observe if $M_{\widetilde{\chi}_1^0} \gtrsim 500$ GeV, there is no constraint on $M_{\widetilde{\chi}_1^{\pm}}$ in the decay topology where $\widetilde{\chi}_1^{\pm}$ decays through sleptons. It is also clear from Fig. 16 that the ATLAS collaboration \cite{ATLAS:2021moa} does not provide any no lower bound on the mass of the lighter eweakinos for $M_{\widetilde{\chi}_1^0}$ beyond 300 GeV in $3l+\cancel{E_T}$ final state.  More information about the search of eweakinos/sleptons can be found in the given references.

    \FloatBarrier
\begin{table}[h!]
    \centering
    \resizebox{0.8\textwidth}{!}{
    \begin{tabular}{||c|c|c|c||}
    \hline
        Pair Production & Signal topology  & \makecell[c]{Bounds on electroweak sparticles \\ for negligible mass of LSP\\ (GeV)} & Ref. \\
        \hline
       \makecell[c]{$\widetilde{\chi}_1^\pm$ $\widetilde{\chi}_1^{\mp}$} & $2l+\cancel{E_T}$  &  1000 & \cite{ATLAS:2019lff} \\
        \hline 
        $\widetilde{l}~ \widetilde{l}^{*}$ & $2l+\cancel{E_T}$ &  700 & \cite{ATLAS:2019lff} \\
        \hline
        \makecell[c]{$\widetilde{\chi}_1^\pm$ $\widetilde{\chi}_2^0$} & $3l+\cancel{E_T}$  & 640 & \cite{ATLAS:2021moa} \\
        \hline
     \end{tabular}}
    \caption{The bounds on eweakinos and sleptons from various search channels at LHC RUN-II.}
    \label{tab:weakconstraints}
\end{table}

\subsubsection{Other Constraints}
\begin{itemize}
    \item The SUSY mass spectra for the benchmark points must satisfy SM-like Higgs boson mass constraints {\it i.e.} $122 < M_h < 128$ GeV, with the central value of 125 GeV \cite{ATLAS:2012yve, CMS:2012qbp}. This can be obtained by proper choice of the trilinear soft breaking term for the top squark ($A_t$ ) and the CP-odd Higgs mass ($M_A$). The theoretical uncertainty \cite{Degrassi:2002fi, Allanach:2004rh} in determining the Higgs mass in a typical SUSY scenario is taken care of by allowing a mass window of $\pm 3$ GeV. 

    \item Tantalizing hint of new physics is coming from the long standing discrepancy of experimental data and SM prediction of the anomalous magnetic moment of muon denoted by $a_{\mu}=\frac{1}{2}(g-2)_{\mu}$. The SUSY contribution denoted as $a^{SUSY}_{\mu}$  becomes significant if the masses of charginos, neutralinos and smuons are relatively light. It also depends on $\tan\beta$, the ratio of vacuum expectation values of the two Higgs doublets. Therefore, one can constrain the SUSY parameter space by comparing the measured value of $\Delta a_{\mu}$, which is the difference between the experimental value of $a_{\mu}$ and the SM prediction. According to the results of Fermilab National Accelerator Laboratory (FNAL), the value of  on $\Delta a_{\mu}$ along with uncertainty is given by \cite{Muong-2:2021ojo} 
    \begin{eqnarray}
        \Delta a_{\mu}=a_{\mu}^{exp} - a_{\mu}^{SM} = (251 \pm 59) \times 10^{-11}
    \end{eqnarray}

    \item In this analysis we also take into account the constraint on the DM relic density as estimated by the WMAP/PLANCK data \cite{Planck:2018vyg} with tiny observational uncertainty: 
    \begin{eqnarray}
        \Omega_\chi h^2 = 0.120 \pm 0.001
    \end{eqnarray}
    where $h=0.733\pm 0.181$ \cite{Wong:2019kwg} is the Hubble constant in the units of 100 km/Mpc-s. 

\end{itemize}

\subsection{The Benchmark Points}
In order to analyze our models and to assess the potential for discovery, we choose two benchmark points (BPs) corresponding to each model described in sec. \ref{sec2}. One is selected from the edge of the gluino exclusion region, while the other is taken from the compressed region where gluino and the LSP are close to each other in terms of mass. 

The BP1 corresponds to heavier mass of the gluino resulting in smaller production cross-section while for the BP2 the situation is just the reverse. On the other hand, BP1 offers more kinematic phase space for the reconstructed objects, while the same becomes more limited in case of BP2. Moreover, the BPs are chosen in such a way so that they can pass the collider constraints mentioned in sec. \ref{sec4.1}. It is evident from sec. \ref{sec4.1} that if mass of the LSP is set to be above 500 GeV, the BPs satisfy the mass bounds obtained from the electroweak searches at the LHC RUN-II. 

Except for a few models, BPs corresponding to all models satisfy the other constraints like the PLANCK data for DM relic density and the precise value of anomalous magnetic moment of muon within the feasible allowed range of uncertainty.
 The benchmark points are so chosen that the estimated value of $\Delta a_{\mu}$ (which is assumed to be the SUSY contribution) is considered to lie within the $4\sigma$ of the central value.

For the choice of the BPs it is assumed that the third generation squarks are decoupled and all other soft-breaking trilinear terms are set to be zero, except for the trilinear  soft breaking term for the top quark, $A_t$ which is set to be 5 TeV for obtaining the SM-like Higgs mass to be around $M_h=125$ GeV. Throughout the work, $\tan\beta$ is set at 30 and the value of the pseudo scalar Higgs mass $M_A$ is taken to be around 3 TeV. The choice of the value of $M_A$ is not crucial for the collider analysis, but it will have substantial significance in satisfying the PLANCK constraints through $H$-resonance or CP-odd Higgs, $A$-resonance annihilations (for more details see ref. \cite{Mukherjee:2022kff}). This particular value of $M_A$ is chosen for illustration purpose and it does not affect the generality of the study. All other mass parameters along with the value of relic density and $\Delta a_{\mu}$  and cross-sections are presented in Table \ref{tab:BP}.

\FloatBarrier
\begin{table}[htb]
    \centering
    \resizebox{\textwidth}{!}{
    \begin{tabular}{|c|c|c|c|c|c|c|c|c|c|c|c|c|c|}
         \hline
         BP & Model & \makecell{$M_{\widetilde{g}}$ \\ (GeV)} & \makecell{$M_{\widetilde{\chi}_1^0}$ \\ (GeV)} & \makecell{$M_{\widetilde{\chi}_1^\pm}$ \\ (GeV)} & \makecell{$M_{\widetilde{\chi}_2^0}$ \\ (GeV)} & \makecell{$M_{\widetilde{\chi}_3^0}$ \\ (GeV)} & \makecell{$M_{\widetilde{\chi}_4^0}$ \\ (GeV)} & \makecell{$M_{\widetilde{\chi}_2^\pm}$ \\ (GeV)} & \makecell{$M_{\widetilde{l}^\pm}$ \\ (GeV)} & $\Omega_\chi h^2$ & $\Delta a_\mu \times 10^{-11}$ & \multicolumn{2}{c|}{\makecell{Production \\ Cross-section\\ (fb)}}  \\
         \cline{13-14}
          &  &  &  &  &  &  &  &  &  &  &  & LLRL & LLRH\\
         \hline 
         \multirow{3}{*}{BP1} & LELSW & 2300 & 500 & 820 & 820 & 1621 & 1624 & 1624 & 550 & 0.021 & 73.4 & \multirow{3}{*}{0.153} & \multirow{3}{*}{0.184}\\
         \cline{2-12}
         & Wiggsino & 2300 & 500 & 800 & 800 & 879 & 920 & 919 & 550 & 0.126 & 77.4 &  &  \\
         \cline{2-12}
         & LELSH & 2300 & 500 & 580 & 589 & 591 & 1000 & 1000 & 540 & 0.325 & 79.0 &  & \\
         \hline 
         \multirow{3}{*}{BP2} & LELSW & 1800 & 1400 & 1455 & 1455 & 1770 & 1783 & 1779 & 1430 & 0.589 &  15.9 & \multirow{3}{*}{2.15} & \multirow{3}{*}{2.56} \\
         \cline{2-12}
         & Wiggsino & 1800 & 1400 & 1550 & 1552 & 1630 & 1672 & 1670 & 1450 & 0.112 & 15.7 &  &  \\
         \cline{2-12}
         & LELSH & 1800 & 1400 & 1460 & 1475 & 1476 & 1700 & 1700 & 1430 & 0.254 & 15.9 &  &  \\
         \hline
    \end{tabular}}
    \caption{The sparticle mass spectra corresponding to different BPs chosen from different models. The last three columns contain the information about DM relic density, $\Delta a_{\mu}$ and cross-sections for Light Left Light Right (LLLR) and Light Left Right Heavy (LLRH) models corresponding to each BP.}
    \label{tab:BP}
\end{table}

\subsection{The Simulation} \label{sec4.2}
In this sub sec. we will discus the general procedure of event generation and the subsequent collider simulation. The sparticle mass spectra and their decay BRs have been generated using {\tt SUSY-HIT}\cite{Djouadi:2006bz}. The relic density and the contribution to the $\Delta a_{\mu}$ of these parameter points are then calculated using {\tt micrOMEGAs 5.2}\cite{Belanger:2020gnr}. For the purpose of event generation we have used {\tt MG5 aMC@NLO} \cite{Alwall:2014hca} for both the SUSY signals and the SM backgrounds. Events upto two jets are generated by implementing the MLM matching scheme. All the events are generated at the center-of-mass energy of 13.6 TeV, which is the produced energy of the recent LHC RUN-III. We have generated as many as $6 \times10^6$ events for the backgrounds and $3\times 10^5$ events for the signal events. All events (signal and backgrounds) are generated weighted to the integrated luminosity. While generating the background events caution have been taken such that a large sample of backgrounds are being generated where the signal resides. This is done by generating events with suitable binning with respect to the transverse momentum of the partons. Moreover, we have taken both on-shell and off-shell contribution of the diboson and triboson backgrounds. The events are statistically sufficient to draw any conclusion. The events are generated with the PDF set {\tt NNPDF2.3LO}\cite{Ball:2012cx}. The cross-sections of the SM background events have been taken from the {\tt MG5 aMC@NLO}. In addition, {\tt Prospino2.0}\cite{Plehn:2004rp} is used to take into account the next-to-leading order (NLO) cross-sections for the gluino pair production.

The parton-level events are passed through {\tt PYTHIA 8.2} \cite{Sjostrand:2014zea} to implement subsequent decays of unstable particles, as well as to take into account the initial and final state radiations (ISR and FSR), showering, fragmentation, and hadronization etc. For detector-level simulation, we have used {\tt Delphes 3.4}\cite{deFavereau:2013fsa}, a fast detector simulation package. 

We have utilised the ATLAS card to reconstruct jets, leptons (electrons and muons), and missing energy within {\tt Delphes 3.4} \cite{deFavereau:2013fsa}. The anti-$k_T$ algorithm has been used for clustering the jets with a radius parameter $R = 0.4$, using the {\tt FastJet} \cite{Cacciari:2011ma} package. The pseudo rapidity of the reconstructed jets must be in the range $|\eta| > 4.5$ and their transverse momentum must be $p_T > 20$ GeV.

The electron (muon) must have a transverse momentum of $p_T > 10$ GeV and must fall within the range of $|\eta| < 2.47 ~(2.7)$. In order to guarantee that the leptons are isolated, we also set a restriction that the scalar sum of the $p_T$ of all other objects inside a cone of radius 0.2 (0.3) surrounding an electron (muon) must be smaller than 12$\%$ (15$\%$) of its $p_T$. Finally, the transverse momentum imbalance corresponding to all reconstructed objects in an event is used to calculate the missing transverse momentum $p_T^{miss}$ (with magnitude $\cancel{E_T}$). 

The relevant backgrounds can be broadly classified into two main groups. The first category consists of situations where a jet is mistaken for a lepton, or when additional leptons are produced due to photon conversions and decays of heavy-flavor particles during initial- and final-state radiation. To mitigate these backgrounds, we implement an isolation requirement and employ object reconstruction techniques (discussed above), which effectively reduce this type of background.

In addition, we apply a set of pre-selection cuts after the object reconstruction to minimize the occurrence of the second category of background events. These events stem from SM backgrounds, namely $t\bar{t}$, Drell-Yan, dibosons ($VV$), tribosons ($VVV$), as well as $t\Bar{t}V$ and $hV$ processes (where $V$ represents either the $W$ or $Z$ boson). They pre-selection cuts are :
\begin{itemize}
    \item Number of leptons ($n_{l}$) = 2,3,4 for the final states comprising of $2l+jets+\cancel{E_T}$, $3l+jets+\cancel{E_T}$ and $4l+jets+\cancel{E_T}$ respectively.
    \item Number of jets ($n_{j}$) $\ge$ 2
    \item b-veto (Number of b - jets ($n_{bjet}$) = 0)
\end{itemize}
These pre-selection criteria substantially reduce large backgrounds that are coming from those SM backgrounds which are associated with $t\Bar{t}$.

After all these data cleaning processes, the remaining events are passed through a boosted decision tree (BDT) classifier to achieve better discrimination and thus improved significance. We conduct a multivariate analysis (MVA) using the BDT classifier implemented in the Toolkit for multivariate analysis ({\tt TMVA 4.3})\cite{Voss:2007jxm} which is integrated within the {\tt ROOT}\cite{Antcheva:2009zz} analysis framework. We have used some simple variables called features in the input of the BDT to discriminate the signal. These variables constitute a minimal set that possesses (a) strong discrimination power between SUSY signal and SM background, and (b) low correlation among themselves. For a given feature $x$, the separation of feature is defined as
\begin{eqnarray}
    <S^2> = \frac{1}{2} \int\frac{[\hat{x}_{SUSY}(x)-\hat{x}_{SM}(x)]^2}{\hat{x}_{SUSY}(x)+\hat{x}_{SM}(x)} dx
\end{eqnarray}
where $\hat{x}_{SUSY}(x)$ and $\hat{x}_{SM}(x)$ are the probability density functions of $x$ for the SUSY signal and the SM background, respectively. To improve the BDT classification, we use the adaptive boost algorithm and combination of 950 decision trees with a minimum node size of 5$\%$ and a depth of 4 layers per tree into a forest. The Gini index is used as the separation criterion for node splitting. A summary of the relevant BDT hyperparameters is presented in Table \ref{tab:hyperbdt}.
\begin{table}[htb]
    \centering
    \begin{tabular}{|c|c|}
    \hline
        BDT hyperparameter & Optimised choice \\
        \hline
        NTrees & 950\\
        MinNodeSize & 5$\%$\\
        MaxDepth & 4 \\
        BoostType & AdaBoost \\
        AdaBoostBeta & 0.5 \\
        UseBaggedBoost & True \\
        BaggedSampleFraction & 0.5\\
        SeparationType & GiniIndex \\
        nCuts & 10\\
        \hline
    \end{tabular}
    \caption{Summary of the optimised BDT hyperparameters.}
    \label{tab:hyperbdt}
\end{table}

For estimating the median predicted discovery significance, we use the following approximations\cite{Cowan:2010js,Cousins:2007yta,Li:1983fv} : 
\begin{eqnarray}
    Z_{dis} = \sqrt{2} \left( (s+b)\ln{\left[\frac{(s+b)(b+\delta_b^2)}{b^2+(s+b)\delta_b^2} \right]}- \frac{b^2}{\delta_b^2} \ln{ \left[1+\frac{\delta_b^2 s}{b(b+\delta_b^2)} \right]} \right)^{\frac{1}{2}}
\end{eqnarray}
where $s$, $b$ and $\delta_b$ are the normalised signal events, background events and the uncertainty in the measurement of backgrounds respectively. If the background events are negligible, a discovery is considered to have been made when there are 5 signal events passed the classifier optimal cut value. The estimation of the background uncertainty arising from various sources such as reconstruction, identification, isolation, and trigger efficiency, energy scale and resolution of various physics objects,measurements of luminosity,modeling of pile-up and parton-shower etc. is beyond the scope of this work. We rather adopt a conservative approach and assume an overall 10$\%$ total uncertainty for the background uncertainty.


\section{Production and Decay modes of Gluino leading to multi-lepton channels for different models}  \label{sec3}
In this section, we discuss the production cross-section and various relevant decay modes of the gluino. The inclusion of sleptons and heavier eweakinos in the gluino decay cascade \footnote{For more details about gluino decay through cascade see \cite{Baer:1986au,Barnett:1987kn,Baer:1991xs,Guchait:1994zk,Bartl:1996cg,Sokolov:1998yx,Alves:2006df,Kim:2008bnd,Baer:2008kc,Baer:2008ey,Acharya:2009gb,Andreev:2009dz,Turlay:2010bk,Reuter:2012ng}} increases the probability of obtaining multi-lepton signals in the final states. It is noted that the production cross-section of a pair of gluinos and their subsequent decays leading to multi-lepton final states are highly sensitive to the squark mass hierarchy. The gluino pair production cross sections in the LELSW model for the two benchmark points are shown in Table \ref{tab:CWBR}. 
The gluino pair production proceeds mainly via gluon initiated process. Next comes the $q\bar{q}$ initiated processes with s-channel gluon exchange and t/u-channel squark exchange diagrams. An important fact is that the two processes namely, the s-channel gluon exchange and t/u-channel squark exchange, contributing to the quark-antiquark initiated production of gluino-pairs, interfere destructively. This fact explains difference in the magnitude of cross sections between the LLRL and LLRH cases for both the BPs quoted in Table \ref{tab:BP}. Destructive interference takes place in both the scenarios. However, in LLRH case, heavier R-squarks are decoupled, and hence, are less likely to appear in the production process. Thus the contribution of t/u channel squark exchange process in the gluino pair production is less, which proceeds only via L-squark exchange. Therefore, the destructive interference between the the s-channel gluon exchange and t/u-channel squark exchange is less, thereby making the cross section larger in LLRH case. On the other hand, if both L- and R-squarks take part in the production of gluino pair, then the destructive interference between two sub processes becomes severe, resulting in reduced cross section in LLRL case.
In the $M_{\widetilde{q}} > M_{\widetilde{g}} $ scenario, the three-body decay BRs of gluinos depend on the composition of the gauginos and higgsinos in eweakinos as well as on the L/R compositions of the squarks. 

\begin{table}[htb]
    \centering
    \begin{tabular}{|c|c|c|c|c|}
    \hline
    \multirow{2}{*}{\makecell[c]{Decay Modes \\ and\\cross section}} & \multicolumn{2}{|c|}{BP1}  & \multicolumn{2}{|c|}{BP2}    \\ 
    \cline{2-5} 
                        & \multicolumn{2}{|c|}{Branching Ratio} &  \multicolumn{2}{|c|}{Branching Ratio}\\
                        \cline{2-5}
                        & LLRL & LLRH & LLRL & LLRH   \\  \hline
    $\widetilde{g} \rightarrow q \Bar{q} \widetilde{\chi}_1^0$ & 0.156  & 0.019 & 0.333 & 0.030\\ 
    $ \rightarrow q \Bar{q}' \widetilde{\chi}_1^{\pm}$ & 0.554  & 0.645 & 0.443 & 0.646\\   
    $ \rightarrow q \Bar{q} \widetilde{\chi}_2^0$ & 0.277 & 0.322 & 0.214 & 0.306 \\ \hline 
    $\widetilde{\chi}_1^{\pm} \rightarrow \widetilde{\nu} l^{\pm}$ & 0.500 & 0.500 & 0.457 & 0.407 \\ 
    $ \rightarrow \widetilde{l}^{\pm} \nu$ & 0.498 & 0.497 & 0.499 & 0.589 \\      \hline
    $\widetilde{\chi}_2^{0} \rightarrow \widetilde{l}^{\pm} l^{\mp}$ & 0.503 & 0.502 & 0.550 & 0.637 \\
    $ \rightarrow \widetilde{\nu} \nu$ & 0.495 & 0.494 & 0.435 & 0.358 \\    \hline
    $\sigma (p p \rightarrow \widetilde{g} \widetilde{g}) $ fb & 0.153 & 0.184 & 2.15 & 2.56 \\
    \hline
    \end{tabular}
    \caption{Decay branching ratio of relevant sparticles and gluino pair production cross section (in $fb$) for the left light right light (LLRL) Light Electroweakino Light Slepton Wino (LELSW) model and left light right heavy (LLRH) Light Electroweakino Light Slepton Wino (LELSW) model are shown. The BRs less than 1$\%$ have been ignored. The last row contains the information about cross-section of the BPs.}
    \label{tab:CWBR}
\end{table}

The decay BR of gluino and eweakinos are presented in Tables \ref{tab:CWBR}, \ref{tab:CHBR} and \ref{tab:MBR}. From Table \ref{tab:CWBR} we can easily see that the gluino predominantly decays into the wino-type lighter charginos  via off shell L-squarks in wino class of models. The lighter R-squarks in LLRL scenario take the gluino directly to $q \Bar{q} \widetilde{\chi}_1^0$. 
The various multi-lepton signals arising  in {\it wino type} models are given below :
\begin{itemize}
    \item Both the lighter eweakinos ($ \widetilde{\chi}_1^{\pm}, \widetilde{\chi}_2^{0}$) decay into sleptons/sneutrinos with almost 100$\%$ BR as shown in Table \ref{tab:CWBR}. $\widetilde{\chi}_1^{\pm}$ produces one lepton via any of its two decay modes, whereas, $\widetilde{\chi}_2^{0}$ produces two leptons almost half the times via lepton-slepton decay mode. Interestingly, gluino being a Majorana particle, it can produce same-sign dilepton (SSDL) final state via $\widetilde{\chi}_1^{\pm}$ decay mode. The $\widetilde{\chi}_2^{0}$ decay mode of gluino, though, mostly produces opposite-sign dilepton (OSDL)) final state.

    \item A significant number of trilepton signal i.e. $3l+jets+\cancel{E_T}$ final state can be produced if one of the gluino decays into $q \Bar{q}' \widetilde{\chi}_1^{\pm}$ and the other in $q \Bar{q} \widetilde{\chi}_2^{0}$.

    \item If both the gluinos decay into $q \Bar{q} \widetilde{\chi}_2^{0}$, this may result in a typical $4l+jets+\cancel{E_T}$ final state. 
\end{itemize} 

On the other hand, in LLRH case of the wino type model, the gluino decaying to $q \Bar{q} \widetilde{\chi}_1^0$ mode is disfavored due to heavier R-squarks, which, in turn, enhances the BRs to $\widetilde{\chi}_1^{\pm}~\rm{and}~\widetilde{\chi}_2^{0}$ channels. This is encouraging for the case we are studying as they tend to produce more leptons in the final state. This will have significant impact in the results which will be discussed later. Since the other heavier eweakinos are higgsino-like in this model, the BRs of gluino decaying into the heavier ones are negligible.

\begin{table}[h!]
    \centering
    \begin{tabular}{|c|c|c|c|c|}
    \hline
    \multirow{2}{*}{\makecell[c]{Decay Modes \\ and\\cross section}} & \multicolumn{2}{|c|}{BP1}  & \multicolumn{2}{|c|}{BP2}    \\ 
    \cline{2-5} 
                & \multicolumn{2}{|c|}{Branching Ratio} &  \multicolumn{2}{|c|}{Branching Ratio}\\
                        \cline{2-5}
                & LLRL & LLRH & LLRL & LLRH  \\  \hline
    $\widetilde{g} \rightarrow q \Bar{q} \widetilde{\chi}_1^0$ &  0.164 & 0.022 & 0.696 & 0.277 \\ 
    $ \rightarrow q \Bar{q}' \widetilde{\chi}_1^{\pm}$ &  0.017 & 0.020 & 0.127 & 0.397 \\  
    $ \rightarrow q \Bar{q}' \widetilde{\chi}_2^{\pm} $  & 0.489 & 0.589 & 0.007 & 0.023 \\  
    $ \rightarrow q \Bar{q} \widetilde{\chi}_3^0$ &  0.032 & 0.012 & 0.120 & 0.170 \\ 
    $ \rightarrow q \Bar{q} \widetilde{\chi}_4^0$ &  0.244 & 0.294 & - & 0.011 \\ \hline
    $\widetilde{\chi}_1^{\pm} \rightarrow \widetilde{\nu} l^{\pm} $ & 0.439  & 0.438 & 0.342 & 0.341 \\  
    $  \rightarrow \widetilde{l}^{\pm} \nu$ & 0.546  & 0.548 & 0.651 & 0.651\\\hline
    $\widetilde{\chi}_2^{\pm} \rightarrow \widetilde{\nu} l^{\pm}$ & 0.318  & 0.318 & 0.205 & 0.205 \\  
    $  \rightarrow \widetilde{l}^{\pm} \nu$ &  0.321 & 0.321 & 0.200 & 0.201\\  
    $  \rightarrow \widetilde{\chi}_i^{0} W^{\pm}$ &  0.176 & 0.176 & 0.292 & 0.293 \\ 
    $  \rightarrow \widetilde{\chi}_1^{\pm} Z$ & 0.091  & 0.091 & 0.156 & 0.156 \\ 
    $  \rightarrow \widetilde{\chi}_1^{\pm} h$ &  0.089 & 0.089 & 0.135 & 0.137 \\  \hline
    $\widetilde{\chi}_3^{0} \rightarrow \widetilde{l}^{\pm} l^{\mp}$ & 0.97  & 0.97 & 0.98 & 0.99  \\ \hline
    $\widetilde{\chi}_4^{0} \rightarrow \widetilde{\chi}_i^{0} Z$ & 0.088  & 0.088 & 0.137 & 0.138 \\
    $  \rightarrow \widetilde{\chi}_i^{0} h$ & 0.085  & 0.087 & 0.136 & 0.136\\ 
    $  \rightarrow \widetilde{\chi}_1^{\pm} W^{\pm}$ & 0.179  & 0.178 & 0.307 & 0.306 \\
    $\rightarrow l^{\pm} \widetilde{l}^{\mp}$  & 0.315  & 0.312 & 0.196 & 0.196 \\  
    $\rightarrow \nu \widetilde{\nu}$  & 0.322  & 0.324 & 0.214 & 0.214 \\   \hline 
    $\sigma (p p \rightarrow \widetilde{g} \widetilde{g}) $ fb & 0.153 & 0.184 & 2.15 & 2.56 \\
    \hline
    \end{tabular}
    \caption{Decay branching ratio of relevant sparticles for the left light right light (LLRL) Light Electroweakino Light Slepton Higgsino (LELSH) model and left light right heavy (LLRH) Light Electroweakino Light Slepton Higgsino (LELSH) model are displayed. Gluino pair production cross sections (in $fb$) for these models are also repeated just for the sake of completeness. The BRs less than 1$\%$ have been ignored. The last row contains the information about cross-section of the BPs}
    \label{tab:CHBR}
\end{table}

In the {\it higgsino} type model, however, the leptonic final states are not so straight forward. In fact, the gluino decay modes are quite different for the two selected BPs as can be seen from Table \ref{tab:CHBR}. For the heavier $M_{\widetilde{g}}$ corresponding to BP1, it predominantly decays into the heavier eweakinos ($\widetilde{\chi}_2^{\pm}, \widetilde{\chi}_4^{0}$). Both of these eweakinos can generate leptonic final states. The heaviest chargino can produce a single lepton via any of its dominant decay channels, whereas its neutral counterpart can give rise to a pair of leptons. The multi-lepton final state may appear in the following ways:

\begin{itemize}
    \item  A significant amount of $2l+jets+\cancel{E_T}$ final state (both SSDL and OSDL) can be obtained in BP1, if both the gluinos decay into the heavier chargino. In case of BP2, however, multi-lepton is hit hard in LLRL scenario as the gluino decays directly to LSP. In LLRH scenario, a handful of dilepton final state may be expected.
    
    \item If one of the gluino decays to  $\widetilde{\chi}_2^{\pm}$ and the other to $\widetilde{\chi}_4^{0}$, the trilepton signal can be anticipated for BP1. The same will be very meager in number for BP2.
    
    \item In BP1, $4l+jets+\cancel{E_T}$ final state can arise if both the gluino decay to $\widetilde{\chi}_4^{0}$ along with a quark-antiquark pair. The possibility of this final state, however, is very slim in BP2.
    
\end{itemize}   

\begin{table}[htb]
    \centering
    \begin{tabular}{|c|c|c|c|c|}
    \hline
    \multirow{2}{*}{\makecell[c]{Decay Modes \\ and\\cross section}} & \multicolumn{2}{|c|}{BP1}  & \multicolumn{2}{|c|}{BP2}    \\ 
    \cline{2-5} 
               & \multicolumn{2}{|c|}{Branching Ratio} &  \multicolumn{2}{|c|}{Branching Ratio}\\
                \cline{2-5} 
               & LLRL & LLRH & LLRL & LLRH    \\  \hline
    $\widetilde{g} \rightarrow q \Bar{q} \widetilde{\chi}_1^0$ & 0.142  & 0.020 & 0.594 & 0.135 \\ 
    $ \rightarrow q \Bar{q}' \widetilde{\chi}_1^{\pm}$ & 0.366  & 0.438 & 0.260 & 0.544 \\  
    $ \rightarrow q \Bar{q}' \widetilde{\chi}_2^{\pm} $  & 0.164  & 0.192 & - & 0.026\\ 
    $ \rightarrow q \Bar{q} \widetilde{\chi}_2^0$ & 0.179 & 0.220 & 0.121 & 0.248 \\
    $ \rightarrow q \Bar{q} \widetilde{\chi}_4^0$ & 0.143 & 0.094 & - & 0.012 \\        \hline
    $\widetilde{\chi}_1^{\pm} \rightarrow q \Bar{q}' W^{\pm}$ & 0.059 & 0.058 & 0.189 & 0.238\\ 
    $ \rightarrow \widetilde{\nu} l^{\pm}$ & 0.478 & 0.480 & 0.395 & 0.356\\ 
    $ \rightarrow \widetilde{l}^{\pm} \nu$ & 0.457 & 0.458 & 0.393 & 0.377 \\      \hline
    $\widetilde{\chi}_2^{\pm} \rightarrow \widetilde{\nu} l^{\pm}$ & 0.373 & 0.373 & 0.289 & 0.323 \\  
    $  \rightarrow \widetilde{l}^{\pm} \nu$ & 0.396 & 0.397 & 0.254 & 0.282 \\  
    $  \rightarrow \widetilde{\chi}_i^{0} W^{\pm}$ & 0.168 & 0.167 & 0.312 & 0.268 \\ 
    $  \rightarrow \widetilde{\chi}_1^{\pm} Z$ & 0.060 & 0.059 & 0.143 & 0.120 \\           \hline
    $\widetilde{\chi}_2^{0} \rightarrow \widetilde{\chi}_1^{0} h $ & 0.061 & 0.060 & 0.197 & 0.234 \\ 
    $ \rightarrow \widetilde{l}^{\pm} l^{\mp}$ & 0.531 & 0.531 & 0.489 & 0.484 \\
    $ \rightarrow \widetilde{\nu} \nu$ & 0.401 & 0.401 & 0.303 & 0.267 \\ \hline
    $\widetilde{\chi}_4^{0} \rightarrow \widetilde{\chi}_i^{0} h$ & 0.080  & 0.080 & 0.116 & 0.116 \\ 
    $  \rightarrow \widetilde{\chi}_1^{\pm} W^{\pm}$ & 0.169 & 0.169 & 0.337 & 0.336\\
    $\rightarrow l^{\pm} \widetilde{l}^{\mp}$  & 0.357  & 0.357 & 0.215 & 0.215 \\  
    $\rightarrow \nu \widetilde{\nu}$  &  0.381 & 0.381 & 0.304 & 0.306\\   \hline
    $\sigma (p p \rightarrow \widetilde{g} \widetilde{g}) $ fb & 0.153 & 0.184 & 2.15 & 2.56 \\
    \hline
    \end{tabular}
    \caption{Decay branching ratio of the relevant sparticles for the left light right light (LLRL) Wino Higgsino mixed (Wiggsino) model and left light right heavy (LLRH) Wino Higgsino mixed (Wiggsino) model. Gluino pair production cross sections (in $fb$) for these models are also shown. The BRs less than 1$\%$ have been ignored. The last row contains the information about cross-section of the BPs }
    \label{tab:MBR}
\end{table}

In the {\it wino-higgsino (wiggsino)} mixed model, the appearance of leptons in the final state is somewhat similar to higgsino scenario. The slight difference in the decay cascade is due to the replacement of $\widetilde{\chi}_3^{0}$ by $\widetilde{\chi}_2^{0}$. In BP1 for both LLRL and LLRH cases:
\begin{itemize}
    \item One can have dominant final state comprising of $2l+jets+\cancel{E_T}$ signal if charginos originate in both the arms of the gluino pair production.
    \item One of the gluinos decaying to charged eweakinos ( i.e. $\widetilde{\chi}_1^{\pm}$ or $\widetilde{\chi}_2^{\pm}$)  and the other to neutral eweakinos (i.e. $\widetilde{\chi}_2^{0}$ or $\widetilde{\chi}_4^{0}$)  may lead to  $3l+jets+\cancel{E_T}$ final state.
    \item The production of two neutral eweakinos (i.e. $\widetilde{\chi}_2^{0}$ or $\widetilde{\chi}_4^{0}$) in both branches of gluinos may give rise to $4l+jets+\cancel{E_T}$ final state.
\end{itemize} 
However, in BP2, the leptonic final state is badly affected as the gluino decays directly to LSP, thereby giving multijet signal for LLRL scenario. For LLRH case, the situation is again similar to that of BP1 as stated above.


\begin{figure}[htb]
    \centering
    \includegraphics[width=0.5\linewidth]{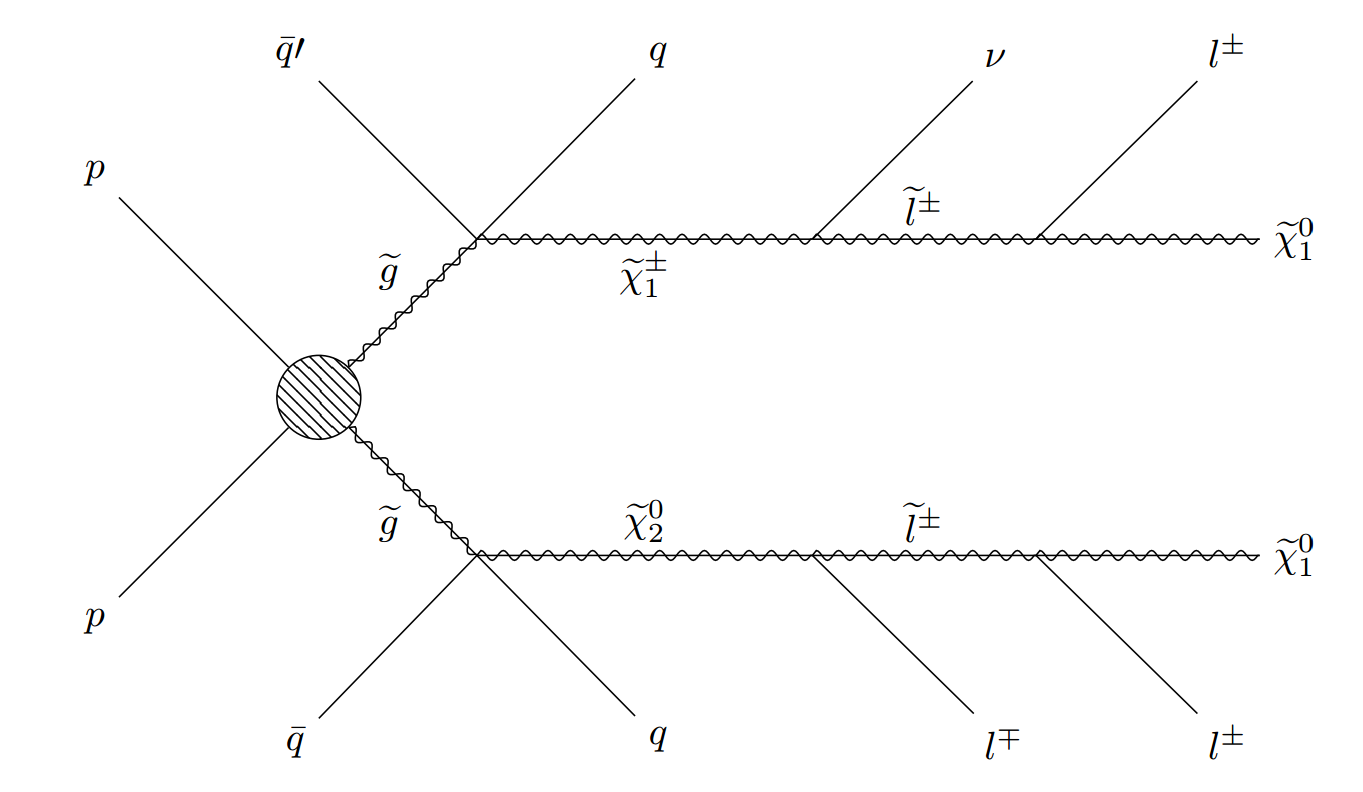}
    \caption{A typical example of the decay topology of a gluino pair giving rise to $3l+jets+\cancel{E_T}$ final state. It has been presented just for illustration.}
    \label{fig:feynman}
\end{figure}

In Fig. \ref{fig:feynman} we present the Feynman diagram for the $3l+jets+\cancel{E_T}$ decay topology for the purpose of illustration. 

From the BR tables we can see that we need to have many branches in the decay topology to obtain a higher lepton multiplicity (e.g. $4l$, $5l$) in the models. These branches significantly reduce the overall probability of obtaining events consists of multi-leptons, especially $5l$. This is why a detailed multivariate analysis (MVA) cannot be conducted for the $5l$ signal. The insufficient number of events may lead to a statistically incorrect interpretation. Therefore, a detailed analyses of our models in the context of the first three signal topologies are the primary focus of this paper.


\section{Results and Analyses}  \label{sec5}
\subsection{Cut and Count Analysis} \label{sec5.1}
The traditional cut and count analyses (CCA) gives a first hand idea on the significance of various signals against relevant backgrounds. Only events that pass pre-selection cuts mentioned in  sub sec. \ref{sec4.2} are considered for CCA. After the data cleaning, we apply several suitable kinematic cuts to differentiate the signal events from the background ones depending on the desired final states comprised of multiple leptons. We have performed CCA technique for $2l$, $3l$ and $4l$ final states. However, we must mention in passing that the results obtained from CCA are not encouraging at all. Hence, we decide against discussing the details of the results of CCA. However, just for the purpose of illustration we present the most promising results obtained for the $3l+jets+\cancel{E_T}$ final state in the following sub sec.

\subsubsection{Results of CCA for $3l+jets+\cancel{E_T}$} \label{sec.5.1.1}
In this sub sec., we present the CCA results for the final state comprising of $3l+jets+\cancel{E_T}$ at LHC RUN-III in the context of various models as described in sec. \ref{sec2}. The main SM backgrounds considered for this particular final state are $t\Bar{t}$, dibosons such as $WZ$, $ZZ$, tribosons like $WWW$, $WWZ$, $WZZ$, $ZZZ$, as well as $t\Bar{t}Z$, $t\Bar{t}W$, $t\Bar{t}h$, $hW$, and $hZ$. In order to distinguish the signal from the SM backgrounds, we have used appropriate kinematic variables for computing the significance using the traditional CCA method. Various kinematic variables and the corresponding selection cuts for BP1 (BP2) are mentioned below:

\FloatBarrier
\begin{table}[h!]
    \centering
    \resizebox{0.7\textwidth}{!}{
    \begin{tabular}{||c|c|c|c|c|c||}
        \hline
        \multirow{2}{*}{BP} & \multirow{3}{*}{Model} & \multicolumn{2}{|c|}{LLRL} & \multicolumn{2}{|c|}{LLRH} \\
         \cline{3-6}
         &  & \makecell[c]{$N_S$\\ after all cuts} & Significance & \makecell[c]{$N_S$\\ after all cuts} & Significance \\
         \hline
         \multirow{3}{*}{BP1} & LELSW & 0.53 & 0.24 & 0.89 & 0.39  \\
         \cline{2-6}
         & Wiggsino & 0.46 & 0.20 & 0.76 & 0.33  \\
         \cline{2-6}
         & LELSH & 0.34 & 0.15 & 0.54 & 0.24 \\
         \hline
         \multicolumn{2}{|c|}{\makecell[c]{Backgrounds ($N_b$)\\ after all cuts }} & \multicolumn{4}{|c|}{4.67} \\
         \hline
         \multirow{3}{*}{BP2} & LELSW  & 2.58 & 0.57 & 5.99 & 0.77 \\
         \cline{2-6}
         & Wiggsino & 0.87 & 0.19 & 4.52 & 0.97 \\
         \cline{2-6}
         & LELSH & 0.39 & 0.09 & 1.36 & 0.30 \\
         \hline
         \multicolumn{2}{|c|}{\makecell[c]{Backgrounds ($N_b$)\\ after all cuts }} & \multicolumn{4}{|c|}{16.76} \\
         \hline
    \end{tabular}}
    \caption{The table displays the normalised event number ($N_S$) and the significance for an integrated luminosity of 139 $fb^{-1}$ for the signal (BPs) and the cumulative (weighted) backgrounds after all cuts ($N_b$) in respect of the final state comprising of $3l+jets+\cancel{E_T}$.}
    \label{tab:cut-3l}
\end{table}

\begin{itemize}
    \item Missing transverse energy ($\cancel{E_T}$) : $\cancel{E_T} >$  150 (100) GeV
    \item The $p_T$ of the leading lepton ($p_T ^{l_1}$) : $p_T ^{l_1}>$ 60 (25) GeV
    \item The $p_T$ of the leading jet ($p_T ^{j_1}$) :  $p_T ^{j_1}>$ 150 (100) GeV
    \item The transverse mass ($m_T$) : $m_T >$ 100 (100) GeV 
    \item The scalar sum of the $p_T$ of the three final state leptons ($L_T$) : $L_T>$ 150 (75) GeV
    \item Effective mass ($M_{eff}$) : $M_{eff} >$ 500 (250) GeV
\end{itemize}
                
One of the three leptons originating from the decay of $W$ boson for different SM backgrounds  exhibits a Jacobian peak around $M_{W}/2$. Such backgrounds containing $W$ boson can be suppressed by applying a suitable cut on $p_T$ of the leading lepton.  

The variable ($m_T$) is defined as 
\begin{eqnarray}
    m_T = \sqrt{2 p_T^{miss} p_T^l [1 - \cos(\Delta\Phi_{m_T})]},
\end{eqnarray} 
where $p_T^l$ refers to the $p_T$ of the lepton which is not a part of the opposite sign (OS) lepton pair closest to the mass of $Z$ boson and $\Delta\Phi_{m_T}$ is the difference of azimuthal angle between $\vec{p}_T^{~miss}$ and $\vec{p}_T^{~l}$. The SM backgrounds show a peak for the kinematic  variable $m_T$ around $M_W$. By imposing a cut on $m_T$ ($m_T > 100$ GeV), the backgrounds are reduced significantly.

The variable $L_T$ is defined as   
         \begin{eqnarray}
	           L_T = \sum_{l=e,\mu} p_{T}^{l}. 
            \end{eqnarray}
In this multi-lepton final state the variable $L_T$ is a good discriminator between the signal events and the corresponding SM background ones. This happens because of the magnitude of $L_T$ obtained from this particular signal is remarkably different than that from the SM backgrounds. 

The variable effective mass ($M_{eff}$) is defined as
         \begin{eqnarray}
	           M_{eff} = \sum_{i} p_{T}^{j_i} + \sum_{i} p_{T}^{l_i} + \cancel{E_T}.
            \end{eqnarray}
Effective mass is a good discriminator for isolating the SUSY events from the corresponding backgrounds. The massive SUSY particles can produce jets and leptons with high $p_T$ along with a massive LSP which always produces a significant amount of transverse missing energy. Thus the combination of them, called effective mass, is strikingly different from the SM backgrounds. Hence it helps us to discriminate the signal from the potentially significant backgrounds.

The results of the traditional CCA for the final state comprising of $3l+jets+\cancel{E_T}$ are presented in Table \ref{tab:cut-3l}, which includes the normalized number of signal and background events with respect to an integrated luminosity of 139 $fb^{-1}$. The signal significances are shown in the fourth and sixth column of Table \ref{tab:cut-3l}, for the models LLRL and LLRH respectively.

The results shown in Table \ref{tab:cut-3l}, points out the fact that a very high luminosity is required for 5$\sigma$ discovery of this particular signal in the CCA method. We have used suitable cuts on some kinematic variables in order to obtain these results. Furthermore, it is noted that a set of new cuts or variables can modify the CCA results. In the following sub sec. \ref{5.2} we carry out the MVA with the same set of kinematic variables adopted in this sub sec. as the input features of the BDT in order to obtain  better significance resulting in the requirement of relatively low luminosity for the discovery of gluino through this particular channel.


\subsection{MVA} \label{5.2}
In this section, an MVA approach is employed for improved signal-to-background differentiation, resulting in increased significance. The BDT algorithm for MVA is implemented in the TMVA framework within ROOT platform. In order to improve the significance from the cut-based analysis previously discussed, finding an optimal cut value for the variables can be a challenging task with traditional rectangular cut methods. The MVA technique serves as a powerful tool for optimizing sensitivity for a given set of input features.

\subsubsection{The discovery prospect of $2l+\cancel{E_T}+jets$ signal} \label{sec5.2.1}
\textbf{Opposite sign same flavour (OSSF) lepton}

The signal comprising of opposite sign same flavour (OSSF) dileptons along with jets and the missing transverse energy stems from the production of a pair of gluinos at 13.6 TeV LHC. The potential SM backgrounds that can mimic the signal are $t\Bar{t}$, Drell-Yan, dibosons such as $WW$, $WZ$, $ZZ$, tribosons like $WWW$,$WWZ$, $WZZ$, $ZZZ$, along with $t\Bar{t}Z$, $t\Bar{t}W$, $t\Bar{t}h$ and $hW$, $hZ$. 

\begin{figure}[h!]
    \centering
    \includegraphics[width=0.7\linewidth]{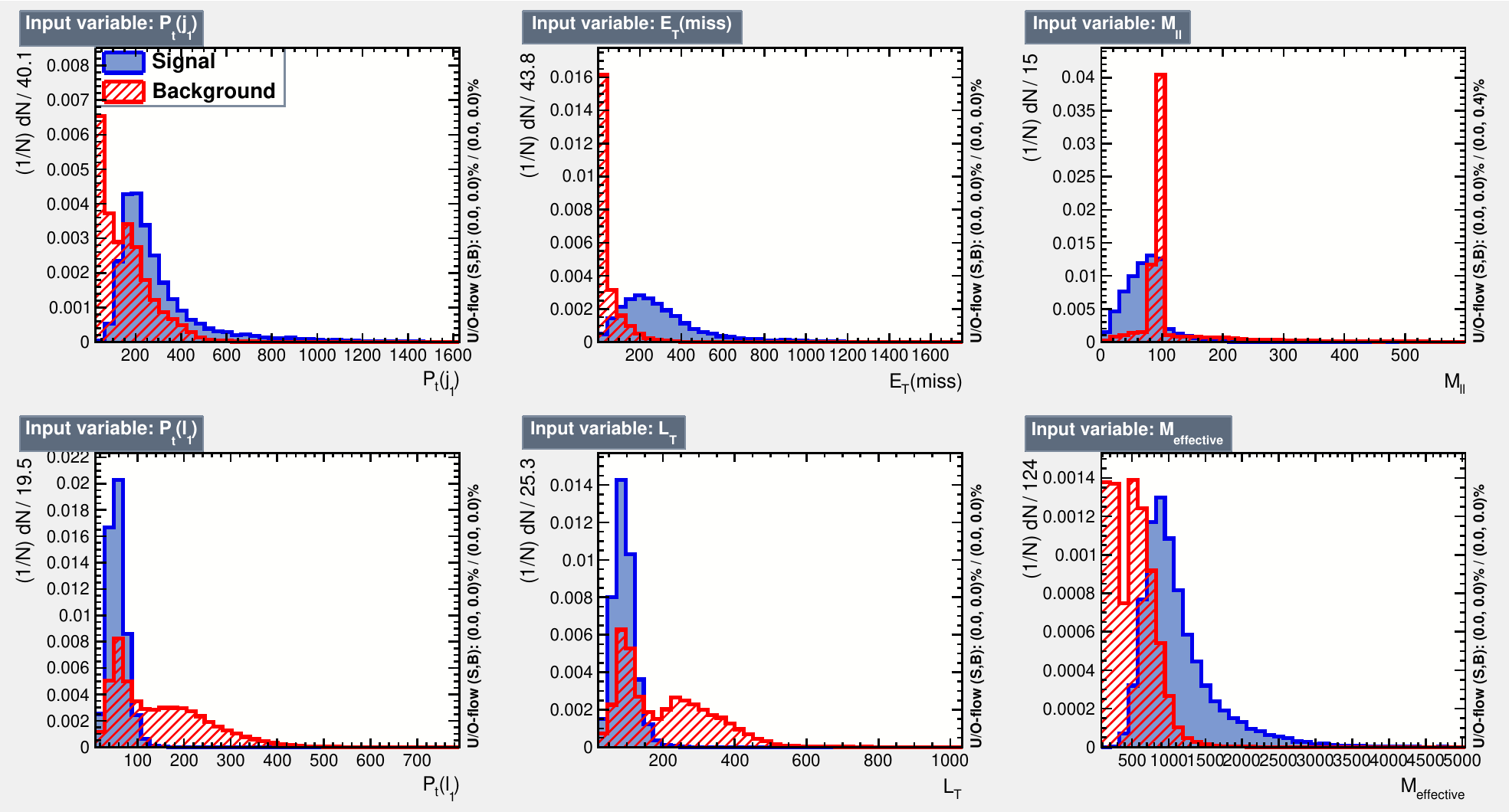}
    \caption{The signal (blue) and background (red) distributions of the input features in the final state comprising of $2l+jets+\cancel{E_T}$ with an opposite sign same flavour (OSSF) dilepton corresponding to BP2 of Light Electroweakino Light Slepton Wino (LELSW) model in left light and right heavy squarks (LLRH) scenario  have been shown in this figure.}
    \label{fig:input2l}
\end{figure}

\begin{figure}[htb]
    \centering
    \begin{minipage}{.45\textwidth}
        \centering
        \includegraphics[width=0.9\linewidth]{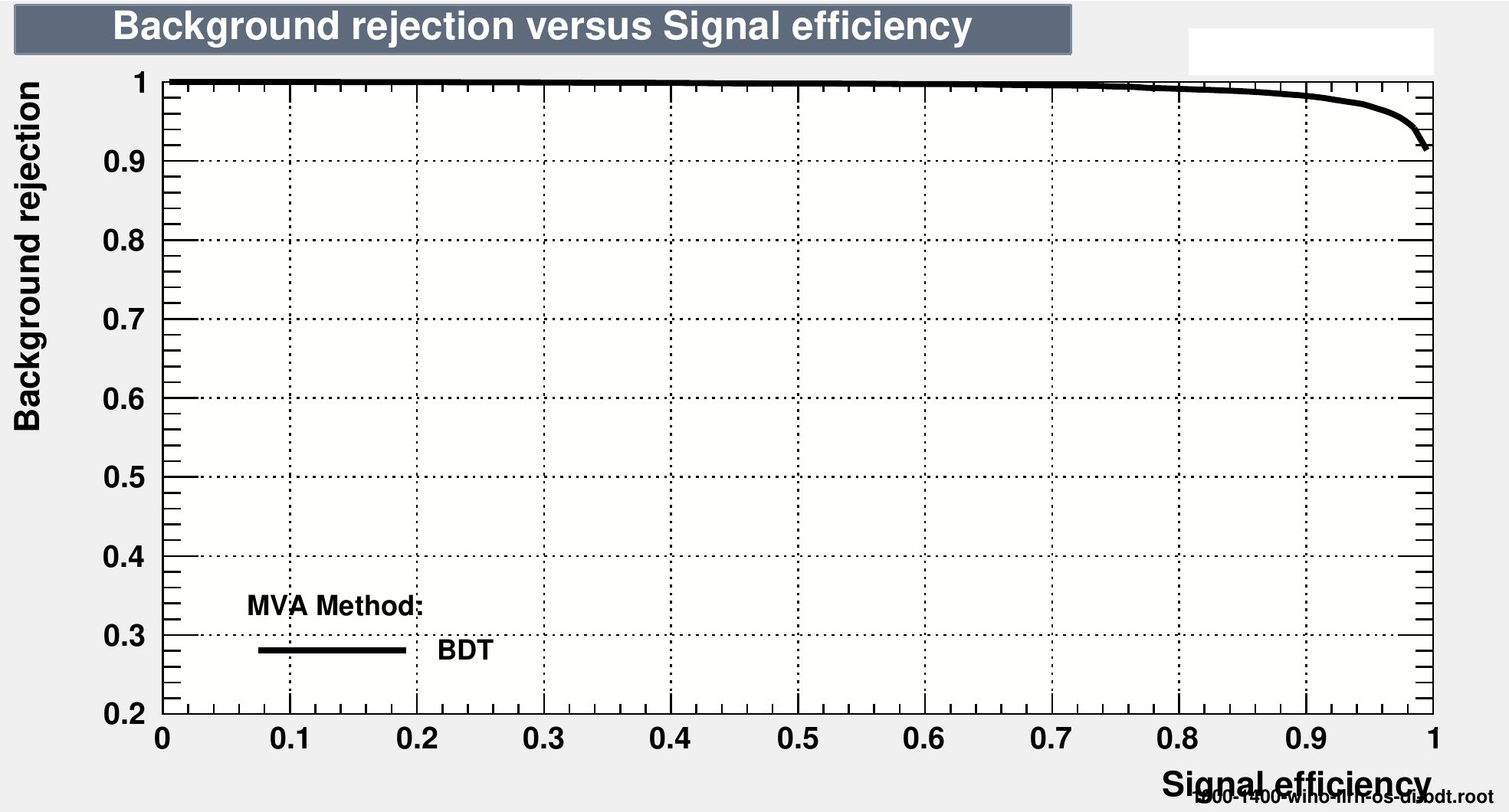}
with    \caption{ROC curve for the signal comprising of $2l+jets+\cancel{E_T}$ with an opposite sign same flavour (OSSF) lepton pair corresponding to BP2 of the same model as in Fig. \ref{fig:input2l} have been shown in this figure.}
    \label{fig:roc2l}
    \end{minipage}%
    \hspace{0.8cm}
    \begin{minipage}{0.45\textwidth}
        \centering
        \includegraphics[width=0.9\linewidth]{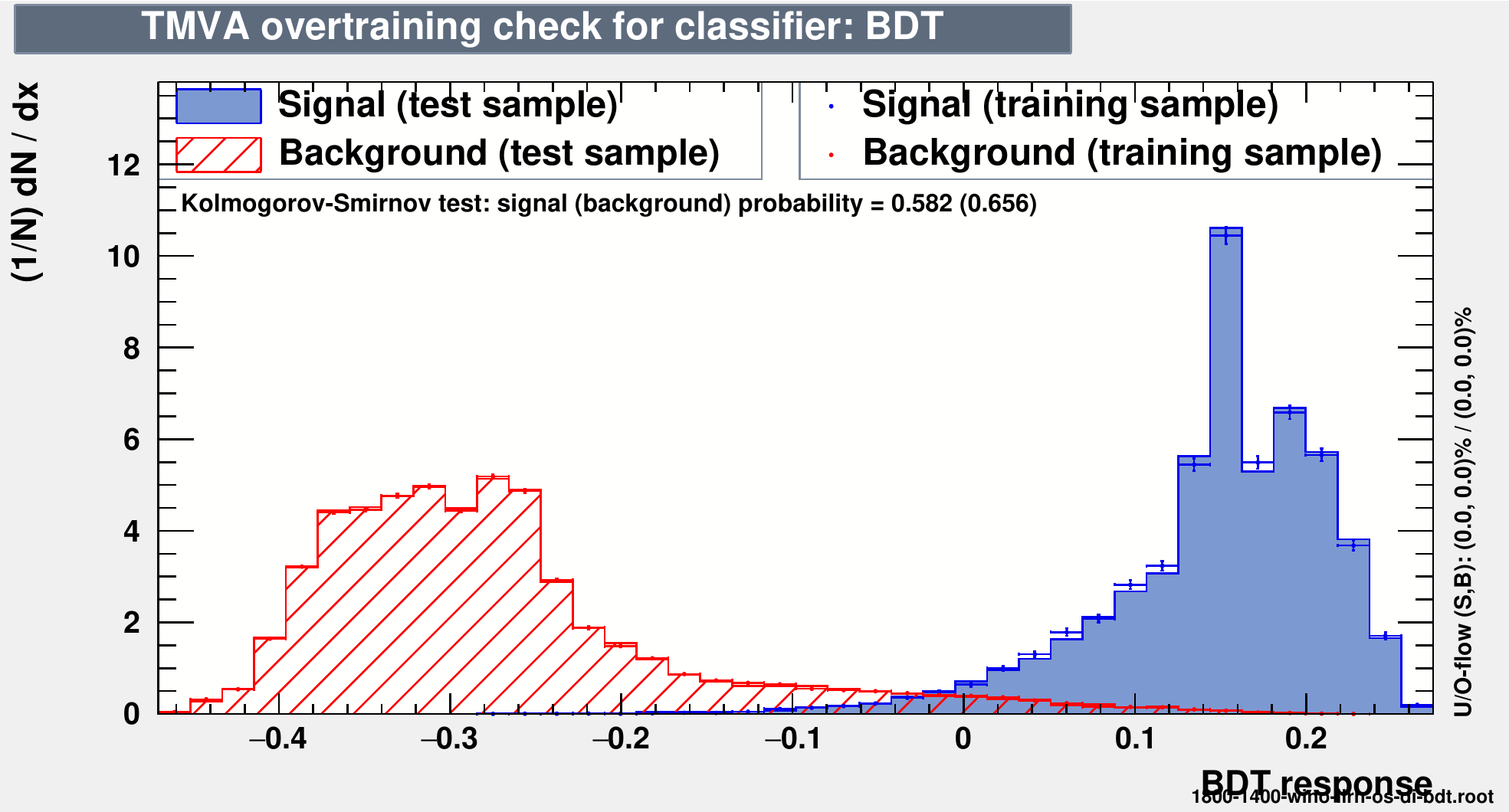}
    \caption{For the purpose of illustration we have presented the over-training check of the BDT response for the model mentioned in Fig. \ref{fig:input2l} using the parameter set of BP2.}
    \label{fig:overtrain2l}
    \end{minipage}
\end{figure}

We have used five kinematic inputs as the features of the BDT classifier to distinguish the SM backgrounds from the corresponding SUSY signal. The features are given below:
\begin{itemize}
    \item The $p_T$ of the leading jet: $p_T ^{j_1}$
    \item Missing transverse energy : $\cancel{E_T}$
    \item The invariant mass of the two opposite sign same flavour leptons : $M_{LL}$
    \item The $p_T$ of the leading lepton : $p_T ^{l_1}$
    \item The scalar sum of the $p_T$ of the two final state leptons : $L_T$\\
    \begin{eqnarray}
	L_T = \sum_{l=e,\mu} p_{T}^{l}. \nonumber
    \end{eqnarray}
    \item Effective mass : $M_{eff}$
\end{itemize}
\begin{table}[htb]
    \centering
    \begin{tabular}{||c|c||}
    \hline
         Variable & Variable Importance  \\
         \hline
         $M_{eff}$ & $2.514 \times 10^{-1}$ \\
         \hline
         $p_T^{j_1}$ & $2.257 \times 10^{-1}$ \\
         \hline
         $p_T^{l_1}$ & $1.801 \times 10^{-1}$ \\ 
         \hline
         $\cancel{E_T}$ & $1.510 \times 10^{-1}$ \\ 
         \hline
         $L_T$ & $1.249 \times 10^{-1}$ \\
         \hline
        $M_{LL}$ & $6.687 \times 10^{-2}$ \\ 
         \hline
    \end{tabular}
    \caption{The method-specific ranking of the input features used in the BDT for discriminating $2l+jets+\cancel{E_T}$ final state with an opposite sign same flavour (OSSF) dilepton from the corresponding backgrounds have been shown in the table.}
    \label{tab:var-imp-2l}
\end{table}
In this study, we employ the basic kinematic variables mentioned above because they have less correlation and demonstrate a considerable power to distinguish the SUSY signal over the SM backgrounds. In Table \ref{tab:var-imp-2l}, we have provided these variable together with their method-specific ranking in the BDT response. This may vary slightly depending upon the specific set of parameters corresponding to BP1 and BP2. The event distributions which have been normalised to the luminosity have been presented in Fig. \ref{fig:input2l}. From Fig. \ref{fig:input2l} we can infer that altogether these five variables have a good amount of discriminating power. In Fig. \ref{fig:roc2l} we present the receiver operating characteristic (ROC) curve. The area below the ROC curve provides a clear indication of the considerable separation between the signal and backgrounds and quantify the combined performance of the BDT. 

The Kolmogorov-Smirnov (KS) test can be used to check if a test sample is over-trained. Generally, a KS probability between 0.1 and 0.9 indicates that the test sample is not over-trained. A critical KS probability value greater than 0.01 confirms that the samples are not over-trained in most cases. The Fig. \ref{fig:overtrain2l} showing the KS probability values for the signal and backgrounds of the BDT response indicates that neither the signal nor the background samples are over-trained. 
As seen in Fig. \ref{fig:overtrain2l}, the signal and background samples in this BDT output are well-separated, enabling us to significantly enhance the signal significance by applying an appropriate BDT cut.

In Table \ref{tab:sig-2l} we present the yield of signal events and that of the background events for wino, higgsino and wino-higgsino mixed (wiggsino) type models with two variants of each model namely LLRL and LLRH (see sec. \ref{sec2}). The BDT cut values for different models for discriminating signals over the backgrounds are shown in columns 3 and 7. The projected luminosity for the discovery of gluinos in terms of significances through this channel are presented in columns 6 and 10. 

From the results it is clear that for wino type models the discovery prospect at the LHC RUN-III is promising compared to other higgsino and the wiggsino ones as the required luminosity is low for the former one. This is because, in wino type models gluino mostly decays into lighter eweakinos as those are wino dominated. Moreover, as the sleptons are lighter than the eweakinos,  they further decay into either sleptons or sneutrinos with 100$\%$ BR. This results in the final state to be comprised of two OSSF leptons, whereas in higgsino type scenarios the gluinos decay into the heavier eweakinos (wino dominated) with a substantial amount of BR (see Table \ref{tab:CHBR}). The heavier eweakinos further decay into the SM particle along with LSP through a long cascades resulting in depreciation in the probability of getting two leptons in the final state. However, for mixed case the results are intermediate between the two.

\FloatBarrier
\begin{table}[htb]
    \centering
    \resizebox{\textwidth}{!}{
    \begin{tabular}{||c|c|c|c|c|c|c|c|c|c||}
        \hline
        \multirow{2}{*}{BP} & \multirow{2}{*}{Model} & \multicolumn{4}{|c|}{LLRL} & \multicolumn{4}{|c|}{LLRH} \\
         \cline{3-10}
          &  &\makecell[c]{BDT \\cut value} & $N_S$ & $N_B$  & \makecell[c]{Required Luminosity\\ for discovery \\(fb$^{-1}$)} & \makecell[c]{BDT \\cut value} & $N_S$ & $N_B$ & \makecell[c]{Required Luminosity\\ for discovery \\(fb$^{-1}$)} \\
         \hline
         \multirow{3}{*}{BP1} & LELSW & 0.352 & 0.98 & -  & 710 & 0.323 & 1.82 & - & 380 \\
         \cline{2-10}
         & Wiggsino & 0.353 & 0.33 & - & 2110 & 0.345 & 0.73 & - & 950\\
         \cline{2-10}
         & LELSH & 0.314 & 0.21 & - & 3300 & 0.329 & 0.29 & - & 2380 \\
         \hline
         \multirow{3}{*}{BP2} & LELSW & 0.212 & 2.29 & 1.35 & 1320 & 0.225 & 1.91 & 0.26 & 640\\
         \cline{2-10}
         & Wiggsino & 0.224 & 0.41 & - & 1700 & 0.220 & 1.86 & 0.48 & 840\\
         \cline{2-10}
         & LELSH & 0.227 & 0.39 & - & 1750 & 0.256 & 0.72 & - & 965 \\
         \hline
    \end{tabular}}
    \caption{The number of $2l+\cancel{E_T}+jets$ (OSSF) signal ($N_S$) events and the corresponding cumulative background ($N_B$) events after passing the BDT cut normalised to integrated luminosity 139 $fb^{-1}$ with centre of mass energy 13.6 TeV at the LHC have been displayed in the table. In addition with this we have considered 10$\%$ systematic uncertainty  to the overall backgrounds. Here we have also shown the required luminosity to achieve a potential for discovery. '-' denotes the negligible background ($N_B\rightarrow0$) events compared to the signal events. In case of the backgrounds to be negligible,  5 signal events are considered to be the discovery criteria.}
    \label{tab:sig-2l}
\end{table}

Moreover, such models when subcategorised in terms of squark mass hierarchy, namely, LLRL and LLRH variants, it is observed that the LLRH case gives the better significance across all class of models. This is due to the fact that if R-squarks are made heavier, the BR of gluinos into bino-like LSP decreases and the BR of chargino enhances thereby increasing the strength of $2l+jets+\cancel{E_T}$ signal. As a result for observing the signal in LLRH wino model, the requirement of integrated luminosity is always smaller compared to the other one. The best result for this particular signal is obtained for BP1 in wino type LLRH scenario since the the required luminosity is about 380 $fb^{-1}$, which may be achieved at an early stage of LHC RUN-III. However, there are other scenarios which can give hint for discovery of gluinos for relatively low luminosity. 

For higgsino type models in LLRH scenario, the BP2 corresponds to integrated luminosity of 965 $fb^{-1}$ for discovery which could be a promising search channel to be probed at the early stage of HL-LHC.

\textbf{ Same sign same flavour (SSSF) dilepton}

In this sub sec., we explore the potential for discovering gluinos in the final state consisting of a pair of SSSF leptons in association with jets and $\cancel{E_T}$. The SM backgrounds for this final state are  $t\Bar{t}$, dibosons such as $WZ$, tribosons like $WWW$,$WWZ$, $WZZ$, along with $t\Bar{t}Z$, $t\Bar{t}W$, and $hW$. Here the backgrounds being relatively weaker compared to the OSSF final states. Hence the presence of SSSF lepton pairs from the Majorana fermions in the SUSY signal makes this search channel worth investigating.  

The following four features considered in the previous search channel are used here to distinguish the required signal from the backgrounds.  The four distinguishing features are given below:
\begin{itemize}
    \item The $p_T$ of the leading jet: $p_T^ {j_1}$
    \item Missing transverse energy : $\cancel{E_T}$
    \item The $p_T$ of the leading lepton : $p_T ^{l_1}$
    \item The scalar sum of the $p_T$ of the two final state leptons : $L_T$
    \item Effective mass : $M_{eff}$
\end{itemize}

\begin{figure}[h!]
    \centering
    \includegraphics[width=0.7\linewidth]{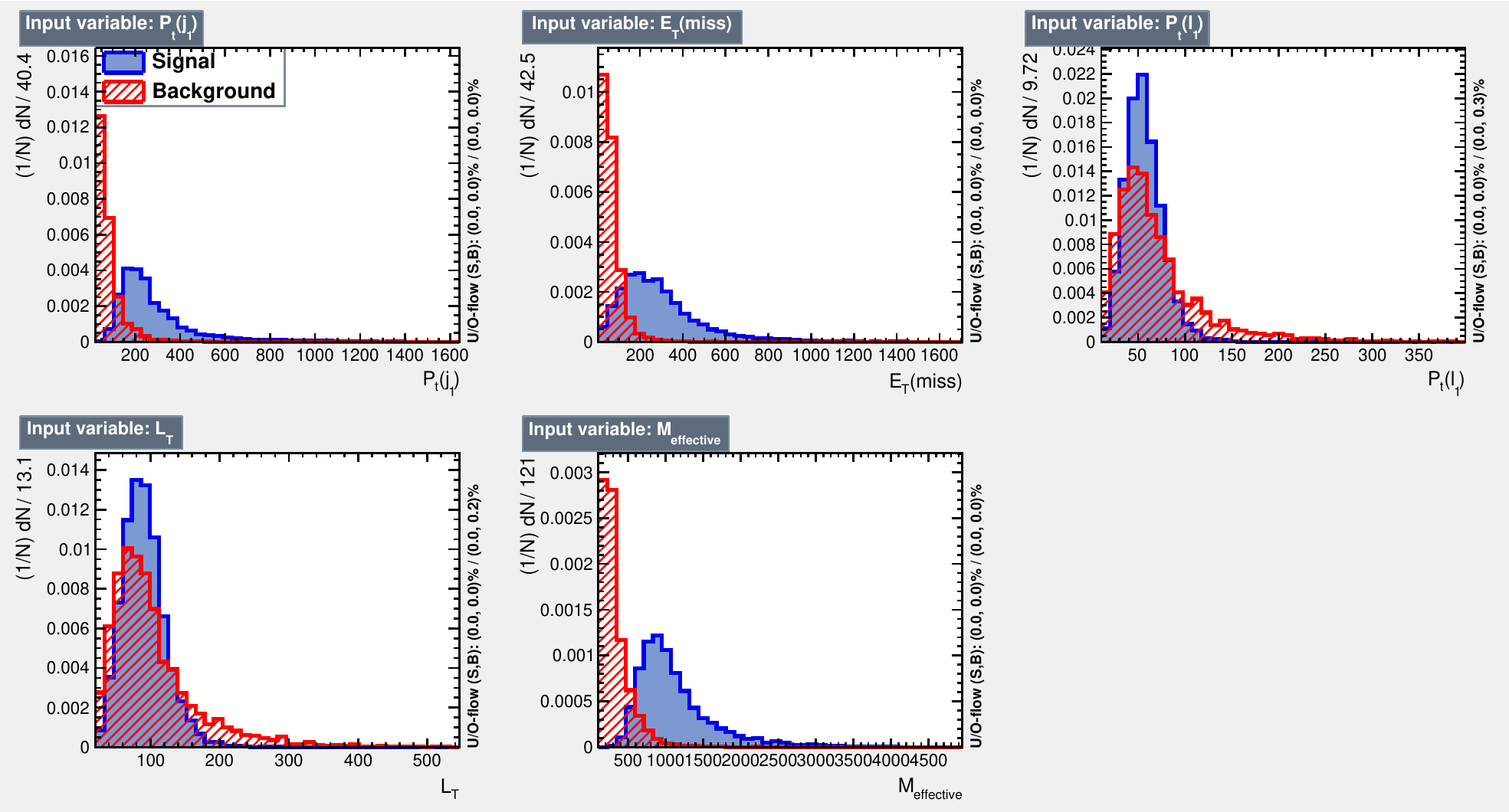}
    \caption{The signal (blue) and background (red) distributions of the input features in the final state of comprising $2l+jets+\cancel{E_T}$ with same sign same flavour (SSSF) lepton pair corresponding to BP2 of the same model mentioned in Fig. \ref{fig:input2l} have been shown in this figure.}
    \label{fig:input2lss}
\end{figure}

\begin{table}[h!]
    \centering
    \begin{tabular}{||c|c||}
    \hline
         Variable & Variable Importance  \\
         \hline
         $M_{eff}$ & $2.290 \times 10^{-1}$ \\
         \hline
         $\cancel{E_T}$ & $2.105 \times 10^{-1}$ \\ 
         \hline
         $p_T ^{j_1}$ & $1.966 \times 10^{-1}$ \\
         \hline
         $L_T$ & $1.844 \times 10^{-1}$ \\
         \hline
         $p_T ^{l_1}$ & $1.795 \times 10^{-1}$ \\
         \hline
    \end{tabular}
    \caption{The method-specific ranking of the input features that are used in the BDT to discriminate the $2l++jets+\cancel{E_T}$ SSSF final state have been shown in the table.}
    \label{tab:var-imp-2lss}
\end{table}
We present the importance of these variables in terms of their ranking in terms of BDT response in Table \ref{tab:var-imp-2lss}. The event distributions, normalized to the luminosity, are shown in Fig. \ref{fig:input2lss}. From this figure, we can see that these four variables possess a considerable amount of discriminating power.

\begin{figure}[h!]
\begin{minipage}{.45\textwidth}
    \centering
    \includegraphics[width=0.9\linewidth]{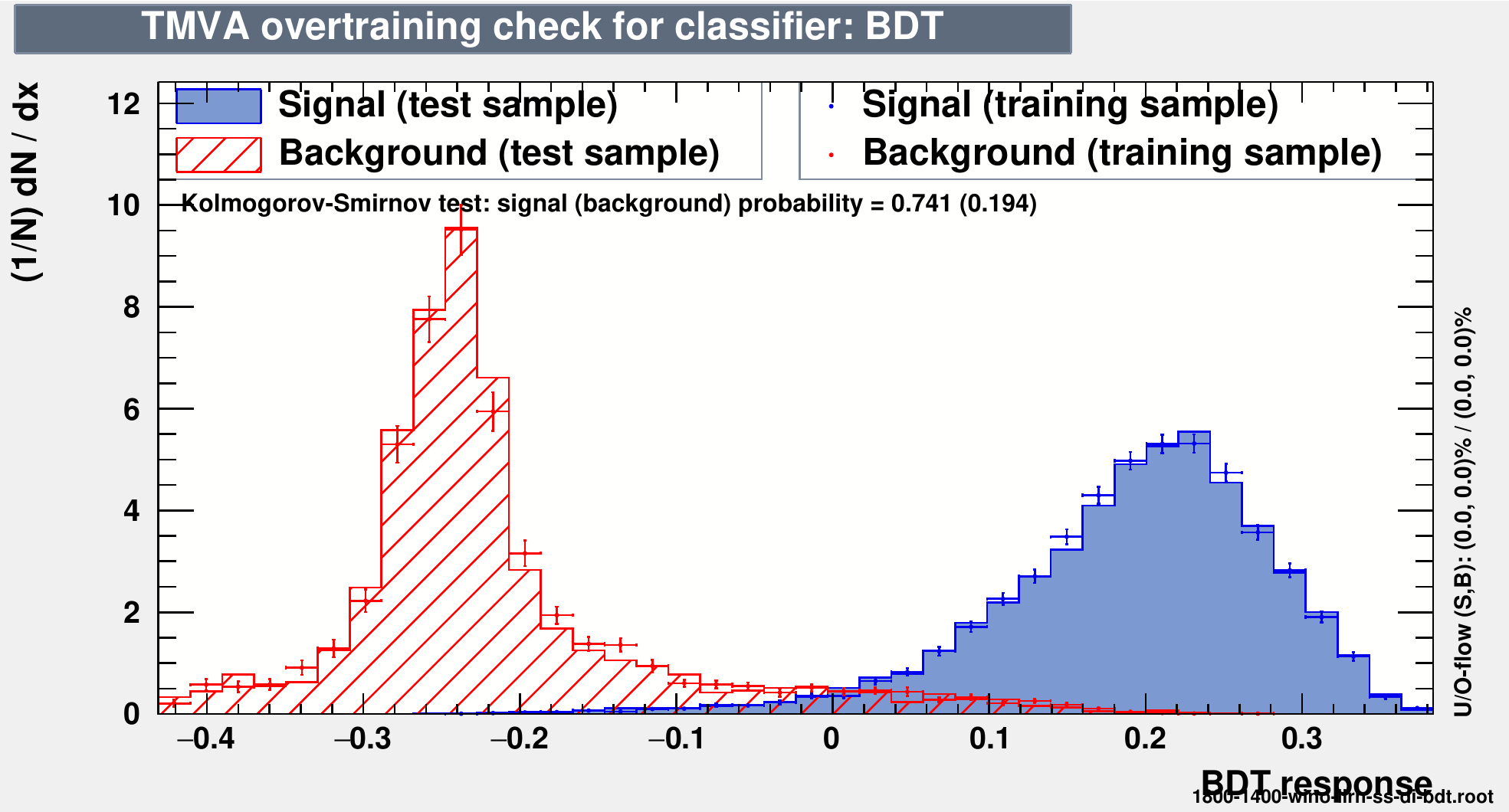}
    \caption{For the purpose of illustration we have presented the over-training check of the BDT response for the model mentioned in Fig. \ref{fig:input2l} with the final state comprising of same sign same flavour (SSSF) pair of lepton in $2l+jets+\cancel{E_T}$ signal using the parameter set of BP2.}
    \label{fig:overtrain2lss}
\end{minipage}
\hspace{0.8cm}
\begin{minipage}{.45\textwidth}
    \centering
    \includegraphics[width=0.9\linewidth]{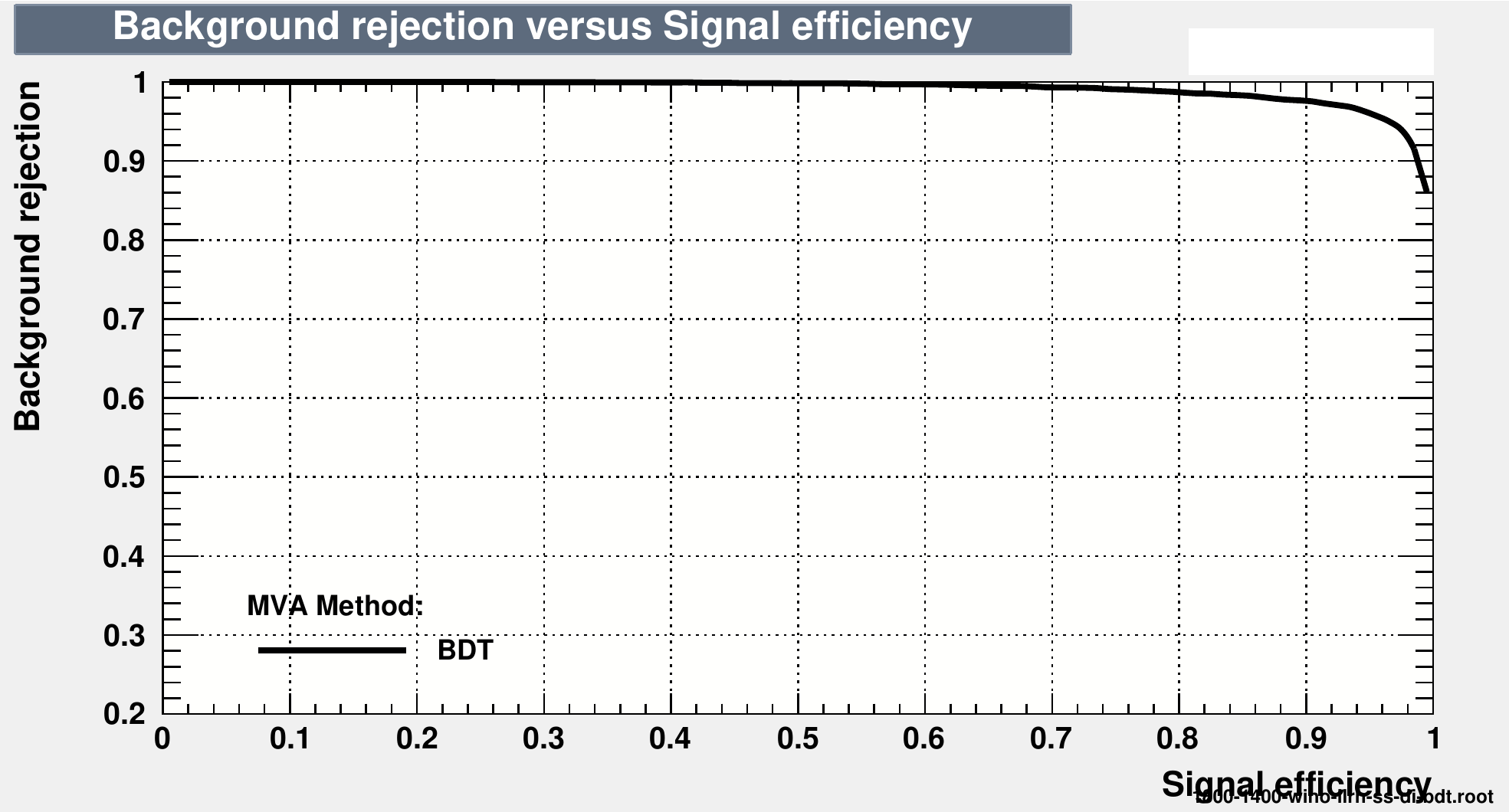}
    \caption{ROC curve for the model mentioned in Fig. \ref{fig:input2l} corresponding to the final state comprising of $2l+jets+\cancel{E_T}$ with a SSSF pair.}
    \label{fig:roc2lss}
\end{minipage}
\end{figure}

Finally, to ensure the classifier to be not over-trained, we perform a KS test, which compares the BDT response curves of the training and testing sub-samples. From Fig. \ref{fig:overtrain2lss}, we can infer that the response curves do not exhibit any significant over-training. Additionally for this topology of the signal we have presented the ROC curve in the Fig. \ref{fig:roc2lss}.

\FloatBarrier
\begin{table}[h!]
    \centering
    \resizebox{\textwidth}{!}{
    \begin{tabular}{||c|c|c|c|c|c|c|c|c|c||}
        \hline
        \multirow{2}{*}{BP} & \multirow{2}{*}{Model} & \multicolumn{4}{|c|}{LLRL} & \multicolumn{4}{|c|}{LLRH} \\
         \cline{3-10}
          &  &\makecell[c]{BDT \\cut value} & $N_S$ & $N_B$ & \makecell[c]{Required Luminosity\\ for discovery \\(fb$^{-1}$)} & \makecell[c]{BDT \\cut value} & $N_S$ & $N_B$ & \makecell[c]{Required Luminosity\\ for discovery \\(fb$^{-1}$)} \\
         \hline
         \multirow{3}{*}{BP1} & LELSW & 0.430 & 0.87 & 0.31 & 2430 & 0.471 & 0.31 & - & 2240  \\
         \cline{2-10}
         & Wiggsino & 0.449 & 0.63 & 0.21 & 3300 & 0.436 & 1.48 & 2.10 & 4000 \\
         \cline{2-10}
         & LELSH & 0.506 & 0.18 & - & 3860 & 0.548 & 0.23  & - & 3050 \\
         \hline
         \multirow{3}{*}{BP2} & LELSW & 0.227 & 3.10 & 5.93 & 2500 & 0.211 & 8.44 & 7.29 & 490\\
         \cline{2-10}
         & Wiggsino & 0.272 & 0.18 & - & 3860 & 0.220 & 3.21 & 2.68 & 1230 \\
         \cline{2-10}
         & LELSH & 0.243 & 0.03 & - & $--$ & 0.238 &  0.85 & 2.14 & $--$ \\
         \hline
    \end{tabular}}
    \caption{The number of $2l+jets+\cancel{E_T}$ (SSSF) signal ($N_S$) and the corresponding cumulative background ($N_B$) events after passing the BDT cut normalised to integrated luminosity 139 $fb^{-1}$ with centre of mass energy 13.6 TeV at the LHC have been displayed in the table. In addition with this we have considered 10$\%$ systematic uncertainty  to the overall backgrounds. In this table we have also shown the required luminosity to achieve a potential for discovery. `- ' denotes the negligible background ($N_B\rightarrow0$) compared to the signal. In case the background is negligible 5 signal events are considered as the requirement for the discovery. `$--$' denotes that the required luminosity is equal or higher than 5000 $fb^{-1}$ and hence the prospect of gluino discovery even in the HL-LHC is not so optimistic for these BPs.}
    \label{tab:sig-2lss}
\end{table}

In Table \ref{tab:sig-2lss} we have shown the normalised event number which have passed the BDT cut value. From the table we can infer that gluinos have the best discovery potential corresponding to an integrated luminosity of 490 $fb^{-1}$ in the final state comprising of $2l+jets+\cancel{E_T}$ with a pair of SSSF lepton at the HL-LHC corresponding to BP2 for LLRH sub-variant of LELSW model. This happens because the strength of the leptonic signal is greater for LLRH sub-variant compared to the LLRL of the wino type model among all the models considered by us. Furthermore, due to the greater OSSF leptonic signal for this particular signal, the prospect of the gluino is more promising than the corresponding SSSF signal which is readily observable from Tables \ref{tab:sig-2l} and \ref{tab:sig-2lss}.

\subsubsection{The discovery prospect of $3l+jets+\cancel{E_T}$ signal} \label{5.2.2}
The pair production of strongly interacting gluinos can result in final state containing $3l+jets+\cancel{E_T}$ signature in various models described in sec. \ref{sec2} at LHC RUN-III. The important SM backgrounds for this particular decay topology have been mentioned in the sub sec. \ref{sec.5.1.1}.


In this sub sec. we have also used five simple kinematic features in the BDT to discriminate the  signal from the SM backgrounds. The features are :
\begin{itemize}
    \item The $p_T$ of the leading jet: $p_T ^{j_1}$
    \item Missing transverse energy : $\cancel{E_T}$
    \item The transverse mass : $m_T$\\
        The variable ($m_T$) is defined as $m_T = \sqrt{2 p_T^{miss} p_T^l [1 - \cos(\Delta\Phi_{m_T})]}$, where $p_T^l$ refers to the $p_T$ of the lepton that is not part of the OS pair closest to the Z boson mass and $\Delta\Phi_{m_T}$ is the difference in azimuth angle between missing transverse momentum ($\vec{p}_T^{~miss}$) and $\vec{p}_T^{~l}$.
    \item The $p_T$ of the leading lepton : $p_T ^{l_1}$
    \item The scalar sum of the $p_T$ of the three final state leptons : $L_T$
    \item Effective mass : $M_{eff}$
\end{itemize}

\begin{figure}[htb]
    \centering
    \includegraphics[width=0.7\linewidth]{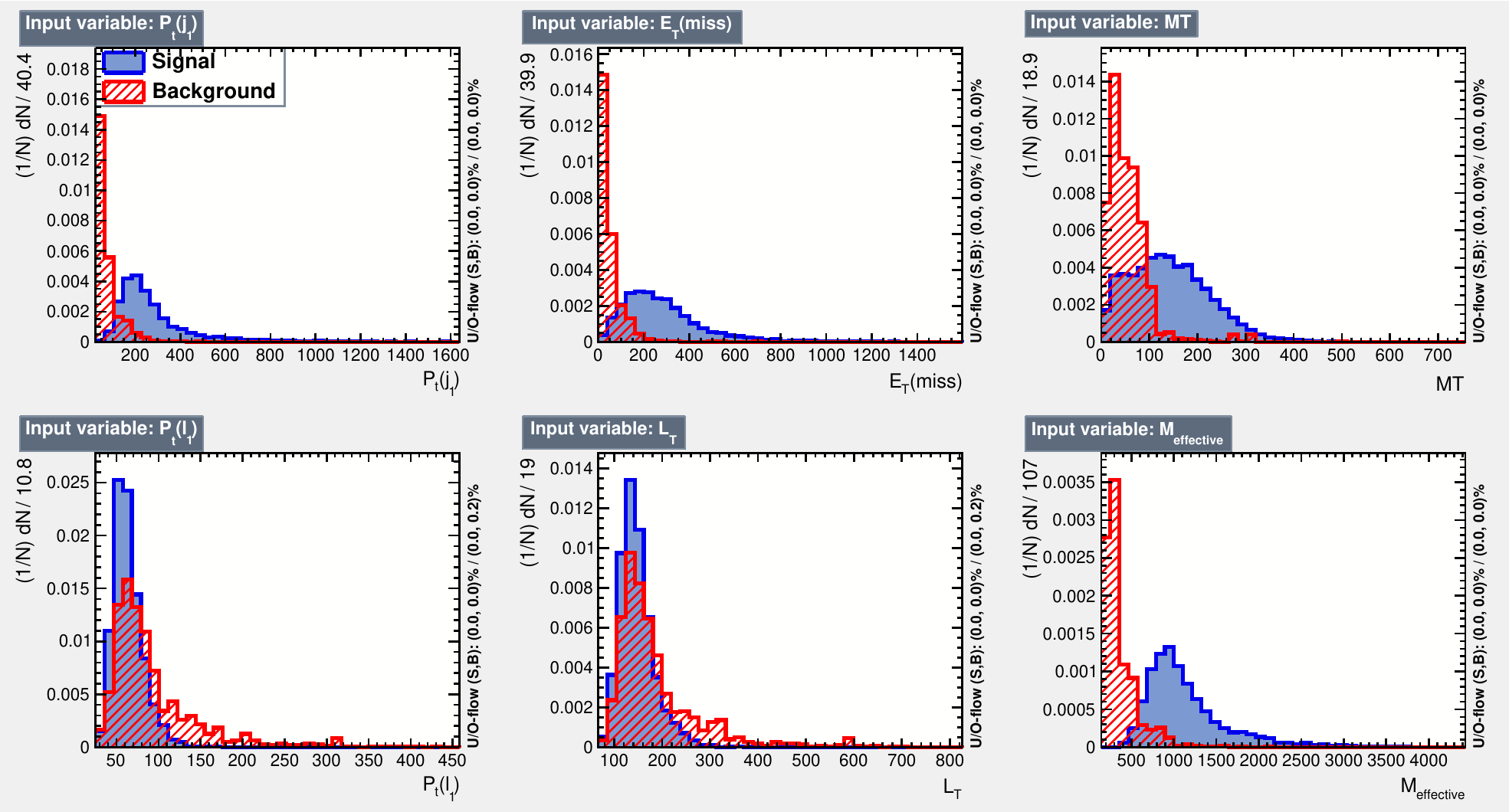}
    \caption{The signal (blue) and background (red) distributions of the input features in the final state of comprising of $3l+jets+\cancel{E_T}$ corresponding to BP2 of the Light Electroweakino Light Slepton Wino model (LELSW) where R-squarks are heavy (LLRH) have been shown in this figure.}
    \label{fig:input3l}
\end{figure}

In addition to the four features which we have used in the last sub sec., we consider the transverse mass as the distinguishing feature for taming down the potential backgrounds significantly. It is further noted from Fig. \ref{fig:input3l} that the transverse mass feature has a significant power to differentiate the signal from the large SM backgrounds containing $W$ boson.

\begin{table}[htb]
    \centering
    \begin{tabular}{||c|c||}
    \hline
         Variable & Variable Importance  \\
         \hline
         $M_{eff}$  & $1.929 \times 10^{-1}$ \\
         \hline
         $m_{T}$ & $1.737 \times 10^{-1}$ \\
         \hline
         $p_T^{l_1}$ & $1.647 \times 10^{-1}$ \\
         \hline
         $\cancel{E_T}$ & $1.610 \times 10^{-1}$ \\
         \hline
         $L_T$ & $1.579 \times 10^{-1}$ \\
         \hline
         $p_T^{j_1}$ & $1.498 \times 10^{-1}$ \\ 
         \hline
    \end{tabular}
    \caption{The method-specific ranking of the input features that are used in the BDT to discriminate the $3l+\cancel{E_T}+jets$ final state have been shown in the table.}
    \label{tab:var-imp-3l}
\end{table}

\begin{figure}[h!]
\begin{minipage}{0.45\textwidth}
    \centering
    \includegraphics[width=0.9\linewidth]{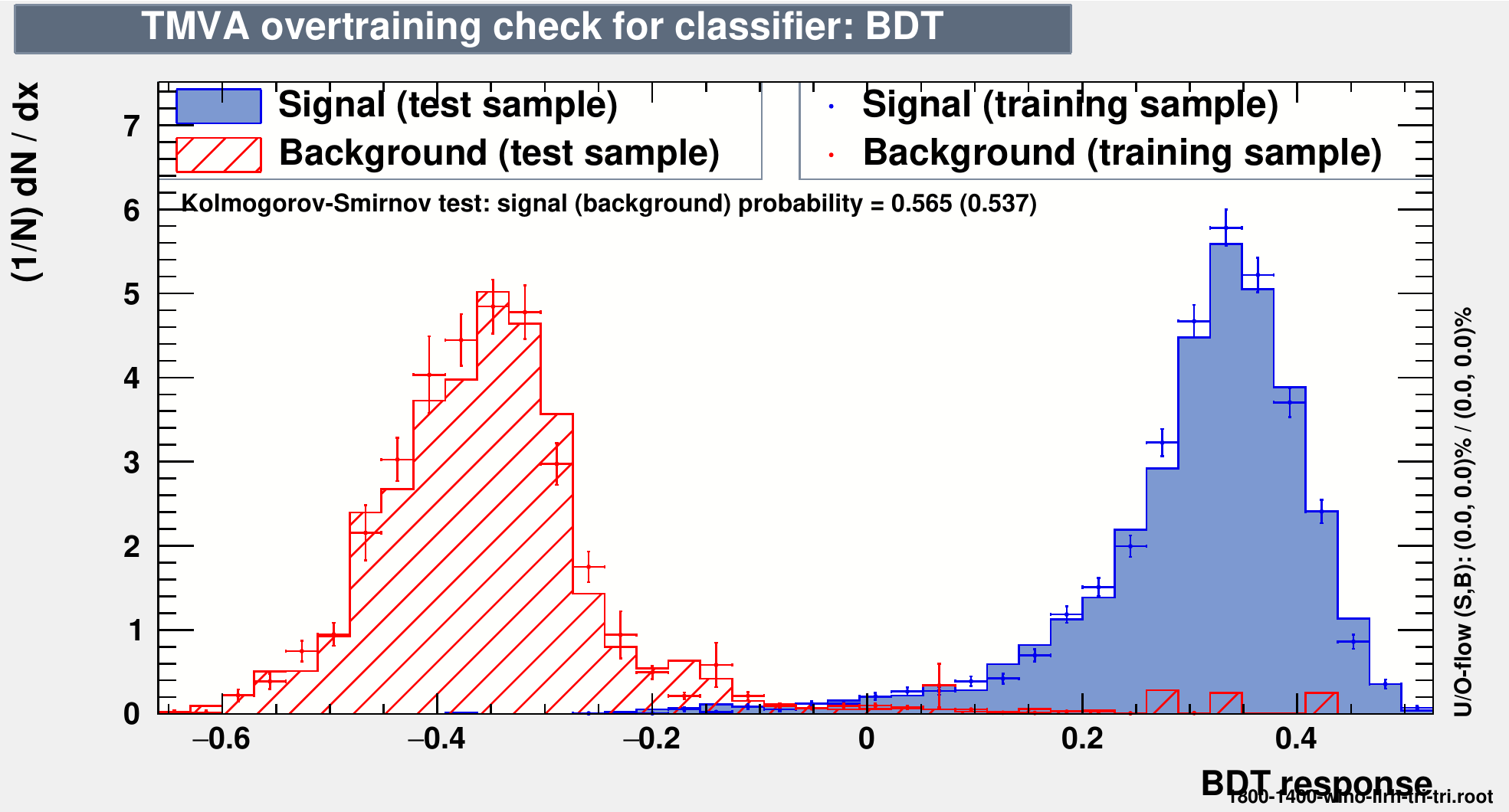}
    \caption{For the purpose of illustration we have presented the over-training check of the BDT response for the model mentioned in Fig. \ref{fig:input3l} with final state comprising of $3l+jets+\cancel{E_T}$ signal using the parameter set of BP2.}
    \label{fig:overtrain3l}
\end{minipage}
\hspace{0.8cm}
\begin{minipage}{0.45\textwidth}
    \centering
    \includegraphics[width=0.9\linewidth]{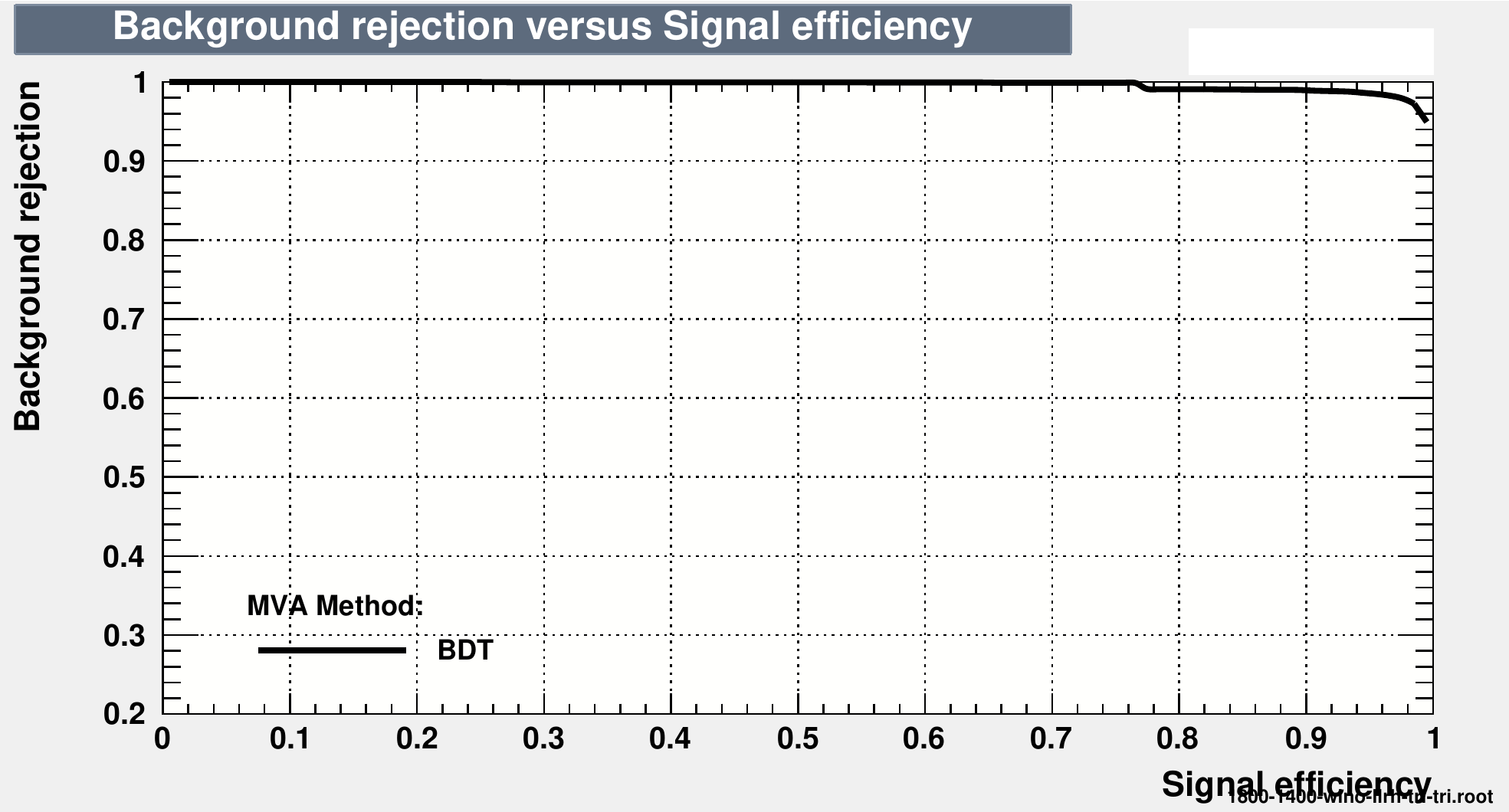}
    \caption{ROC curve corresponding to the model mentioned in Fig. \ref{fig:input3l} for the signal state comprising of $3l+jets+\cancel{E_T}$ have been shown in this figure.}
    \label{fig:roc3l}
\end{minipage}
\end{figure}

From Table \ref{tab:var-imp-3l}, where we have presented the method specific variable importance in the BDT output response, it can easily be seen that these features have remarkable importance to identify the signal. Additionally, in Fig. \ref{fig:input3l} we have presented the distribution of weighted number of the signal and background events against each feature used here.   
Next, to check whether the classifier has over-trained or not, we perform the KS test, which compares the BDT response curves of the training and testing sub samples, as seen in Fig. \ref{fig:overtrain3l}. The response curves are well within the tolerance of the over-training. In addition with the KS test we have presented the ROC curve in Fig. \ref{fig:roc3l} for this decay topology. The area under the ROC curve infer that the chosen features are altogether very good discriminator between the signal and the SM backgrounds.

Table \ref{tab:sig-3l} displays the number of signal events and the background events in response to the classifier cut value. The description of the table is as described in the previous section. Now we discuss the important results obtained in this analysis.

BP2 corresponding to the lighter gluino mass has larger production cross section compared to BP1. From the table it is observed that Wino type models  are more promising than other models in terms of discovery prospects of gluino. 

The three lepton signal can primarily arise from gluinos decaying into lighter charginos and second lightest neutralino leading to final state comprising of three leptons in wino type scenario. In other models (e.g., higgsino type) the three lepton signal may come from the decays of heavier eweakinos. However, the probability of obtaining this signal is small as the leptons arise through long cascades resulting in reduction in the signal strength. Hence luminosity requirement for observability of signal is high for such models compared to wino type models.

It is further noticed from Table \ref{tab:sig-3l} that for BP2 in LLRH model even for 270 fb$^{-1}$ luminosity the signal reaches the 5$\sigma$ discovery limit. This is the best probe for gluino search at HL-LHC which corresponds to the lowest integrated luminosity. The LLRL models correspond to high luminosity compared to LLRH models for discovery prospects which we have discussed in sub sec. \ref{sec5.2.1}. Furthermore, due to the low SM backgrounds corresponding to this signal topology, the significance is greater resulting in the required luminosity to be less compared to the signal comprising of $2l+jets+\cancel{E_T}$.
 

\FloatBarrier
\begin{table}[h!]
    \centering
    \resizebox{\textwidth}{!}{
    \begin{tabular}{||c|c|c|c|c|c|c|c|c|c||}
        \hline
        \multirow{2}{*}{BP} & \multirow{2}{*}{Model} & \multicolumn{4}{|c|}{LLRL} & \multicolumn{4}{|c|}{LLRH} \\
         \cline{3-10}
          &  &\makecell[c]{BDT \\cut value} & $N_S$ & $N_B$  & \makecell[c]{Required Luminosity\\ for discovery \\(fb$^{-1}$)} & \makecell[c]{BDT \\cut value} & $N_S$ & $N_B$ & \makecell[c]{Required Luminosity\\ for discovery \\(fb$^{-1}$)} \\
         \hline
         \multirow{3}{*}{BP1} & LELSW & 0.429 & 0.61 & 0.11  & 2550 & 0.401 & 1.06 & 0.33 & 1855 \\
         \cline{2-10}
         & Wiggsino & 0.488 & 0.52 & 0.18 & 3850 & 0.455 & 0.90 & 0.21 & 2050\\
         \cline{2-10}
         & LELSH & 0.345 & 0.46 & 0.11 & 3550 & 0.411 & 0.68 & 0.18 & 2600 \\
         \hline
         \multirow{2}{*}{BP2} & LELSW & 0.344 & 2.30 & 0.59 & 720 & 0.270 & 7.46 & 2.34 & 270\\
         \cline{2-10}
         & Wiggsino & 0.356 & 0.21 & - & 3310 & 0.226 & 4.45 & 1.56 & 480\\
         \cline{2-10}
         & LELSH & 0.379 & 0.28 & - & 2480 & 0.247 & 1.76 & 1.38 & 2090 \\
         \hline
    \end{tabular}}
    \caption{Numbers of $3l+jets+\cancel{E_T}$ signal ($N_S$) and the corresponding cumulative background ($N_B$) events after passing the BDT cut normalised to integrated luminosity 139 $fb^{-1}$ with centre of mass energy 13.6 TeV at the LHC have been displayed in the table. In addition with this we have considered 10$\%$ systematic uncertainty  to the overall backgrounds. In this table we have also shown the required luminosity to achieve a potential for discovery. '- ' denotes the negligible background ($N_B\rightarrow0$) compared to the signal. In case the background is negligible 5 signal events are considered as the requirement for the discovery. }
    \label{tab:sig-3l}
\end{table}

\subsubsection{The discovery prospect of $4l+jets+\cancel{E_T}$ signal}
Gluinos can be probed through the $4l+jets+\cancel{E_T}$ channel by reducing the backgrounds to a negligibly small level using MVA technique. 
 The cascade decay of heavier eweakinos leading to the signal comprising of $4l+\cancel{E_T}+jets$, causes the signal strength to be weaker compared to the other signatures discussed previously. For the four lepton final state, the dominant backgrounds from the SM are processes involving three or two vector bosons such as $ZZ$, $ZZZ$, $WZZ$, $WWZ$, top quark pair production in association with a $Z$ boson $(t\Bar{t}Z)$, top quark production in association with a Higgs boson $(t\Bar{t}h)$ and Higgs boson production in association with a $Z$ boson $(hZ)$. 
 The five kinematic variables that are used as the input features of the BDT are :
\begin{itemize}
    \item The $p_T$ of the leading jet: $p_T^{j_1}$
    \item Missing energy : $\cancel{E_T}$
    \item The invariant mass of the four leptons : $M_{4L}$
    \item The $p_T$ of the leading lepton : $p_T^{l_1}$
    \item The scalar sum of the $P_T$ of the four final state lepton : $L_T$
    \item Effective mass : $M_{eff}$
\end{itemize}
\begin{table}[htb]
    \centering
    \begin{tabular}{||c|c||}
    \hline
         Variable & Variable importance  \\
         \hline
         $M_{eff}$ & $2.377 \times 10^{-1}$ \\ 
         \hline
         $p_T^{l_1}$ & $1.682 \times 10^{-1}$ \\
         \hline
         $L_T$ & $1.634 \times 10^{-1}$ \\ 
         \hline
         $\cancel{E_T}$ & $1.593 \times 10^{-1}$ \\  
         \hline
         $M_{4l}$ & $1.430 \times 10^{-1}$ \\
         \hline
         $p_T^{j_1}$ & $1.284 \times 10^{-1}$ \\  
         \hline
    \end{tabular}
    \caption{The method-specific ranking of the input features that are used in the BDT to discriminate the $4l+jets+\cancel{E_T}$ final state have been shown in the table.}
    \label{tab:var-imp-4l}
\end{table}

\begin{figure}[h!]
    \centering
    \includegraphics[width=0.7\linewidth]{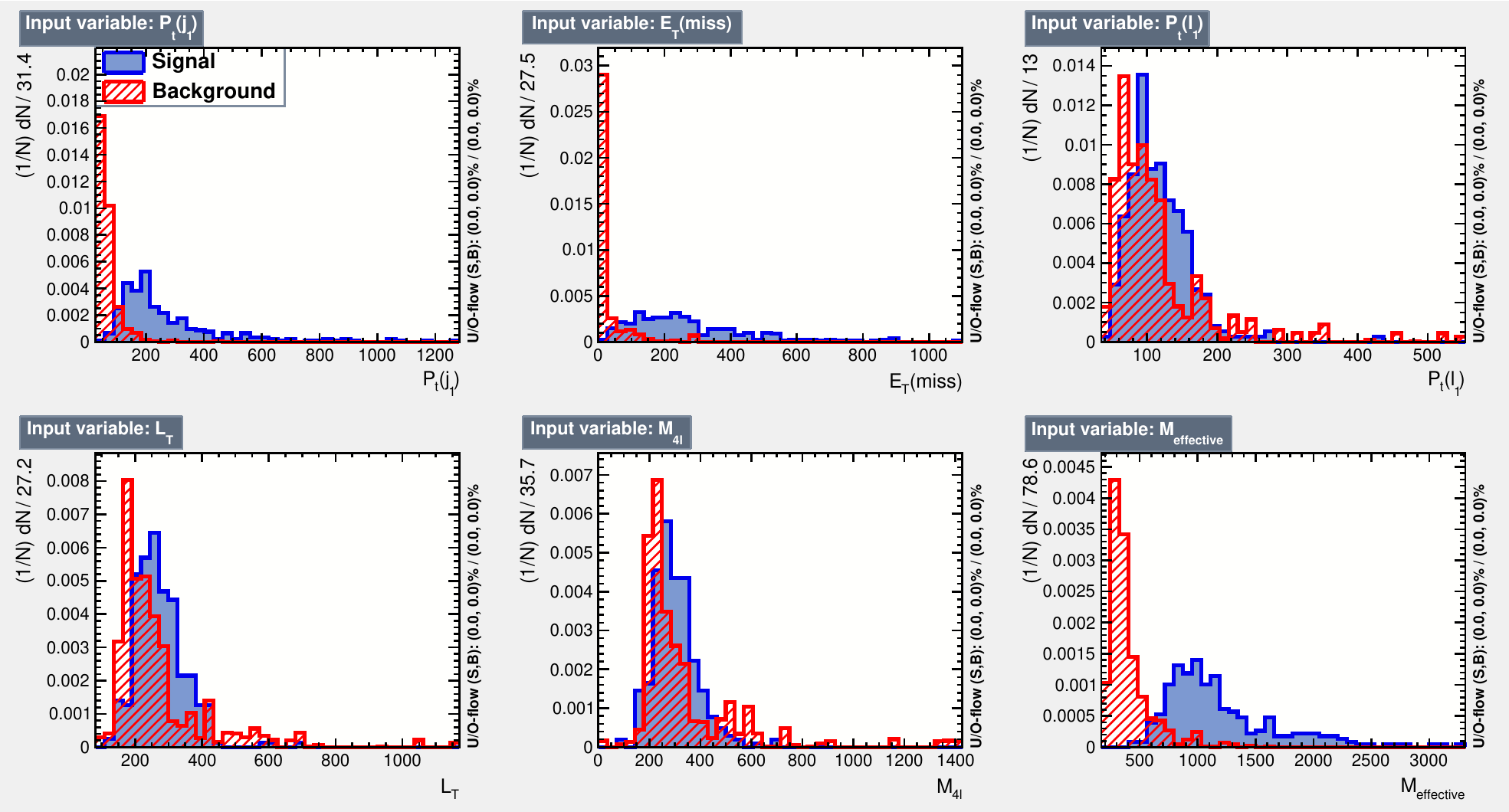}
    \caption{The signal (blue) and background (red) distributions of the input features in the final state of comprising of $4l+jets+\cancel{E_T}$ corresponding to BP2 of the Wino Higgsino mixed type model (WIGGSINO) in Left Light and Right Heavy squarks (LLRH) scenario have been shown in this figure.}
    \label{fig:input4l}
\end{figure}

\begin{figure}[h!]
\begin{minipage}{0.45\textwidth}
    \centering
    \includegraphics[width=0.9\linewidth]{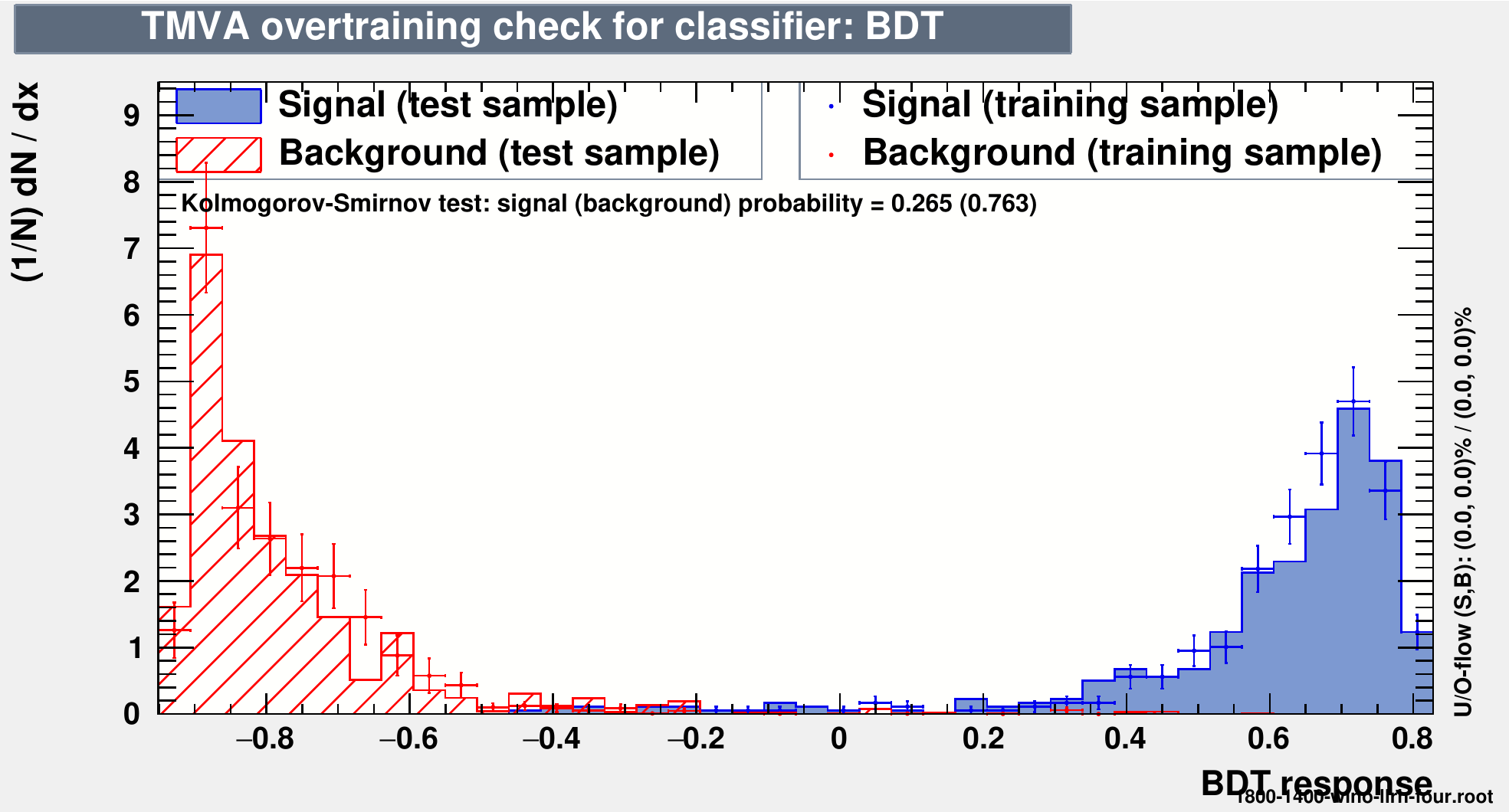}
    \caption{For the purpose of illustration we have presented the over-training check of the BDT response for the  model mentioned in Fig. \ref{fig:input4l} with the final state comprising of $4l+jets+\cancel{E_T}$ signal using the parameter set of BP1.}
    \label{fig:overtrain4l}
\end{minipage}
\hspace{0.8cm}
\begin{minipage}{0.45\textwidth}
    \centering
    \includegraphics[width=0.9\linewidth]{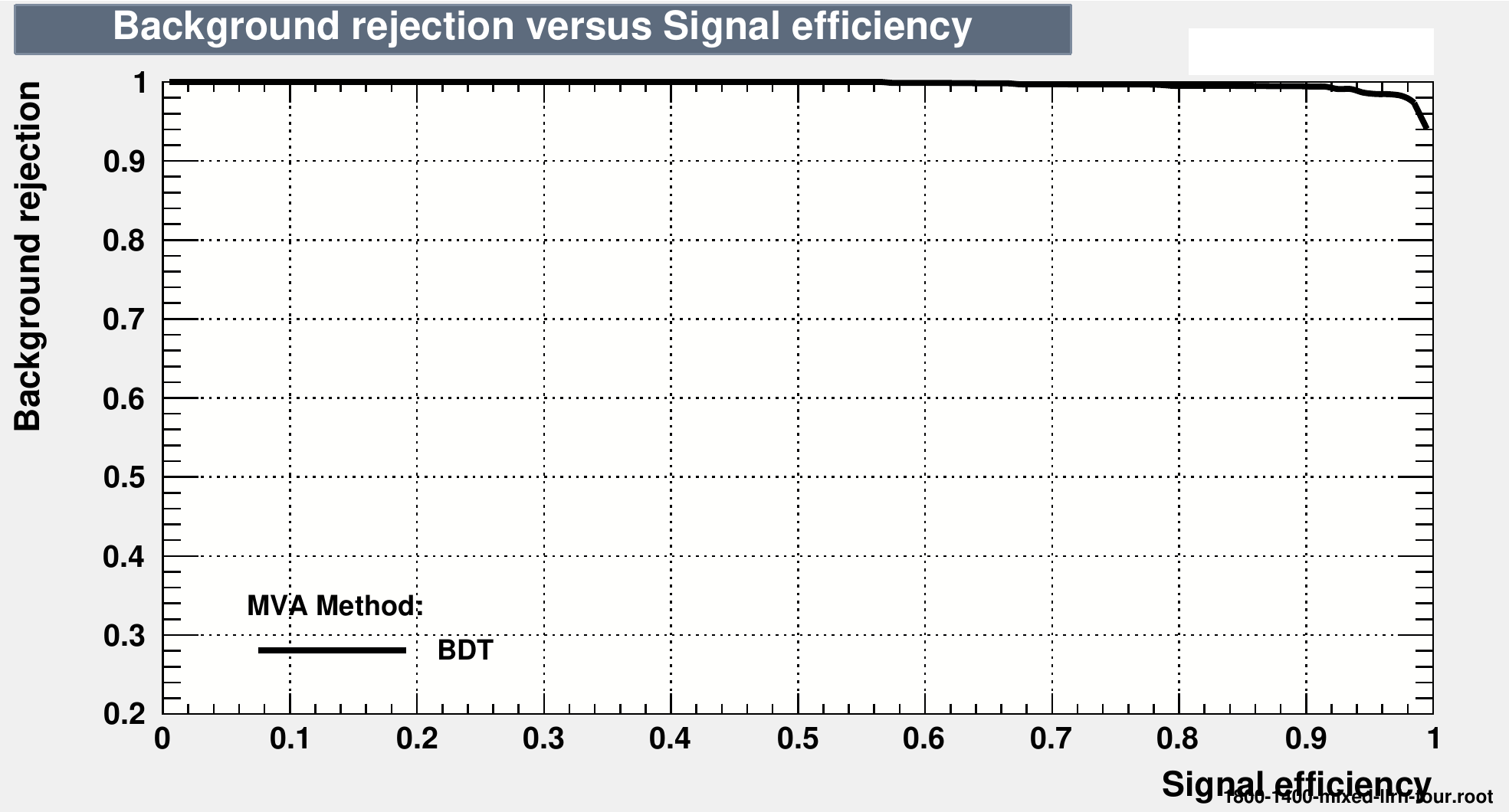}
    \caption{ROC curve corresponding to the model mentioned in Fig. \ref{fig:input4l} for the signal state comprising of $4l+jets+\cancel{E_T}$ have been presented in this figure.}
    \label{fig:roc4l}
\end{minipage}
\end{figure}

The method-specific variable importance have been presented in Table \ref{tab:var-imp-4l} for the five input features mentioned above. Furthermore, in Fig. \ref{fig:input4l} we have shown the distribution of the number of signal and background events against each feature used, with appropriate weightage. KS test is performed to ensure that the classifier is not overtrained. This test compares the response curves of the BDT for the training and testing sub-samples, as shown in Fig. \ref{fig:overtrain4l}. These response curves shows that there is no significant over-training and the classifier is able to effectively differentiate the signal and backgrounds. In addition with this we have presented the ROC curve in Fig. \ref{fig:roc4l} for this decay topology.

Table \ref{tab:sig-4l} shows the number of events for the signal and backgrounds after applying the classifier specific optimised cut value. Two sets of BPs for various pMSSM scenarios are displayed in the table. 

It is noticed from  Table \ref{tab:sig-4l} that BP2 corresponding to LLRL and LLRH of wino like scenarios and LLRH for higgsino like scenario can be probed in the ongoing HL-LHC. In all the cases the backgrounds being vanishingly small, only five signal events are considered to be the discovery criteria. For BP2 in LLRH higgsino type model the required luminosity is 1580 $fb^{-1}$ which is expected to be achieved in the ongoing LHC RUN-III operation.  However, for this particular signal the BP2 corresponding to wino type scenario for LLRH case the required luminosity is the least which is 1350 $fb^{-1}$. Comparing Tables \ref{tab:sig-2l}, \ref{tab:sig-2lss}, \ref{tab:sig-3l} and \ref{tab:sig-4l} we can observe that for probing gluino searches in higgsino type models the required luminosity is least corresponding to $4l+jets+\cancel{E_T}$ signal. It can be seen from the results that the $4l+jets+\cancel{E_T}$ signal significance is smaller compared to the other signal topologies we have considered. This is due to the small BRs of gluinos decaying into the final state comprising of $4l+jets+\cancel{E_T}$ signal in wino type models. The same happens due to the long cascade decay for the other models discussed here.  

\FloatBarrier
\begin{table}[htb]
    \centering
    \resizebox{\textwidth}{!}{
    \begin{tabular}{||c|c|c|c|c|c|c|c|c|c||}
        \hline
        \multirow{2}{*}{BP} & \multirow{2}{*}{Model} & \multicolumn{4}{|c|}{LLRL} & \multicolumn{4}{|c|}{LLRH} \\
         \cline{3-10}
          &  &\makecell[c]{BDT \\cut value} & $N_S$ & $N_B$  & \makecell[c]{Required Luminosity\\ for discovery \\(fb$^{-1}$)} & \makecell[c]{BDT \\cut value} & $N_S$ & $N_B$ & \makecell[c]{Required Luminosity\\ for discovery \\(fb$^{-1}$)} \\
         \hline
         \multirow{3}{*}{BP1} & LELSW & 0.205 & 0.06 & - & $--$ & 0.293 & 0.11 & - & $--$ \\
         \cline{2-10}
         & Wiggsino & 0.303 & 0.06 & - & $--$ & 0.242 & 0.10 & - & $--$ \\
         \cline{2-10}
         & LELSH & 0.235 & 0.08 & - & $--$ & 0.293 & 0.11 & - & $--$ \\
         \hline
         \multirow{2}{*}{BP2} & LELSW & 0.301 & 0.33 & 0.02 & 2580 & 0.248 & 0.90 & 13 & 1350\\
         \cline{2-10}
         & Wiggsino & 0.334 & 0.08 & 0.02 & $--$ & 0.391 & 0.39 & -  & 1780\\
         \cline{2-10}
         & LELSH & 0.628 & 0.13 & - & $--$ & 0.548 & 0.44 & - & 1580 \\
         \hline
    \end{tabular}}
    \caption{Numbers of $4l+jets+\cancel{E_T}$ signal ($N_S$) and the corresponding cumulative background ($N_B$) events after passing the BDT cut normalised to integrated luminosity 139 $fb^{-1}$ with centre of mass energy 13.6 TeV at the LHC have been displayed in the table. In addition with this we have considered 10$\%$ systematic uncertainty  to the overall backgrounds. In this table we have also shown the required luminosity to achieve a potential for discovery. `- ' denotes the negligible background ($N_B\rightarrow0$) compared to the signal. In case the background is negligible 5 signal events are considered as the requirement for the discovery. These numbers are normalized to a luminosity of 139 $fb^{-1}$. `$--$' denotes that the required luminosity is equal or higher than 5000 $fb^{-1}$ and hence the prospect of gluino discovery even in the HL-LHC is not so optimistic for these scenarios.}
    \label{tab:sig-4l}
\end{table}

\section{Conclusion}  \label{sec6}
We explore the possibility of probing the strong sector of the SUSY through gluino searches in different channels with multiple leptons in various pMSSM scenarios at the ongoing LHC RUN-III experiment with center-of-mass energy 13.6 TeV. This study provides the first results on gluino searches through multi-lepton channels on the most recent and, possibly, the highest center-of-mass energy of the LHC so far.

All the BPs considered in this work are consistent with the LHC RUN-II data. We also consider other low energy observables like, the relic density of the SUSY DM (the lightest neutralino) and the anomalous magnetic moment of muon. The BPs lie well within the $4\sigma$ range about the experimentally quoted central value of $(g-2)_{\mu}$. However, in most of the pMSSM scenarios the BPs satisfy the PLANCK/WMAP data. 

In sec. \ref{sec2} we define primarily three different models in the pMSSM framework based on different composition of gaugino and higgsino components in eweakinos. Moreover, those models can be dissected further over the L- and R-type squarks and their mass hierarchy. In addition, slepton masses are set at value less than eweakinos masses. Such mass hierarchies are required to achieve multi-lepton final state from gluino pair production. Instead of conventional multijet and one/two lepton searches for supersymmetric gluino, we try to put forward a proposition where multi-lepton ($\ge 2$ leptons) search can become a potential discovery channel at the LHC RUN-III.

First, we discuss the traditional cut and count analysis (CCA) in sub sec. \ref{sec5.1}  where  the results corresponding to the signal $3l+jets+\cancel{E_T}$ are displayed in Table \ref{tab:cut-3l}.  Since, CCA method does not exhibit a statistically significant surplus of events compared to the SM prediction, we take recourse to MVA technique. In sub sec. \ref{5.2} we present the results of the MVA.


We use MVA technique with BDT as classifier to discriminate signal events over the background events.The important kinematic variables which are chosen as features for discriminating signal and background are shown in Tables \ref{tab:var-imp-2l}, \ref{tab:var-imp-2lss}, \ref{tab:var-imp-3l} and \ref{tab:var-imp-4l}. The ROC curves are also displayed in order to establish the sanctity of the results. The results of the BDT analysis are summarized in Tables \ref{tab:sig-2l}, \ref{tab:sig-2lss}, \ref{tab:sig-3l} and \ref{tab:sig-4l}.

Although we have considered two BPs satisfying the ATLAS data \cite{Aad:2020aze,Aad:2021zyy,ATLAS:2022zwa}, the conclusions drawn in this work through multi lepton channels of gluino searches over the region of parameter space in between those BPs, do not change. The first BP is chosen in such a way that the gluino mass is higher ($\simeq 2.3$ TeV) and the LSP is lighter. This choice allows for a relaxed kinematic phase space including other sparticles in between the gluino and the LSP. However, the gluino pair production cross section, in this case, is relatively smaller due to heavier mass.

On the other hand, the second BP is located in a more compressed region where the mass difference between the gluino and LSP is only about $400$ GeV, with the other sparticles packed in between. In this case, the cross section for gluino pair production is larger compared to the first BP, as the gluino mass, in this case, is much smaller. However, the mass splitting is rather small which results in rather smaller cut efficiency.

In our decay topology, the inclusion of sleptons and other heavier eweakinos lead to a final state rich in leptons, making it observable in the upcoming LHC RUN III. By analyzing these BPs, we can gain insight into the discovery potential of the gluino between these two points. The interplay between the cut efficiency and the cross section makes the study interesting. The result can be generalized for parameter space points in between these two extreme BPs.

 




Some important findings  of this study are abridged below:
\begin{itemize}
    \item The most encouraging result is that there is a possibility to observe $3l+jets+\cancel{E_T}$ signal with $5\sigma$ significance at an integrated luminosity of 270 fb$^{-1}$ at the LHC RUN-III, which can potentially come from a pair of gluinos having $M_{\widetilde{g}} \approx 1.8$ TeV corresponding to wino type model in LLRH case. 

   \item LHC RUN-III also has the potential to discover a gluino of mass $2.3$ TeV in the same wino type model in LLRH case in two OSSF leptons $+jets+\cancel{E_T}$ final state after collecting and analysing 380 fb$^{-1}$ data. 

  \item Same sign dilepton signals are usually considered as the propitious search channels, particularly at hadron collider. However, in the SSSF dilepton$+jets+ \cancel{E_T}$ final state, the LHC RUN-III requires to collect 490 fb$^{-1}$ data to reach the discovery significance for a gluino of mass around 1.8 TeV in the wino type model in LLRH case. 

   \item On the other hand, observing signals in higgsino type models is rather challenging. It requires almost 965 fb$^{-1}$ luminosity for the LHC RUN-III to claim discovery of gluino having mass 1.8 TeV in OSSF dilepton$+jets+ \cancel{E_T}$ final state. 
   \item Final states with lepton multiplicity greater than three, however, reduce the signal significance significantly although the corresponding backgrounds are vanishingly small. For example, 
 $4l+jets+\cancel{E_T}$ signal originating from gluino pair having $M_{\widetilde{g}}\approx 1.8$ TeV corresponding to wino type scenario for LLRH case, the required luminosity is 1350 $fb^{-1}$. The situation for the rest of the scenarios need HL-LHC.   
\end{itemize}

In summary, the LLRH wino-type model in pMSSM offers the most encouraging route to discover gluino, requiring the least integrated luminosity which may be expected to obtain at an early stage of LHC RUN-III. This model resembles the simplified model that the ATLAS collaboration normally examine. The crucial factor is to include the sleptons and all other eweakinos in the gluino decay chain, resulting in an enhancement in the multi-lepton signal in the final state. The analysis carried out in this paper is a generic one and hence can be employed in similar type of BSM scenarios. We hope that following our line of analysis, particularly the MVA technique for such multi-lepton final states, the ongoing run of the LHC shall be able to resolve the long-standing impasse in particle physics.


\section*{Acknowledgement}
 Abhi Mukherjee would like to acknowledge DST for providing the INSPIRE Fellowship[IF170693]. Jyoti Prasad Saha would like to thanks University of Kalyani for providing personal research grant (PRG) for doing research.




    \bibliographystyle{hephys}
	\bibliography{references}

\begin{thebibliography}{100}
\newcommand{\enquote}[1]{``#1''}

\bibitem{Nilles:1983ge}
H.~P. Nilles, \enquote{{Supersymmetry, Supergravity and Particle Physics}},
  \href{http://dx.doi.org/10.1016/0370-1573(84)90008-5}{\emph{Phys. Rept.}
  \textbf{110} (1984) 1}.

\bibitem{bagger1992theory}
J.~Bagger and J.~Wess,
  \href{https://press.princeton.edu/books/paperback/9780691025308/supersymmetry-and-supergravity}{Supersymmetry
  and Supergravity: Revised Ed}, Princeton University Press Mar 23, 1992.

\bibitem{Lykken:1996xt}
J.~D. Lykken, \enquote{{Introduction to supersymmetry}}, \emph{in} {Theoretical
  Advanced Study Institute in Elementary Particle Physics (TASI 96): Fields,
  Strings, and Duality} 1996, \href{http://arxiv.org/abs/hep-th/9612114}{{\tt
  arXiv:hep-th/9612114}}.

\bibitem{Martin:1997ns}
S.~P. Martin, \enquote{{A Supersymmetry primer}},
  \href{http://dx.doi.org/10.1142/9789812839657_0001}{\emph{Adv. Ser. Direct.
  High Energy Phys.} \textbf{18} (1998) 1},
  \href{http://arxiv.org/abs/hep-ph/9709356}{{\tt arXiv:hep-ph/9709356}}.

\bibitem{Chung:2003fi}
D.~J.~H. Chung, L.~L. Everett, G.~L. Kane, S.~F. King, J.~D. Lykken and L.-T.
  Wang, \enquote{{The Soft supersymmetry breaking Lagrangian: Theory and
  applications}},
  \href{http://dx.doi.org/10.1016/j.physrep.2004.08.032}{\emph{Phys. Rept.}
  \textbf{407} (2005) 1}, \href{http://arxiv.org/abs/hep-ph/0312378}{{\tt
  arXiv:hep-ph/0312378}}.

\bibitem{drees2005theory}
M.~Drees, R.~Godbole and P.~Roy,
  \href{https://books.google.co.in/books?id=U3jVCgAAQBAJ}{Theory And
  Phenomenology Of Sparticles: An Account Of Four-dimensional N=1 Supersymmetry
  In High Energy Physics}, World Scientific Publishing Company 2005.

\bibitem{Fayet:1974pd}
P.~Fayet, \enquote{{Supergauge Invariant Extension of the Higgs Mechanism and a
  Model for the electron and Its Neutrino}},
  \href{http://dx.doi.org/10.1016/0550-3213(75)90636-7}{\emph{Nucl. Phys. B}
  \textbf{90} (1975) 104}.

\bibitem{Fayet:1977yz}
P.~Fayet, \enquote{{Introduction to Supersymmetry and Its Applications to
  Particle Interactions}}, \emph{in} {Fundamentals of Microprocessing} 1977.

\bibitem{Fayet:1978qc}
P.~Fayet, \enquote{{MASSIVE GLUINOS}},
  \href{http://dx.doi.org/10.1016/0370-2693(78)90474-4}{\emph{Phys. Lett. B}
  \textbf{78} (1978) 417}.

\bibitem{Fayet:1978rb}
P.~Fayet, \enquote{{Supersymmetric Theories of Particles}},
  \href{http://dx.doi.org/10.1007/978-1-4613-2865-0_17}{\emph{Stud. Nat. Sci.}
  \textbf{14} (1978) 413}.

\bibitem{Baer:2006rs}
H.~Baer and X.~Tata, {Weak scale supersymmetry: From superfields to scattering
  events}, Cambridge University Press 2006.

\bibitem{Gildener:1976ih}
E.~Gildener and S.~Weinberg, \enquote{{Symmetry Breaking and Scalar Bosons}},
  \href{http://dx.doi.org/10.1103/PhysRevD.13.3333}{\emph{Phys. Rev. D}
  \textbf{13} (1976) 3333}.

\bibitem{PhysRevD.14.1667}
E.~Gildener, \enquote{Gauge-symmetry hierarchies},
  \href{http://dx.doi.org/10.1103/PhysRevD.14.1667}{\emph{Phys. Rev. D}
  \textbf{14} (1976) 1667}.

\bibitem{Witten:1981nf}
E.~Witten, \enquote{{Dynamical Breaking of Supersymmetry}},
  \href{http://dx.doi.org/10.1016/0550-3213(81)90006-7}{\emph{Nucl. Phys. B}
  \textbf{188} (1981) 513}.

\bibitem{Dimopoulos:1982af}
S.~Dimopoulos and H.~Georgi, \enquote{{Solution of the Gauge Hierarchy
  Problem}}, \href{http://dx.doi.org/10.1016/0370-2693(82)90720-1}{\emph{Phys.
  Lett. B} \textbf{117} (1982) 287}.

\bibitem{Dimopoulos:1995mi}
S.~Dimopoulos and G.~F. Giudice, \enquote{{Naturalness constraints in
  supersymmetric theories with nonuniversal soft terms}},
  \href{http://dx.doi.org/10.1016/0370-2693(95)00961-J}{\emph{Phys. Lett. B}
  \textbf{357} (1995) 573}, \href{http://arxiv.org/abs/hep-ph/9507282}{{\tt
  arXiv:hep-ph/9507282}}.

\bibitem{Arkani-Hamed:1998jmv}
N.~Arkani-Hamed, S.~Dimopoulos and G.~R. Dvali, \enquote{{The Hierarchy problem
  and new dimensions at a millimeter}},
  \href{http://dx.doi.org/10.1016/S0370-2693(98)00466-3}{\emph{Phys. Lett. B}
  \textbf{429} (1998) 263}, \href{http://arxiv.org/abs/hep-ph/9803315}{{\tt
  arXiv:hep-ph/9803315}}.

\bibitem{Ellis:1990wk}
J.~R. Ellis, S.~Kelley and D.~V. Nanopoulos, \enquote{{Probing the desert using
  gauge coupling unification}},
  \href{http://dx.doi.org/10.1016/0370-2693(91)90980-5}{\emph{Phys. Lett. B}
  \textbf{260} (1991) 131}.

\bibitem{Amaldi:1991cn}
U.~Amaldi, W.~de~Boer and H.~Furstenau, \enquote{{Comparison of grand unified
  theories with electroweak and strong coupling constants measured at LEP}},
  \href{http://dx.doi.org/10.1016/0370-2693(91)91641-8}{\emph{Phys. Lett. B}
  \textbf{260} (1991) 447}.

\bibitem{Jungman:1995df}
G.~Jungman, M.~Kamionkowski and K.~Griest, \enquote{{Supersymmetric dark
  matter}}, \href{http://dx.doi.org/10.1016/0370-1573(95)00058-5}{\emph{Phys.
  Rept.} \textbf{267} (1996) 195},
  \href{http://arxiv.org/abs/hep-ph/9506380}{{\tt arXiv:hep-ph/9506380}}.

\bibitem{Lahanas:2003bh}
A.~B. Lahanas, N.~E. Mavromatos and D.~V. Nanopoulos, \enquote{{WMAPing the
  universe: Supersymmetry, dark matter, dark energy, proton decay and collider
  physics}}, \href{http://dx.doi.org/10.1142/S0218271803004286}{\emph{Int. J.
  Mod. Phys. D} \textbf{12} (2003) 1529},
  \href{http://arxiv.org/abs/hep-ph/0308251}{{\tt arXiv:hep-ph/0308251}}.

\bibitem{Freedman:2003ys}
W.~L. Freedman and M.~S. Turner, \enquote{{Measuring and understanding the
  universe}}, \href{http://dx.doi.org/10.1103/RevModPhys.75.1433}{\emph{Rev.
  Mod. Phys.} \textbf{75} (2003) 1433},
  \href{http://arxiv.org/abs/astro-ph/0308418}{{\tt arXiv:astro-ph/0308418}}.

\bibitem{Roszkowski:2004jc}
L.~Roszkowski, \enquote{{Particle dark matter: A Theorist's perspective}},
  \href{http://dx.doi.org/10.1007/BF02705097}{\emph{Pramana} \textbf{62} (2004)
  389}, \href{http://arxiv.org/abs/hep-ph/0404052}{{\tt arXiv:hep-ph/0404052}}.

\bibitem{Bertone:2004pz}
G.~Bertone, D.~Hooper and J.~Silk, \enquote{{Particle dark matter: Evidence,
  candidates and constraints}},
  \href{http://dx.doi.org/10.1016/j.physrep.2004.08.031}{\emph{Phys. Rept.}
  \textbf{405} (2005) 279}, \href{http://arxiv.org/abs/hep-ph/0404175}{{\tt
  arXiv:hep-ph/0404175}}.

\bibitem{Olive:2005qz}
K.~A. Olive, \enquote{{TASI lectures on astroparticle physics}}, \emph{in}
  {Theoretical Advanced Study Institute in Elementary Particle Physics}:
  {Physics in D $\geqq$ 4} 2005,
  \href{http://arxiv.org/abs/astro-ph/0503065}{{\tt arXiv:astro-ph/0503065}}.

\bibitem{Baer:2008uu}
H.~Baer and X.~Tata, {Dark matter and the LHC} 2009,
  \href{http://arxiv.org/abs/0805.1905}{{\tt arXiv:0805.1905 [hep-ph]}}.

\bibitem{Drees:2012ji}
M.~Drees and G.~Gerbier, \enquote{{Mini-Review of Dark Matter: 2012}},
  \href{http://arxiv.org/abs/1204.2373}{{\tt arXiv:1204.2373 [hep-ph]}}.

\bibitem{Arrenberg:2013rzp}
S.~Arrenberg et~al., \enquote{{Working Group Report: Dark Matter
  Complementarity}}, \emph{in} {Community Summer Study 2013}: {Snowmass on the
  Mississippi} 2013, \href{http://arxiv.org/abs/1310.8621}{{\tt arXiv:1310.8621
  [hep-ph]}}.

\bibitem{aad2014search}
G.~Aad, B.~Abbott, J.~Abdallah, S.~Abdel~Khalek, R.~Aben, B.~Abi, M.~Abolins,
  O.~AbouZeid, H.~Abramowicz, H.~Abreu et~al., \enquote{Search for strong
  production of supersymmetric particles in final states with missing
  transverse momentum and at least three b-jets at $sqrt $\{$s$\}$= 8$ TeV
  proton-proton collisions with the ATLAS detector}, \emph{Journal of High
  Energy Physics} \textbf{2014[10]} (2014) 1.

\bibitem{ATLAS:2014kpx}
G.~Aad et~al. (ATLAS), \enquote{{Search for supersymmetry at $\sqrt{s}$=8 TeV
  in final states with jets and two same-sign leptons or three leptons with the
  ATLAS detector}},
  \href{http://dx.doi.org/10.1007/JHEP06(2014)035}{\emph{JHEP} \textbf{06}
  (2014) 035}, \href{http://arxiv.org/abs/1404.2500}{{\tt arXiv:1404.2500
  [hep-ex]}}.

\bibitem{ATLAS:2014eel}
G.~Aad et~al. (ATLAS), \enquote{{Search for supersymmetry in events with large
  missing transverse momentum, jets, and at least one tau lepton in 20
  fb$^{-1}$ of $\sqrt{s}=$ 8 TeV proton-proton collision data with the ATLAS
  detector}}, \href{http://dx.doi.org/10.1007/JHEP09(2014)103}{\emph{JHEP}
  \textbf{09} (2014) 103}, \href{http://arxiv.org/abs/1407.0603}{{\tt
  arXiv:1407.0603 [hep-ex]}}.

\bibitem{Aad:2013wta}
G.~Aad et~al. (ATLAS), \enquote{{Search for new phenomena in final states with
  large jet multiplicities and missing transverse momentum at $\sqrt{s}$=8 TeV
  proton-proton collisions using the ATLAS experiment}},
  \href{http://dx.doi.org/10.1007/JHEP10(2013)130}{\emph{JHEP} \textbf{10}
  (2013) 130}, [Erratum: JHEP 01, 109 (2014)],
  \href{http://arxiv.org/abs/1308.1841}{{\tt arXiv:1308.1841 [hep-ex]}}.

\bibitem{CMS:2015adc}
V.~Khachatryan et~al. (CMS), \enquote{{Search for Supersymmetry Using Razor
  Variables in Events with $b$-Tagged Jets in $pp$ Collisions at $\sqrt{s} =$ 8
  TeV}}, \href{http://dx.doi.org/10.1103/PhysRevD.91.052018}{\emph{Phys. Rev.
  D} \textbf{91} (2015) 052018}, \href{http://arxiv.org/abs/1502.00300}{{\tt
  arXiv:1502.00300 [hep-ex]}}.

\bibitem{aad2015search}
G.~Aad, B.~Abbott, J.~Abdallah, S.~Abdel~Khalek, R.~Aben, B.~Abi, M.~Abolins,
  O.~AbouZeid, H.~Abramowicz, H.~Abreu et~al., \enquote{Search for squarks and
  gluinos in events with isolated leptons, jets and missing transverse momentum
  at $sqrt $\{$s$\}$= 8$ TeV with the ATLAS detector}, \emph{Journal of High
  Energy Physics} \textbf{2015[4]} (2015) 1.

\bibitem{javurek2016search}
T.~Javurek, A.~Collaboration et~al., \enquote{Search for squarks and gluinos
  with the ATLAS detector in final states with jets and missing transverse
  momentum and 20.3 fb- 1 of s= 8 TeV proton--proton collision data},
  \emph{Nuclear and particle physics proceedings} \textbf{273} (2016) 2418.

\bibitem{Aad:2016eki}
G.~Aad et~al. (ATLAS), \enquote{{Search for pair production of gluinos decaying
  via stop and sbottom in events with $b$-jets and large missing transverse
  momentum in $pp$ collisions at $\sqrt{s} = 13$ TeV with the ATLAS detector}},
  \href{http://dx.doi.org/10.1103/PhysRevD.94.032003}{\emph{Phys. Rev. D}
  \textbf{94[3]} (2016) 032003}, \href{http://arxiv.org/abs/1605.09318}{{\tt
  arXiv:1605.09318 [hep-ex]}}.

\bibitem{Aad:2016qqk}
G.~Aad et~al. (ATLAS), \enquote{{Search for gluinos in events with an isolated
  lepton, jets and missing transverse momentum at $\sqrt{s}$ = 13 TeV with the
  ATLAS detector}},
  \href{http://dx.doi.org/10.1140/epjc/s10052-016-4397-x}{\emph{Eur. Phys. J.
  C} \textbf{76[10]} (2016) 565}, \href{http://arxiv.org/abs/1605.04285}{{\tt
  arXiv:1605.04285 [hep-ex]}}.

\bibitem{Aaboud:2016zdn}
M.~Aaboud et~al. (ATLAS), \enquote{{Search for squarks and gluinos in final
  states with jets and missing transverse momentum at $\sqrt{s} =$ 13 TeV with
  the ATLAS detector}},
  \href{http://dx.doi.org/10.1140/epjc/s10052-016-4184-8}{\emph{Eur. Phys. J.
  C} \textbf{76[7]} (2016) 392}, \href{http://arxiv.org/abs/1605.03814}{{\tt
  arXiv:1605.03814 [hep-ex]}}.

\bibitem{Aad:2016tuk}
G.~Aad et~al. (ATLAS), \enquote{{Search for supersymmetry at $\sqrt{s}=13$ TeV
  in final states with jets and two same-sign leptons or three leptons with the
  ATLAS detector}},
  \href{http://dx.doi.org/10.1140/epjc/s10052-016-4095-8}{\emph{Eur. Phys. J.
  C} \textbf{76[5]} (2016) 259}, \href{http://arxiv.org/abs/1602.09058}{{\tt
  arXiv:1602.09058 [hep-ex]}}.

\bibitem{Aaboud:2017bac}
M.~Aaboud et~al. (ATLAS), \enquote{{Search for squarks and gluinos in events
  with an isolated lepton, jets, and missing transverse momentum at
  $\sqrt{s}=13$ TeV with the ATLAS detector}},
  \href{http://dx.doi.org/10.1103/PhysRevD.96.112010}{\emph{Phys. Rev. D}
  \textbf{96[11]} (2017) 112010}, \href{http://arxiv.org/abs/1708.08232}{{\tt
  arXiv:1708.08232 [hep-ex]}}.

\bibitem{Aaboud:2017ayj}
M.~Aaboud et~al. (ATLAS), \enquote{{Search for a scalar partner of the top
  quark in the jets plus missing transverse momentum final state at
  $\sqrt{s}$=13 TeV with the ATLAS detector}},
  \href{http://dx.doi.org/10.1007/JHEP12(2017)085}{\emph{JHEP} \textbf{12}
  (2017) 085}, \href{http://arxiv.org/abs/1709.04183}{{\tt arXiv:1709.04183
  [hep-ex]}}.

\bibitem{Aaboud:2017dmy}
M.~Aaboud et~al. (ATLAS), \enquote{{Search for supersymmetry in final states
  with two same-sign or three leptons and jets using 36 fb$^{-1}$ of
  $\sqrt{s}=13$ TeV $pp$ collision data with the ATLAS detector}},
  \href{http://dx.doi.org/10.1007/JHEP09(2017)084}{\emph{JHEP} \textbf{09}
  (2017) 084}, [Erratum: JHEP 08, 121 (2019)],
  \href{http://arxiv.org/abs/1706.03731}{{\tt arXiv:1706.03731 [hep-ex]}}.

\bibitem{Aaboud:2017vwy}
M.~Aaboud et~al. (ATLAS), \enquote{{Search for squarks and gluinos in final
  states with jets and missing transverse momentum using 36 fb$^{-1}$ of
  $\sqrt{s}=13$ TeV pp collision data with the ATLAS detector}},
  \href{http://dx.doi.org/10.1103/PhysRevD.97.112001}{\emph{Phys. Rev. D}
  \textbf{97[11]} (2018) 112001}, \href{http://arxiv.org/abs/1712.02332}{{\tt
  arXiv:1712.02332 [hep-ex]}}.

\bibitem{Aaboud:2017hrg}
M.~Aaboud et~al. (ATLAS), \enquote{{Search for supersymmetry in final states
  with missing transverse momentum and multiple $b$-jets in proton-proton
  collisions at $ \sqrt{s}=13 $ TeV with the ATLAS detector}},
  \href{http://dx.doi.org/10.1007/JHEP06(2018)107}{\emph{JHEP} \textbf{06}
  (2018) 107}, \href{http://arxiv.org/abs/1711.01901}{{\tt arXiv:1711.01901
  [hep-ex]}}.

\bibitem{Aaboud:2018ujj}
M.~Aaboud et~al. (ATLAS), \enquote{{Search for new phenomena using the
  invariant mass distribution of same-flavour opposite-sign dilepton pairs in
  events with missing transverse momentum in $\sqrt{s}=13$ $\text {Te}\text
  {V}$ pp collisions with the ATLAS detector}},
  \href{http://dx.doi.org/10.1140/epjc/s10052-018-6081-9}{\emph{Eur. Phys. J.
  C} \textbf{78[8]} (2018) 625}, \href{http://arxiv.org/abs/1805.11381}{{\tt
  arXiv:1805.11381 [hep-ex]}}.

\bibitem{Aaboud:2018mna}
M.~Aaboud et~al. (ATLAS), \enquote{{Search for squarks and gluinos in final
  states with hadronically decaying $\tau$-leptons, jets, and missing
  transverse momentum using $pp$ collisions at $\sqrt{s}$ = 13 TeV with the
  ATLAS detector}},
  \href{http://dx.doi.org/10.1103/PhysRevD.99.012009}{\emph{Phys. Rev. D}
  \textbf{99[1]} (2019) 012009}, \href{http://arxiv.org/abs/1808.06358}{{\tt
  arXiv:1808.06358 [hep-ex]}}.

\bibitem{Aad:2019ftg}
G.~Aad et~al. (ATLAS), \enquote{{Search for squarks and gluinos in final states
  with same-sign leptons and jets using 139 fb$^{-1}$ of data collected with
  the ATLAS detector}},
  \href{http://dx.doi.org/10.1007/JHEP06(2020)046}{\emph{JHEP} \textbf{06}
  (2020) 046}, \href{http://arxiv.org/abs/1909.08457}{{\tt arXiv:1909.08457
  [hep-ex]}}.

\bibitem{Aad:2020nyj}
G.~Aad et~al. (ATLAS), \enquote{{Search for new phenomena in final states with
  large jet multiplicities and missing transverse momentum using $ \sqrt{s} $ =
  13 TeV proton-proton collisions recorded by ATLAS in Run 2 of the LHC}},
  \href{http://dx.doi.org/10.1007/JHEP10(2020)062}{\emph{JHEP} \textbf{10}
  (2020) 062}, \href{http://arxiv.org/abs/2008.06032}{{\tt arXiv:2008.06032
  [hep-ex]}}.

\bibitem{Aad:2021egl}
G.~Aad et~al. (ATLAS), \enquote{{Search for new phenomena in events with an
  energetic jet and missing transverse momentum in $pp$ collisions at $\sqrt{s}
  = 13$ TeV with the ATLAS detector}},
  \href{http://arxiv.org/abs/2102.10874}{{\tt arXiv:2102.10874 [hep-ex]}}.

\bibitem{Aad:2021jmg}
G.~Aad et~al. (ATLAS), \enquote{{Search for new phenomena in final states with
  $b$-jets and missing transverse momentum in $\sqrt{s}=13$ TeV $pp$ collisions
  with the ATLAS detector}}, \href{http://arxiv.org/abs/2101.12527}{{\tt
  arXiv:2101.12527 [hep-ex]}}.

\bibitem{Fayet:1977yc}
P.~Fayet, \enquote{{Spontaneously Broken Supersymmetric Theories of Weak,
  Electromagnetic and Strong Interactions}},
  \href{http://dx.doi.org/10.1016/0370-2693(77)90852-8}{\emph{Phys. Lett. B}
  \textbf{69} (1977) 489}.

\bibitem{Inoue:1983pp}
K.~Inoue, A.~Kakuto, H.~Komatsu and S.~Takeshita, \enquote{{Renormalization of
  Supersymmetry Breaking Parameters Revisited}},
  \href{http://dx.doi.org/10.1143/PTP.71.413}{\emph{Prog. Theor. Phys.}
  \textbf{71} (1984) 413}.

\bibitem{Ibanez:1983di}
L.~E. Ibanez and C.~Lopez, \enquote{{N=1 Supergravity, the Weak Scale and the
  Low-Energy Particle Spectrum}},
  \href{http://dx.doi.org/10.1016/0550-3213(84)90581-9}{\emph{Nucl. Phys. B}
  \textbf{233} (1984) 511}.

\bibitem{Aad:2020aze}
G.~Aad et~al. (ATLAS), \enquote{{Search for squarks and gluinos in final states
  with jets and missing transverse momentum using 139 fb$^{-1}$ of $\sqrt{s}$
  =13 TeV $pp$ collision data with the ATLAS detector}},
  \href{http://dx.doi.org/10.1007/JHEP02(2021)143}{\emph{JHEP} \textbf{02}
  (2021) 143}, \href{http://arxiv.org/abs/2010.14293}{{\tt arXiv:2010.14293
  [hep-ex]}}.

\bibitem{Aad:2021zyy}
G.~Aad et~al. (ATLAS), \enquote{{Search for squarks and gluinos in final states
  with one isolated lepton, jets, and missing transverse momentum at
  $\sqrt{s}=13$ TeV with the ATLAS detector}},
  \href{http://arxiv.org/abs/2101.01629}{{\tt arXiv:2101.01629 [hep-ex]}}.

\bibitem{ATLAS:2022zwa}
\enquote{{Searches for new phenomena in events with two leptons, jets, and
  missing transverse momentum in $139~\text{fb}^{-1}$ of $\sqrt{s}=13~$TeV $pp$
  collisions with the ATLAS detector}},
  \href{http://arxiv.org/abs/2204.13072}{{\tt arXiv:2204.13072 [hep-ex]}}.

\bibitem{Mukherjee:2022kff}
A.~Mukherjee, S.~Niyogi and S.~Poddar, \enquote{{Revisiting the gluino mass
  limits in the pMSSM in the light of the latest LHC data and Dark Matter
  constraints}}, \href{http://arxiv.org/abs/2201.02531}{{\tt arXiv:2201.02531
  [hep-ph]}}.

\bibitem{LeCompte:2011fh}
T.~J. LeCompte and S.~P. Martin, \enquote{{Compressed supersymmetry after 1/fb
  at the Large Hadron Collider}},
  \href{http://dx.doi.org/10.1103/PhysRevD.85.035023}{\emph{Phys. Rev. D}
  \textbf{85} (2012) 035023}, \href{http://arxiv.org/abs/1111.6897}{{\tt
  arXiv:1111.6897 [hep-ph]}}.

\bibitem{Dreiner:2012gx}
H.~K. Dreiner, M.~Kramer and J.~Tattersall, \enquote{{How low can SUSY go?
  Matching, monojets and compressed spectra}},
  \href{http://dx.doi.org/10.1209/0295-5075/99/61001}{\emph{EPL} \textbf{99[6]}
  (2012) 61001}, \href{http://arxiv.org/abs/1207.1613}{{\tt arXiv:1207.1613
  [hep-ph]}}.

\bibitem{Bhattacherjee:2012mz}
B.~Bhattacherjee and K.~Ghosh, \enquote{{Degenerate SUSY search at the 8 TeV
  LHC}}, \href{http://arxiv.org/abs/1207.6289}{{\tt arXiv:1207.6289 [hep-ph]}}.

\bibitem{Bhattacherjee:2013wna}
B.~Bhattacherjee, A.~Choudhury, K.~Ghosh and S.~Poddar, \enquote{{Compressed
  supersymmetry at 14 TeV LHC}},
  \href{http://dx.doi.org/10.1103/PhysRevD.89.037702}{\emph{Phys. Rev. D}
  \textbf{89[3]} (2014) 037702}, \href{http://arxiv.org/abs/1308.1526}{{\tt
  arXiv:1308.1526 [hep-ph]}}.

\bibitem{Cohen:2013xda}
T.~Cohen, T.~Golling, M.~Hance, A.~Henrichs, K.~Howe, J.~Loyal, S.~Padhi and
  J.~G. Wacker, \enquote{{SUSY Simplified Models at 14, 33, and 100 TeV Proton
  Colliders}}, \href{http://dx.doi.org/10.1007/JHEP04(2014)117}{\emph{JHEP}
  \textbf{04} (2014) 117}, \href{http://arxiv.org/abs/1311.6480}{{\tt
  arXiv:1311.6480 [hep-ph]}}.

\bibitem{Mukhopadhyay:2014dsa}
S.~Mukhopadhyay, M.~M. Nojiri and T.~T. Yanagida, \enquote{{Compressed SUSY
  search at the 13 TeV LHC using kinematic correlations and structure of ISR
  jets}}, \href{http://dx.doi.org/10.1007/JHEP10(2014)012}{\emph{JHEP}
  \textbf{10} (2014) 012}, \href{http://arxiv.org/abs/1403.6028}{{\tt
  arXiv:1403.6028 [hep-ph]}}.

\bibitem{Low:2014cba}
M.~Low and L.-T. Wang, \enquote{{Neutralino dark matter at 14 TeV and 100
  TeV}}, \href{http://dx.doi.org/10.1007/JHEP08(2014)161}{\emph{JHEP}
  \textbf{08} (2014) 161}, \href{http://arxiv.org/abs/1404.0682}{{\tt
  arXiv:1404.0682 [hep-ph]}}.

\bibitem{Chakraborti:2017vxz}
M.~Chakraborti, A.~Datta, N.~Ganguly and S.~Poddar, \enquote{{Multilepton
  signals of heavier electroweakinos at the LHC}},
  \href{http://dx.doi.org/10.1007/JHEP11(2017)117}{\emph{JHEP} \textbf{11}
  (2017) 117}, \href{http://arxiv.org/abs/1707.04410}{{\tt arXiv:1707.04410
  [hep-ph]}}.

\bibitem{Chakraborti:2014gea}
M.~Chakraborti, U.~Chattopadhyay, A.~Choudhury, A.~Datta and S.~Poddar,
  \enquote{{The Electroweak Sector of the pMSSM in the Light of LHC - 8 TeV and
  Other Data}}, \href{http://dx.doi.org/10.1007/JHEP07(2014)019}{\emph{JHEP}
  \textbf{07} (2014) 019}, \href{http://arxiv.org/abs/1404.4841}{{\tt
  arXiv:1404.4841 [hep-ph]}}.

\bibitem{Chakraborti:2015mra}
M.~Chakraborti, U.~Chattopadhyay, A.~Choudhury, A.~Datta and S.~Poddar,
  \enquote{{Reduced LHC constraints for higgsino-like heavier
  electroweakinos}},
  \href{http://dx.doi.org/10.1007/JHEP11(2015)050}{\emph{JHEP} \textbf{11}
  (2015) 050}, \href{http://arxiv.org/abs/1507.01395}{{\tt arXiv:1507.01395
  [hep-ph]}}.

\bibitem{Datta:2016ypd}
A.~Datta, N.~Ganguly and S.~Poddar, \enquote{{New Limits on Heavier
  Electroweakinos and their LHC Signatures}},
  \href{http://dx.doi.org/10.1016/j.physletb.2016.10.034}{\emph{Phys. Lett. B}
  \textbf{763} (2016) 213}, \href{http://arxiv.org/abs/1606.04391}{{\tt
  arXiv:1606.04391 [hep-ph]}}.

\bibitem{Datta:2018lup}
A.~Datta and N.~Ganguly, \enquote{{The past, present and future of the heavier
  electroweakinos in the light of LHC and other data}},
  \href{http://dx.doi.org/10.1007/JHEP01(2019)103}{\emph{JHEP} \textbf{01}
  (2019) 103}, \href{http://arxiv.org/abs/1809.05129}{{\tt arXiv:1809.05129
  [hep-ph]}}.

\bibitem{Adam:2021rrw}
W.~Adam and I.~Vivarelli, \enquote{{Status of searches for electroweak-scale
  supersymmetry after LHC Run 2}},
  \href{http://dx.doi.org/10.1142/S0217751X21300222}{\emph{Int. J. Mod. Phys.
  A} \textbf{37[02]} (2022) 2130022},
  \href{http://arxiv.org/abs/2111.10180}{{\tt arXiv:2111.10180 [hep-ex]}}.

\bibitem{Aad:2014vma}
G.~Aad et~al. (ATLAS), \enquote{{Search for direct production of charginos,
  neutralinos and sleptons in final states with two leptons and missing
  transverse momentum in $pp$ collisions at $\sqrt{s} =$ 8 TeV with the ATLAS
  detector}}, \href{http://dx.doi.org/10.1007/JHEP05(2014)071}{\emph{JHEP}
  \textbf{05} (2014) 071}, \href{http://arxiv.org/abs/1403.5294}{{\tt
  arXiv:1403.5294 [hep-ex]}}.

\bibitem{Aad:2014nua}
G.~Aad et~al. (ATLAS), \enquote{{Search for direct production of charginos and
  neutralinos in events with three leptons and missing transverse momentum in
  $\sqrt{s} =$ 8TeV $pp$ collisions with the ATLAS detector}},
  \href{http://dx.doi.org/10.1007/JHEP04(2014)169}{\emph{JHEP} \textbf{04}
  (2014) 169}, \href{http://arxiv.org/abs/1402.7029}{{\tt arXiv:1402.7029
  [hep-ex]}}.

\bibitem{Aad:2014yka}
G.~Aad et~al. (ATLAS), \enquote{{Search for the direct production of charginos,
  neutralinos and staus in final states with at least two hadronically decaying
  taus and missing transverse momentum in $pp$ collisions at $\sqrt{s}$ = 8 TeV
  with the ATLAS detector}},
  \href{http://dx.doi.org/10.1007/JHEP10(2014)096}{\emph{JHEP} \textbf{10}
  (2014) 096}, \href{http://arxiv.org/abs/1407.0350}{{\tt arXiv:1407.0350
  [hep-ex]}}.

\bibitem{Aad:2015jqa}
G.~Aad et~al. (ATLAS), \enquote{{Search for direct pair production of a
  chargino and a neutralino decaying to the 125 GeV Higgs boson in $\sqrt{s} =
  8$ TeV ${pp}$ collisions with the ATLAS detector}},
  \href{http://dx.doi.org/10.1140/epjc/s10052-015-3408-7}{\emph{Eur. Phys. J.
  C} \textbf{75[5]} (2015) 208}, \href{http://arxiv.org/abs/1501.07110}{{\tt
  arXiv:1501.07110 [hep-ex]}}.

\bibitem{Sabatta:2019nfg}
D.~Sabatta, A.~S. Cornell, A.~Goyal, M.~Kumar, B.~Mellado and X.~Ruan,
  \enquote{{Connecting muon anomalous magnetic moment and multi-lepton
  anomalies at LHC}},
  \href{http://dx.doi.org/10.1088/1674-1137/44/6/063103}{\emph{Chin. Phys. C}
  \textbf{44[6]} (2020) 063103}, \href{http://arxiv.org/abs/1909.03969}{{\tt
  arXiv:1909.03969 [hep-ph]}}.

\bibitem{Buddenbrock:2019tua}
S.~Buddenbrock, A.~S. Cornell, Y.~Fang, A.~Fadol~Mohammed, M.~Kumar, B.~Mellado
  and K.~G. Tomiwa, \enquote{{The emergence of multi-lepton anomalies at the
  LHC and their compatibility with new physics at the EW scale}},
  \href{http://dx.doi.org/10.1007/JHEP10(2019)157}{\emph{JHEP} \textbf{10}
  (2019) 157}, \href{http://arxiv.org/abs/1901.05300}{{\tt arXiv:1901.05300
  [hep-ph]}}.

\bibitem{Hernandez:2019geu}
Y.~Hernandez, M.~Kumar, A.~S. Cornell, S.-E. Dahbi, Y.~Fang, B.~Lieberman,
  B.~Mellado, K.~Monnakgotla, X.~Ruan and S.~Xin, \enquote{{The anomalous
  production of multi-lepton and its impact on the measurement of $Wh$
  production at the LHC}},
  \href{http://dx.doi.org/10.1140/epjc/s10052-021-09137-1}{\emph{Eur. Phys. J.
  C} \textbf{81[4]} (2021) 365}, \href{http://arxiv.org/abs/1912.00699}{{\tt
  arXiv:1912.00699 [hep-ph]}}.

\bibitem{Baer:1986au}
H.~Baer, V.~D. Barger, D.~Karatas and X.~Tata, \enquote{{Detecting Gluinos at
  Hadron Supercolliders}},
  \href{http://dx.doi.org/10.1103/PhysRevD.36.96}{\emph{Phys. Rev. D}
  \textbf{36} (1987) 96}.

\bibitem{Barnett:1987kn}
R.~M. Barnett, J.~F. Gunion and H.~E. Haber, \enquote{{Gluino Decay Patterns
  and Signatures}},
  \href{http://dx.doi.org/10.1103/PhysRevD.37.1892}{\emph{Phys. Rev. D}
  \textbf{37} (1988) 1892}.

\bibitem{Baer:1995va}
H.~Baer, C.-h. Chen, F.~Paige and X.~Tata, \enquote{{Signals for minimal
  supergravity at the CERN large hadron collider. 2: Multi - lepton channels}},
  \href{http://dx.doi.org/10.1103/PhysRevD.53.6241}{\emph{Phys. Rev. D}
  \textbf{53} (1996) 6241}, \href{http://arxiv.org/abs/hep-ph/9512383}{{\tt
  arXiv:hep-ph/9512383}}.

\bibitem{Baer:2003wx}
H.~Baer, C.~Balazs, A.~Belyaev, T.~Krupovnickas and X.~Tata, \enquote{{Updated
  reach of the CERN LHC and constraints from relic density, $b \to s \gamma$
  and a($\mu$) in the mSUGRA model}},
  \href{http://dx.doi.org/10.1088/1126-6708/2003/06/054}{\emph{JHEP}
  \textbf{06} (2003) 054}, \href{http://arxiv.org/abs/hep-ph/0304303}{{\tt
  arXiv:hep-ph/0304303}}.

\bibitem{Sakai:1981gr}
N.~Sakai, \enquote{{Naturalness in Supersymmetric Guts}},
  \href{http://dx.doi.org/10.1007/BF01573998}{\emph{Z. Phys. C} \textbf{11}
  (1981) 153}.

\bibitem{Kaul:1981hi}
R.~K. Kaul and P.~Majumdar, \enquote{{Cancellation of Quadratically Divergent
  Mass Corrections in Globally Supersymmetric Spontaneously Broken Gauge
  Theories}}, \href{http://dx.doi.org/10.1016/0550-3213(82)90565-X}{\emph{Nucl.
  Phys. B} \textbf{199} (1982) 36}.

\bibitem{Barbieri:1987fn}
R.~Barbieri and G.~F. Giudice, \enquote{{Upper Bounds on Supersymmetric
  Particle Masses}},
  \href{http://dx.doi.org/10.1016/0550-3213(88)90171-X}{\emph{Nucl. Phys. B}
  \textbf{306} (1988) 63}.

\bibitem{Feng:1999mn}
J.~L. Feng, K.~T. Matchev and T.~Moroi, \enquote{{Multi - TeV scalars are
  natural in minimal supergravity}},
  \href{http://dx.doi.org/10.1103/PhysRevLett.84.2322}{\emph{Phys. Rev. Lett.}
  \textbf{84} (2000) 2322}, \href{http://arxiv.org/abs/hep-ph/9908309}{{\tt
  arXiv:hep-ph/9908309}}.

\bibitem{Arkani-Hamed:2006wnf}
N.~Arkani-Hamed, A.~Delgado and G.~F. Giudice, \enquote{{The Well-tempered
  neutralino}},
  \href{http://dx.doi.org/10.1016/j.nuclphysb.2006.02.010}{\emph{Nucl. Phys. B}
  \textbf{741} (2006) 108}, \href{http://arxiv.org/abs/hep-ph/0601041}{{\tt
  arXiv:hep-ph/0601041}}.

\bibitem{GAMBIT:2018gjo}
P.~Athron et~al. (GAMBIT), \enquote{{Combined collider constraints on
  neutralinos and charginos}},
  \href{http://dx.doi.org/10.1140/epjc/s10052-019-6837-x}{\emph{Eur. Phys. J.
  C} \textbf{79[5]} (2019) 395}, \href{http://arxiv.org/abs/1809.02097}{{\tt
  arXiv:1809.02097 [hep-ph]}}.

\bibitem{Huang:2018xle}
Y.~Huang, G.~E. Addison, J.~L. Weiland and C.~L. Bennett, \enquote{{Assessing
  Consistency Between WMAP 9-year and Planck 2015 Temperature Power Spectra}},
  \href{http://dx.doi.org/10.3847/1538-4357/aaeb1f}{\emph{Astrophys. J.}
  \textbf{869[1]} (2018) 38}, \href{http://arxiv.org/abs/1804.05428}{{\tt
  arXiv:1804.05428 [astro-ph.CO]}}.

\bibitem{Planck:2018vyg}
N.~Aghanim et~al. (Planck), \enquote{{Planck 2018 results. VI. Cosmological
  parameters}},
  \href{http://dx.doi.org/10.1051/0004-6361/201833910}{\emph{Astron.
  Astrophys.} \textbf{641} (2020) A6}, [Erratum: Astron.Astrophys. 652, C4
  (2021)], \href{http://arxiv.org/abs/1807.06209}{{\tt arXiv:1807.06209
  [astro-ph.CO]}}.

\bibitem{Chakraborti:2021dli}
M.~Chakraborti, S.~Heinemeyer and I.~Saha, \enquote{{The new
  \textquotedblleft{}MUON G-2\textquotedblright{} result and supersymmetry}},
  \href{http://dx.doi.org/10.1140/epjc/s10052-021-09900-4}{\emph{Eur. Phys. J.
  C} \textbf{81[12]} (2021) 1114}, \href{http://arxiv.org/abs/2104.03287}{{\tt
  arXiv:2104.03287 [hep-ph]}}.

\bibitem{Chakraborti:2022sbj}
M.~Chakraborti, S.~Heinemeyer and I.~Saha, \enquote{{SUSY Dark Matter Direct
  Detection Prospects Based on $\boldsymbol{(g-2)}_{\boldsymbol{\mu}}$}},
  \href{http://dx.doi.org/10.3103/S0027134922020412}{\emph{Moscow Univ. Phys.
  Bull.} \textbf{77[2]} (2022) 116},
  \href{http://arxiv.org/abs/2201.03390}{{\tt arXiv:2201.03390 [hep-ph]}}.

\bibitem{Chakraborti:2022vds}
M.~Chakraborti, S.~Iwamoto, J.~S. Kim, R.~Mase\l{}ek and K.~Sakurai,
  \enquote{{Supersymmetric explanation of the muon g \textendash{} 2 anomaly
  with and without stable neutralino}},
  \href{http://dx.doi.org/10.1007/JHEP08(2022)124}{\emph{JHEP} \textbf{08}
  (2022) 124}, \href{http://arxiv.org/abs/2202.12928}{{\tt arXiv:2202.12928
  [hep-ph]}}.

\bibitem{Chakraborti:2022wii}
M.~Chakraborti, S.~Heinemeyer and I.~Saha, \enquote{{$(g-2) \mu$ and SUSY}},
  \href{http://dx.doi.org/10.1142/S0217751X22460101}{\emph{Int. J. Mod. Phys.
  A} \textbf{37[30]} (2022) 2246010}.

\bibitem{Muong-2:2021ojo}
B.~Abi et~al. (Muon g-2), \enquote{{Measurement of the Positive Muon Anomalous
  Magnetic Moment to 0.46 ppm}},
  \href{http://dx.doi.org/10.1103/PhysRevLett.126.141801}{\emph{Phys. Rev.
  Lett.} \textbf{126[14]} (2021) 141801},
  \href{http://arxiv.org/abs/2104.03281}{{\tt arXiv:2104.03281 [hep-ex]}}.

\bibitem{Kosower:1983yw}
D.~A. Kosower, L.~M. Krauss and N.~Sakai, \enquote{{Low-Energy Supergravity and
  the Anomalous Magnetic Moment of the Muon}},
  \href{http://dx.doi.org/10.1016/0370-2693(83)90152-1}{\emph{Phys. Lett. B}
  \textbf{133} (1983) 305}.

\bibitem{Yuan:1984ww}
T.~C. Yuan, R.~L. Arnowitt, A.~H. Chamseddine and P.~Nath,
  \enquote{{Supersymmetric Electroweak Effects on G-2 (mu)}},
  \href{http://dx.doi.org/10.1007/BF01452567}{\emph{Z. Phys. C} \textbf{26}
  (1984) 407}.

\bibitem{ATLAS:2019lff}
G.~Aad et~al. (ATLAS), \enquote{{Search for electroweak production of charginos
  and sleptons decaying into final states with two leptons and missing
  transverse momentum in $\sqrt{s}=13$ TeV $pp$ collisions using the ATLAS
  detector}},
  \href{http://dx.doi.org/10.1140/epjc/s10052-019-7594-6}{\emph{Eur. Phys. J.
  C} \textbf{80[2]} (2020) 123}, \href{http://arxiv.org/abs/1908.08215}{{\tt
  arXiv:1908.08215 [hep-ex]}}.

\bibitem{ATLAS:2021moa}
G.~Aad et~al. (ATLAS), \enquote{{Search for chargino\textendash{}neutralino
  pair production in final states with three leptons and missing transverse
  momentum in $\sqrt{s} = 13$~TeV pp collisions with the ATLAS detector}},
  \href{http://dx.doi.org/10.1140/epjc/s10052-021-09749-7}{\emph{Eur. Phys. J.
  C} \textbf{81[12]} (2021) 1118}, \href{http://arxiv.org/abs/2106.01676}{{\tt
  arXiv:2106.01676 [hep-ex]}}.

\bibitem{ATLAS:2012yve}
G.~Aad et~al. (ATLAS), \enquote{{Observation of a new particle in the search
  for the Standard Model Higgs boson with the ATLAS detector at the LHC}},
  \href{http://dx.doi.org/10.1016/j.physletb.2012.08.020}{\emph{Phys. Lett. B}
  \textbf{716} (2012) 1}, \href{http://arxiv.org/abs/1207.7214}{{\tt
  arXiv:1207.7214 [hep-ex]}}.

\bibitem{CMS:2012qbp}
S.~Chatrchyan et~al. (CMS), \enquote{{Observation of a New Boson at a Mass of
  125 GeV with the CMS Experiment at the LHC}},
  \href{http://dx.doi.org/10.1016/j.physletb.2012.08.021}{\emph{Phys. Lett. B}
  \textbf{716} (2012) 30}, \href{http://arxiv.org/abs/1207.7235}{{\tt
  arXiv:1207.7235 [hep-ex]}}.

\bibitem{Degrassi:2002fi}
G.~Degrassi, S.~Heinemeyer, W.~Hollik, P.~Slavich and G.~Weiglein,
  \enquote{{Towards high precision predictions for the MSSM Higgs sector}},
  \href{http://dx.doi.org/10.1140/epjc/s2003-01152-2}{\emph{Eur. Phys. J. C}
  \textbf{28} (2003) 133}, \href{http://arxiv.org/abs/hep-ph/0212020}{{\tt
  arXiv:hep-ph/0212020}}.

\bibitem{Allanach:2004rh}
B.~Allanach, A.~Djouadi, J.~Kneur, W.~Porod and P.~Slavich, \enquote{{Precise
  determination of the neutral Higgs boson masses in the MSSM}},
  \href{http://dx.doi.org/10.1088/1126-6708/2004/09/044}{\emph{JHEP}
  \textbf{09} (2004) 044}, \href{http://arxiv.org/abs/hep-ph/0406166}{{\tt
  arXiv:hep-ph/0406166}}.

\bibitem{Wong:2019kwg}
K.~C. Wong et~al., \enquote{{H0LiCOW \textendash{} XIII. A 2.4 per cent
  measurement of H0 from lensed quasars: 5.3\ensuremath{\sigma} tension between
  early- and late-Universe probes}},
  \href{http://dx.doi.org/10.1093/mnras/stz3094}{\emph{Mon. Not. Roy. Astron.
  Soc.} \textbf{498[1]} (2020) 1420},
  \href{http://arxiv.org/abs/1907.04869}{{\tt arXiv:1907.04869 [astro-ph.CO]}}.

\bibitem{Djouadi:2006bz}
A.~Djouadi, M.~Muhlleitner and M.~Spira, \enquote{{Decays of supersymmetric
  particles: The Program SUSY-HIT (SUspect-SdecaY-Hdecay-InTerface)}},
  \emph{Acta Phys. Polon. B} \textbf{38} (2007) 635,
  \href{http://arxiv.org/abs/hep-ph/0609292}{{\tt arXiv:hep-ph/0609292}}.

\bibitem{Belanger:2020gnr}
G.~Belanger, A.~Mjallal and A.~Pukhov, \enquote{{Recasting direct detection
  limits within micrOMEGAs and implication for non-standard Dark Matter
  scenarios}},
  \href{http://dx.doi.org/10.1140/epjc/s10052-021-09012-z}{\emph{Eur. Phys. J.
  C} \textbf{81[3]} (2021) 239}, \href{http://arxiv.org/abs/2003.08621}{{\tt
  arXiv:2003.08621 [hep-ph]}}.

\bibitem{Alwall:2014hca}
J.~Alwall, R.~Frederix, S.~Frixione, V.~Hirschi, F.~Maltoni, O.~Mattelaer,
  H.~S. Shao, T.~Stelzer, P.~Torrielli and M.~Zaro, \enquote{{The automated
  computation of tree-level and next-to-leading order differential cross
  sections, and their matching to parton shower simulations}},
  \href{http://dx.doi.org/10.1007/JHEP07(2014)079}{\emph{JHEP} \textbf{07}
  (2014) 079}, \href{http://arxiv.org/abs/1405.0301}{{\tt arXiv:1405.0301
  [hep-ph]}}.

\bibitem{Ball:2012cx}
R.~D. Ball et~al., \enquote{{Parton distributions with LHC data}},
  \href{http://dx.doi.org/10.1016/j.nuclphysb.2012.10.003}{\emph{Nucl. Phys. B}
  \textbf{867} (2013) 244}, \href{http://arxiv.org/abs/1207.1303}{{\tt
  arXiv:1207.1303 [hep-ph]}}.

\bibitem{Plehn:2004rp}
T.~Plehn, \enquote{{Measuring the MSSM Lagrangean}}, \emph{Czech. J. Phys.}
  \textbf{55} (2005) B213, \href{http://arxiv.org/abs/hep-ph/0410063}{{\tt
  arXiv:hep-ph/0410063}}.

\bibitem{Sjostrand:2014zea}
T.~Sj\"ostrand, S.~Ask, J.~R. Christiansen, R.~Corke, N.~Desai, P.~Ilten,
  S.~Mrenna, S.~Prestel, C.~O. Rasmussen and P.~Z. Skands, \enquote{{An
  introduction to PYTHIA 8.2}},
  \href{http://dx.doi.org/10.1016/j.cpc.2015.01.024}{\emph{Comput. Phys.
  Commun.} \textbf{191} (2015) 159}, \href{http://arxiv.org/abs/1410.3012}{{\tt
  arXiv:1410.3012 [hep-ph]}}.

\bibitem{deFavereau:2013fsa}
J.~de~Favereau, C.~Delaere, P.~Demin, A.~Giammanco, V.~Lema\^\i{}tre,
  A.~Mertens and M.~Selvaggi (DELPHES 3), \enquote{{DELPHES 3, A modular
  framework for fast simulation of a generic collider experiment}},
  \href{http://dx.doi.org/10.1007/JHEP02(2014)057}{\emph{JHEP} \textbf{02}
  (2014) 057}, \href{http://arxiv.org/abs/1307.6346}{{\tt arXiv:1307.6346
  [hep-ex]}}.

\bibitem{Cacciari:2011ma}
M.~Cacciari, G.~P. Salam and G.~Soyez, \enquote{{FastJet User Manual}},
  \href{http://dx.doi.org/10.1140/epjc/s10052-012-1896-2}{\emph{Eur. Phys. J.
  C} \textbf{72} (2012) 1896}, \href{http://arxiv.org/abs/1111.6097}{{\tt
  arXiv:1111.6097 [hep-ph]}}.

\bibitem{Voss:2007jxm}
H.~Voss, A.~Hocker, J.~Stelzer and F.~Tegenfeldt, \enquote{{TMVA, the Toolkit
  for Multivariate Data Analysis with ROOT}},
  \href{http://dx.doi.org/10.22323/1.050.0040}{\emph{PoS} \textbf{ACAT} (2007)
  040}.

\bibitem{Antcheva:2009zz}
I.~Antcheva et~al., \enquote{{ROOT: A C++ framework for petabyte data storage,
  statistical analysis and visualization}},
  \href{http://dx.doi.org/10.1016/j.cpc.2009.08.005}{\emph{Comput. Phys.
  Commun.} \textbf{180} (2009) 2499},
  \href{http://arxiv.org/abs/1508.07749}{{\tt arXiv:1508.07749
  [physics.data-an]}}.

\bibitem{Cowan:2010js}
G.~Cowan, K.~Cranmer, E.~Gross and O.~Vitells, \enquote{{Asymptotic formulae
  for likelihood-based tests of new physics}},
  \href{http://dx.doi.org/10.1140/epjc/s10052-011-1554-0}{\emph{Eur. Phys. J.
  C} \textbf{71} (2011) 1554}, [Erratum: Eur.Phys.J.C 73, 2501 (2013)],
  \href{http://arxiv.org/abs/1007.1727}{{\tt arXiv:1007.1727
  [physics.data-an]}}.

\bibitem{Cousins:2007yta}
R.~D. Cousins, J.~T. Linnemann and J.~Tucker, \enquote{{Evaluation of three
  methods for calculating statistical significance when incorporating a
  systematic uncertainty into a test of the background-only hypothesis for a
  Poisson process}},
  \href{http://dx.doi.org/10.1016/j.nima.2008.07.086}{\emph{Nucl. Instrum.
  Meth. A} \textbf{595[2]} (2008) 480},
  \href{http://arxiv.org/abs/physics/0702156}{{\tt arXiv:physics/0702156}}.

\bibitem{Li:1983fv}
T.~P. Li and Y.~Q. Ma, \enquote{{Analysis methods for results in gamma-ray
  astronomy}}, \href{http://dx.doi.org/10.1086/161295}{\emph{Astrophys. J.}
  \textbf{272} (1983) 317}.

\bibitem{Baer:1991xs}
H.~Baer, X.~Tata and J.~Woodside, \enquote{{Multi - lepton signals from
  supersymmetry at hadron super colliders}},
  \href{http://dx.doi.org/10.1103/PhysRevD.45.142}{\emph{Phys. Rev. D}
  \textbf{45} (1992) 142}.

\bibitem{Guchait:1994zk}
M.~Guchait and D.~P. Roy, \enquote{{Like sign dilepton signature for gluino
  production at CERN LHC including top quark and Higgs boson effects}},
  \href{http://dx.doi.org/10.1103/PhysRevD.52.133}{\emph{Phys. Rev. D}
  \textbf{52} (1995) 133}, \href{http://arxiv.org/abs/hep-ph/9412329}{{\tt
  arXiv:hep-ph/9412329}}.

\bibitem{Bartl:1996cg}
A.~Bartl, W.~Porod, F.~de~Campos, M.~A. Garcia-Jareno, M.~B. Magro, J.~W.~F.
  Valle and W.~Majerotto, \enquote{{Signatures of spontaneous breaking of
  R-parity in gluino cascade decays at LHC}},
  \href{http://dx.doi.org/10.1016/S0550-3213(97)00434-3}{\emph{Nucl. Phys. B}
  \textbf{502} (1997) 19}, \href{http://arxiv.org/abs/hep-ph/9612436}{{\tt
  arXiv:hep-ph/9612436}}.

\bibitem{Sokolov:1998yx}
A.~A. Sokolov, T.~Lindblad, C.~S. Lindsey, A.~Vanyashin and J.~Waldemark,
  \enquote{{Identification of high mass gluino production in p p interactions
  using a neural network}},
  \href{http://dx.doi.org/10.1016/S0168-9002(98)00616-0}{\emph{Nucl. Instrum.
  Meth. A} \textbf{416} (1998) 391}.

\bibitem{Alves:2006df}
A.~Alves, O.~Eboli and T.~Plehn, \enquote{{It's a gluino}},
  \href{http://dx.doi.org/10.1103/PhysRevD.74.095010}{\emph{Phys. Rev. D}
  \textbf{74} (2006) 095010}, \href{http://arxiv.org/abs/hep-ph/0605067}{{\tt
  arXiv:hep-ph/0605067}}.

\bibitem{Kim:2008bnd}
Y.~G. Kim, \enquote{{SUSY at the LHC}},
  \href{http://dx.doi.org/10.24532/soken.116.1_A77}{\emph{Soryushiron Kenkyu
  Electron.} \textbf{116} (2008) A77}.

\bibitem{Baer:2008kc}
H.~Baer, H.~Prosper and H.~Summy, \enquote{{Early SUSY discovery at CERN LHC
  without missing E(T): The Role of multi-leptons}},
  \href{http://dx.doi.org/10.1103/PhysRevD.77.055017}{\emph{Phys. Rev. D}
  \textbf{77} (2008) 055017}, \href{http://arxiv.org/abs/0801.3799}{{\tt
  arXiv:0801.3799 [hep-ph]}}.

\bibitem{Baer:2008ey}
H.~Baer, A.~Lessa and H.~Summy, \enquote{{Early SUSY discovery at LHC via
  sparticle cascade decays to same-sign and multimuon states}},
  \href{http://dx.doi.org/10.1016/j.physletb.2009.03.002}{\emph{Phys. Lett. B}
  \textbf{674} (2009) 49}, \href{http://arxiv.org/abs/0809.4719}{{\tt
  arXiv:0809.4719 [hep-ph]}}.

\bibitem{Acharya:2009gb}
B.~S. Acharya, P.~Grajek, G.~L. Kane, E.~Kuflik, K.~Suruliz and L.-T. Wang,
  \enquote{{Identifying Multi-Top Events from Gluino Decay at the LHC}},
  \href{http://arxiv.org/abs/0901.3367}{{\tt arXiv:0901.3367 [hep-ph]}}.

\bibitem{Andreev:2009dz}
Y.~M. Andreev, N.~V. Krasnikov and A.~N. Toropin, \enquote{{The MSSM with Large
  Gluino Mass}}, \href{http://dx.doi.org/10.1142/S0217732309030771}{\emph{Mod.
  Phys. Lett. A} \textbf{24} (2009) 1317},
  \href{http://arxiv.org/abs/0902.0502}{{\tt arXiv:0902.0502 [hep-ph]}}.

\bibitem{Turlay:2010bk}
E.~Turlay, R.~Lafaye, T.~Plehn, M.~Rauch and D.~Zerwas, \enquote{{Measuring
  Supersymmetry with Heavy Scalars}},
  \href{http://dx.doi.org/10.1088/0954-3899/38/3/035003}{\emph{J. Phys. G}
  \textbf{38} (2011) 035003}, \href{http://arxiv.org/abs/1011.0759}{{\tt
  arXiv:1011.0759 [hep-ph]}}.

\bibitem{Reuter:2012ng}
J.~Reuter and D.~Wiesler, \enquote{{A Fat Gluino in Disguise}},
  \href{http://dx.doi.org/10.1140/epjc/s10052-013-2355-4}{\emph{Eur. Phys. J.
  C} \textbf{73[3]} (2013) 2355}, \href{http://arxiv.org/abs/1212.5559}{{\tt
  arXiv:1212.5559 [hep-ph]}}.

\end{thebibliography}

\end{document}